## Technische Universität Dresden

# Generalization of a reduced Trefftz type approach

Sulaiman Abo Diab

Faculty of Civil Engineering
Tishreen University
Lattakia
Syria

supported by the "Georg Forster Research Fellowship" of the
Alexander von Humboldt Foundation, number IV-1-1721-SYR/1069911

Host Institute: Department of Civil Engineering,
Institute of Structural Analysis, Dresden University of Technology, Germany

Lehrstuhl für Statik
Technische Universität Dresden
Mommsenstr. 13
D-01062 Dresden
Germany

Telephon: 0049 351 / 4633 43 86
Fax: 0049 351 / 4633 70 86
e-mail: moeller@rcs.urz.tu-dresden.de

**©2001** **Sulaiman Abo Diab**
Hussain Al Bahr
Tartous
Syria
Telephon: 00963 43 / 600 866
e-mail: sabodiab@scs-net.org

سليمان يوسف أبودياب
سوريا طرطوس
حصين البحر

# Contents

## Overview

This work presents variational concepts associated with reduced Trefftz type approaches and discusses the interrelationship between various concepts of the displacement, hybrid and Trefftz methods. The basic concept of the displacement version of the reduced Treffetz method operates on the natural boundary conditions enforced in an integral form whereas the stress version of the reduced Trefftz type approach operates on the essential boundary conditions enforced in an integral sense. The application of the method proposed in the framework of the finite element method is briefly outlined. The methods used by the reduced Trefftz type approach for enforcing conformity and inter-element continuity between neighboured elements are also discussed. Comparisons with other known methods for the same purpose are performed. General procedure for developing finite elements of general geometric form such as quadrilateral elements with invariance properties is presented. The basic idea of this procedure consists in using the natural coordinate system only for defining the element geometry and performing the element integration in the biunit interval. For defining the approximation functions a local coordinate system defined from the directions of the covariant base vectors and the perpendicular contravariant base vectors computed in the geometric centre of the element is used. This procedure can also be used to implement other versions of finite elements and other forms of finite elements. Different sets of numerical calculations and comparisons in the linear statics and kinetics are performed in order to assess the convergence criteria and the numerical performance of finite elements developed by applying the reduced Trefftz type approach.

## Aim of the study

The aim of the work is to introduce in detail the reduced Trefftz type approach and its application in the framework of the finite element method. The classification of the introduced method under variational methods in approximation and its relation to the variational displacement and hybrid concepts should be studied.
A general procedure for formatting finite elements of general geometric form is to be briefly introduced and discussed.
Numerical performance of finite element family developed by applying the reduced Trefftz type approach should be compared with analytical solutions and other well-known finite elements.

## Acknowledgement

The present work is supported by the “Georg Forster Research Fellowship” of the Alexander von Humboldt Foundation, number IV-1-1721-SYR/1069911 at the host institute: Department of Civil Engineering, Institute of Structural Analysis, Dresden University of Technology, Dresden Germany.
This support is gratefully acknowledged.

# 1 Introduction

A survey of the most used conventional and extended variational principles in elasticity and plasticity are involved in /1/ to /6/. The basic idea of the variational formulation is based on the fundamental lemma of calculus of variation. The application starts with approximation functions which fulfil exactly a set of the fundamental equations and / or boundary conditions of the problem under consideration. The remaining part of them is fulfilled approximately through integration over the domain and / or boundary.
The trends of using approximation functions with relaxed continuity requirement dominated for more than thirty years /7/, /8/.

Recently, more attention was paid to variational formulations which apply approximation functions with strict constraints such as Trefftz method /9/. Although the application of the Trefftz finite element method was thoroughly developed just in the same period of developing the hybrid and mixed methods (only few years in delay), the method has not become extensively and widely use. The achievement in this field was internationally demonstrated only in rare events /10/, /11/.

Trefftz published his method for solving boundary value problems in the year 1926 as a counterpart to the Ritz method. The basic idea of his method is to approximate the boundary conditions rather than the Lagrangian equation. The earliest contributions concerning application of the Trefftz method in numerical analysis were introduced by Loof /12/, Stein /13, 14/, Ruof /15/, Desmukh /16/, Quinlan /17/, Tong, Pian /18/, Zienkiewicz, Kelly and Bettess /19/. A practical finite element analysis for solving problems in the theory of elasticity using the Trefftz concept was introduced by Jirousek /20/. The mathematical background for applying Trefftz functions to solve boundary value problems and to investigate their completeness was achieved by Herrera and co-workers in the following few years /21-25/. The research works /26-34/ of Jirousek between (1982) and (1997) has demonstrated the superiority of the displacement and hybrid Trefftz finite element method compared with the conventional displacement and hybrid methods. Numerous researchers have applied the Jirousek strategy for developing finite elements of the Trefftz type, such as Zielinski /35, 36/, Piltner /37/, Peters /38/ and others. Some of the element developments are based on extended variational principles /39-41/. The state of recent developments in Trefftz finite element application was reported at the first international workshop on the Trefftz method in Cracow, Poland (1996). Another workshop about the Trefftz-method was held in Sintra, Portugal (1999).

The reduced Trefftz-type approach with a basic concept that operates on the natural boundary conditions enforced in an integral form has already been introduced in /42-44/ and thoroughly developed in /45-49/. The trial functions used are chosen so as to satisfy *a priori* the governing differential equations of the problem under consideration. However, the concept used here differs from the Trefftz method in the way the essential boundary conditions are approximated. Like some formulations which apply the Trefftz method, the essential boundary conditions are used in the approximation at the finite element level. But what distinguishes the approach proposed here from its counterparts is the technique used to enforce conformity and inter-element continuity. The essential boundary conditions at the finite element level are applied in form of a “modified geometrical interpolation technique” to construct ‘interpolation’ functions over an element, which approximate the internal displacement field. The constructed field contains both homogeneous shape functions and particular shape functions. The added particular shape functions depend on the geometry and the loading of the element and are explicitly linked to the homogeneous shape functions through the interpolation technique used. Two procedures are also presented to enforce conformity and inter-element continuity. The ‘frame function’ concept is used in both cases. The first procedure is hybrid and consists simply in applying the Gauss divergence theorem for the stress variation at the finite element level, after imposing the essential boundary conditions of the element and imposing the equations of equilibrium in order to eliminate the undetermined parameters. The second procedure is boundary equivalent. This results from the application of the identity of virtual work over the entire boundary and the virtual work over the kinematic and static boundaries.

The displacement concept of the reduced Trefftz type approach explained above recovers the conventional displacement and hybrid stress finite element concept. Recently, another concept of the reduced Trefftz type approach, namely the stress concept, is introduced in /45/. This concept operates on the essential boundary conditions enforced in an integral sense. It recovers the stress and hybrid displacement finite element concept. The element approximation basis can be constructed in an analogous way used by applying the displacement formulation in form of stress components.

A combination of the modified interpolation technique and the use of the Gauss divergence theorem or the equivalency of virtual work is possible.

The obstacles in formatting finite elements of general geometric form considering invariance, convergence and insensitivity to nodal point numbering are reported in many publications /50-55/. In addition to these difficulties encountered in the application of the usual displacement and hybrid version of the finite element approach the requirement of the reduced Trefftz type approach on the element approximation basis involves the satisfaction of the Lagrangian equation of the problem under consideration. Such requirements make the approach nearly inapplicable when the approximation basis is selected in a natural coordinate system.

In order to apply the reduced Trefftz type approach in the framework of the finite element method it is possible to select the approximation functions in an element local Cartesian coordinate system defined using the directions of the covariant base vectors and the perpendicular contravariant base vectors of the element. This facilitates considerably the satisfaction of the Lagrangian equation. We can facilitate the integration procedure by transforming the approximation functions and their derivatives to a natural coordinate system using the isoparametric transformation and after that performing the element integrals in these coordinates. The origin of the local coordinate system is located at the element centre in order to ensure element invariance.

This work deals with the theoretical background of the reduced Trefftz-type approach and lists some numerical results produced recently by the 12-DOF quadrilateral thin-plate bending elements and plain stress elements developed on its basis. The work also reports the formatting of such elements so that they satisfy the convergence requirement and describes the derivation technique used to circumvent the difficulties associated with the generalization of the 12-DOF non-conforming thin-plate bending elements. Such difficulties (instability, frame invariance and sensitivity to nodal point numbering) have been reported in many publications, see for example /50-55/.

Section 2 starts with global survey of the fundamental equations of the problem to be solved and the variational concepts can be used to deal with the described problem. Two main concepts of the reduced Trefftz type approach are recovered; the displacement version and the stress version. The interrelationship between these versions and other known displacement and stress versions are discussed.

In section 3 different basic ideas for constructing the finite element approximation basis are introduced. All approximation basis use the parametric form for approximating and constructing the function basis applied in the finite element approach. The undetermined parameters are eliminated using one of three introduced procedures:

In the first procedure a “modified geometrical interpolation technique” for linking the undetermined parameters with the nodal degrees of freedom of the element is adopted. The second procedure makes use of the Gauss divergence theorem and the equilibrium equation at the finite element level for this purpose. In the third procedure, the identity of virtual work over the entire boundary and the kinematic and static boundaries is used.

In section 4 a general procedure for formatting finite elements of general geometric form is suggested. The basic idea of this procedure consists in using the natural coordinate system only for defining the element geometry and performing the element integration in the biunit interval. For defining the approximation functions a local coordinate system defined from the directions of the covariant basis vectors and the perpendicular contravariant basis vectors is used. This procedure can also be used to implement other versions of finite elements and other forms of finite elements. The application is shown on displacement, hybrid and boundary quadrilateral finite elements for plate bending and plain stress problems.

Numerical calculations and comparisons in the linear static and kinetic of circular, rhombic and trapezoidal structures are performed in section 5 and 6 to asses convergence, invariance and insensitivity to nodal point numbering and numerical performance of the finite elements derived by applying the reduced Trefftz type approach.

# 2 Reduced Trefftz type approach

## 2.1 Fundamental equations

Consider a linear problem in the elastodynamics described by the following fundamental equations:
Equation of motion (dynamic field equations)

$$\sigma^{ij}{}_{,j} + \bar{f}^i - \rho\, \ddot{u}^i = 0 \quad \text{in} \quad V \tag{2.1}$$

Kinematic field equation

$$\varepsilon_{ij} = \frac{1}{2}(u_{i,j} + u_{j,i}) \qquad \text{in} \quad V \tag{2.2}$$

Constitutive equation

$$\sigma^{ij} = c^{ijkl}\varepsilon_{kl} \qquad \text{in} \quad V \tag{2.3}$$

and the specified boundary conditions
Kinematic boundary conditions (essential boundary conditions)

$$u_i = \bar{u}_i \quad \text{on } s_u \subseteq s \tag{2.4}$$

Mechanical boundary conditions (natural boundary conditions)

$$\sigma^{ij} n_j = \bar{T}^i \quad \text{in } s_\sigma \quad s / s_u \tag{2.5}$$

with the prescribed displacements $\bar{u}_i$ and the prescribed boundary tractions $\bar{T}^i$, where $n_j$ denotes the components of the outward normal vector to the boundary.
Initial boundary conditions

$$u_i(t = t_0) = u_0 \tag{2.6}$$

$$\dot{u}_i(t = t_0) = \dot{u}_0 \tag{2.7}$$

In the equations above, $u_i$ are the displacement functions, $\varepsilon_{ij}$ is the strain tensor, $\sigma^{ij}$ is the stress tensor and $\bar{f}^i$, $\rho\ddot{u}^i$ are the body and inertia forces, respectively. $\sigma^{ij} n_j$ denotes the boundary tractions on the part of boundary $s_\sigma$ where the forces $\bar{T}^i$ are prescribed.
$\bar{u}_i$ are the prescribed displacements on the part of boundary $s_u$ where the displacement are prescribed. V denotes the domain volume and s the domain surface. $u_i(t = t_0)$ and $\dot{u}_i(t = t_0)$ are the initial displacement and velocity at the intial time point $t_0$ of the vibration process.
The quantities $\sigma^{ij}$, $\varepsilon_{ij}$, $u_i$, $\bar{f}^i$, $T^i$ are functions of space coordinates and of time t.

## 2.2 Displacement concept

Suppose that the variation $\delta u_i$ of the actual displacement field $u_i$ satisfies at any time point the following equations:

Governing differential equations
Kinematic boundary conditions
Initial boundary conditions

Then the variational expression must lead to an approximation of the natural (mechanical) boundary conditions between two fixed time points of the vibration process

$$\int_{t_1}^{t_2} \{ \int_{s_\sigma} (\sigma^{ij} n_i - \overline{T}^i) \delta u_i \, ds \} dt = 0 \tag{2.8}$$

By introducing the first surface integral in (2.8) in form of equation (2.9) and observing the essential boundary conditions $\delta u_i = \delta \overline{u}_i = 0$ on $s_u$, we obtain the extended variational basis (2.10) of the natural boundary conditions

$$\int_{s_\sigma} \sigma^{ij} n_j \, \delta u_i \, ds = \int_{s} \sigma^{ij} n_j \, \delta u_i \, ds - \int_{s_u} \sigma^{ij} n_j \, \delta u_i \, ds \tag{2.9}$$

$$\delta I = \int_{t_1}^{t_2} \{ \int_{s} \sigma^{ij} n_j \, \delta u_i \, ds - \int_{s_\sigma} \overline{T}^i \, \delta u_i \, ds \} dt = 0 \tag{2.10}$$

By introducing some modifications to (2.10) and applying partial integration it is possible to change over to the most hybrid and displacement philosophy. For instance, the variational basis (2.13) may be obtained by applying the partial integration (2.11) with an imposed Gauss divergence theorem for the displacement variation and observing the integral form of equation of motion (2.12) as well as the essential boundary conditions.

$$\int_{s} \sigma^{ij} n_j \, \delta u_i \, ds = \int_{V} \sigma^{ij}{}_{,j} \, \delta u_i \, dV - \int_{V} \sigma^{ij} \, \delta u_{i,j} \, dV \tag{2.11}$$

$$\int_{V} \sigma^{ij}{}_{,j} \, \delta u_i \, dV = \int_{V} (-\overline{f}^i + \rho \, \ddot{u}^i) \, \delta u_i \, dV \tag{2.12}$$

$$\int_{t_1}^{t_2} \{ \int_{V} \rho \ddot{u}^i \delta u_i dV + \int_{V} \sigma^{ij} \, \delta \varepsilon_{ij} dV - \int_{V} \overline{f}^i \delta u_i \, dV - \int_{s_\sigma} \overline{T}^i \delta u_i \, ds \} dt = 0 \tag{2.13}$$

The last expression is mathematically similar to the principle of virtual work but distinct in what concerns the subsidiary conditions of the variation. In (2.13) $\delta u_i$ are kinematically admissible and the $\delta\sigma^{ij}$ associated with $\delta u_i$ satisfy the dynamic field equation in the entire domain V.

Another possible transformation of (2.10) is also obtained by introducing the modification (2.14) and replacing the surface integral of the stress variation (last term of (2.14)) with its domain equivalents after making use of the variation of equations of motion.

$$\int_s \sigma^{ij} n_i \delta u_i \, ds = \delta \int_s \sigma^{ij} n_j u_i \, ds - \int_s (\delta \sigma^{ij} n_j) u_i \, ds \tag{2.14}$$

This yields the result (2.15)

$$\int_{t_1}^{t_2} \{ \int_V \rho \delta \ddot{u}^i u_i dV + \int_V \varepsilon_{ij} \, \delta\sigma^{ij} \, dV - \int_{s_u} \bar{u}_i \, \delta\sigma^{ij} n_j ds - \delta \int_{s_\sigma} u_i (\sigma^{ij} n_j - \bar{T}^i) ds \} dt = 0 \tag{2.15}$$

This expression is mathematically similar to the modified principle of complementary virtual work but distinct in what concerns the subsidiary conditions of the variation. In (2.15) $\delta\sigma^{ij}$ satisfy the dynamic field equation and $\delta u_i$ associated with $\delta\sigma^{ij}$ are kinematically admissible. Moreover, $u_i$ on $s_\sigma$ are not selected independently.

## 2.3 Stress concept

Suppose that the variation $\delta\sigma^{ij}$ of the actual stress field $\sigma^{ij}$ satisfies at any time point the following equations:

Governing differential equations
Mechanical boundary conditions
Initial boundary conditions (including stress initial conditions)

In such assumptions, all equations of the elasticity problem are satisfied except that of the kinematic boundary conditions. The variational expression (2.16) leads to a satisfaction of these conditions approximately between two fixed time points of the vibration process according to the fundamental lemma of calculus of variation

$$\int_{t_1}^{t_2} \{ \int_{s_u} (u_i - \bar{u}_i) \delta\sigma^{ij} n_j \, ds \} dt = 0 \tag{2.16}$$

This variational basis can be transformed in an analogous way used for transforming the displacement concept. By introducing the first surface integral in equation (2.16) in form of equation (2.17) we obtain by observing the natural boundary conditions the extended variational basis (2.18) of the essential boundary conditions

$$\int_{s_u} u_i \, \delta\sigma^{ij} n_j \, ds = \int_s u_i \, \delta\sigma^{ij} n_j \, ds - \int_{s_\sigma} u_i \, \delta\sigma^{ij} n_j \, ds \tag{2.17}$$

$$\int_{t_1}^{t_2} \{ \int_s (u_i \, \delta\sigma^{ij} n_j \, ds - \int_{s_u} \bar{u}_i \, \delta\sigma^{ij} n_j \, ds \} dt = 0 \tag{2.18}$$

By applying the partial integration to the first term of equation (2.18) and using the Gaus divergence theorem for the stress variation in form of equation (2.19) we obtain, noting the variation of equation of motions and the kinematic field equations, the variational basis (2.20)

$$\int_s (\delta\sigma^{ij} n_j)\, u_i\, ds = \int_V (\delta\, \sigma^{ij}{}_{,j}) u_i\, dV - \int_V (\delta\sigma^{ij})\ \ u_{i,j}\, dV \tag{2.19}$$

$$\int_{t_1}^{t_2} \{ \int_V \rho\, \delta\ddot{u}^i\, u_i dV + \int_V \varepsilon_{ij}\, \delta\sigma^{ij}\, dV - \int_{s_u} \bar{u}_i\, \delta\sigma^{ij}\, n_j ds \} dt = 0 \tag{2.20}$$

The expression (2.20) is mathematically similar to the principle of complementary virtual work but distinct in what concerns the subsidiary conditions of the variation. In it, $\delta\sigma^{ij}$ satisfy the equation of motion. They are, also, associated with $\delta u_i$. $\delta\sigma^{ij}$ on $s_\sigma$ satisfy the mechanical boundary conditions.

In a similar way we can introduce the modification (2.21) in equation (2.18) to obtain (2.22) by observing the natural boundary conditions.

$$\int_s (\delta\sigma^{ij} n_i) u_i\, ds = \delta \int_s \sigma^{ij} n_j\, u_i\, ds - \int_s (\sigma^{ij} n_j) \delta u_i\, ds \tag{2.21}$$

$$\int_{t_1}^{t_2} \{\ \delta \int_s (u_i\ \ \sigma^{ij}\, n_j\, ds - \int_s (\sigma^{ij} n_j) \delta u_i\, ds - \int_{s_u} \bar{u}_i\ \ \delta\sigma^{ij}\, n_j\, ds\ \} dt = 0 \tag{2.22}$$

The last equation may also be transformed to the form (2.23) by replacing the second surface integral by its domain equivalents. This leads after some manipulations to the following variational expression.

$$\int_{t_1}^{t_2} \{ \int_V \rho\, \ddot{u}^i\ \ \delta u_i dV + \int_V \sigma^{ij}\, \delta\varepsilon_{ij}\ \ dV - \int_V \bar{f}^i \delta u_i\, dV - \int_{s_\sigma} \bar{T}^i \delta u_i\, ds + \delta \int_{s_u} \sigma^{ij} n_j (u_i - \bar{u}_i) ds \} dt = 0 \tag{2.23}$$

The expression (2.23) is mathematically similar to the modified principle of virtual work but distinct in what concerns the subsidiary conditions of the variation. In it, $\delta\sigma^{ij}$ satisfy the dynamic field equation and $\delta u_i$ associated with $\delta\sigma^{ij}$ are kinematically admissible in the entire domain. Moreover, $\sigma^{ij}$ on $s_u$ are not selected independently.

## 2.4. Interrelation between displacement, stress, hybrid and Trefftz concepts

As may be seen, both displacement and stress approach can be applied as a boundary method in form of (2.10), (2.18) or as a domain and mixed method in form of (2.13), (2.20) and (2.15), (2.23) following a computational technique similar to that of the conventional displacement and hybrid finite element approach. It is expected, also, that the different computational techniques yield the same results provided that the conditions assumed in the variation are strictly observed.

In table 2.1 all these variational concepts are listed beside the Trefftz-concept and the interrelationship between them is explained.

As known, the original Trefftz method exploits trial approximating functions with the property that they satisfied exactly the governing differential equations within the domain defining the problem (for the problem defined in section 2.1 the governing differential equations can be derived formally by eliminating the strain tensor from the constitutive equation (2.3) with the aid of the kinematic field equation (2.2) and substituting the result into the dynamic field equation (2.1)). The trial approximating functions did not need to satisfy neither the essential boundary conditions nor the natural boundary conditions. These conditions are satisfied approximately in an integral sense using the fundamental lemma of calculus of variation by the following variational statement of the Trefftz method

$$\int_{t_1}^{t_2} \Big\{ \int_{s_\sigma} (\sigma^{ij} n_i - \overline{T}^i)\,\delta u_i \,ds + \int_{s_u} (u_i - \overline{u}_i)\,\delta\sigma^{ij} n_j \,ds \Big\}\, dt = 0 \qquad (2.24)$$

It is obvious from table 2.1 that the variational statement of the displacement version of the reduced Trefftz type approach is recovered from the variational statement of the Trefftz approach when the Trefftz trial functions are so chosen as to satisfy the essential boundary conditions. On the other hand the variational statement of the Trefftz approach reduces to that associated with the stress version of the reduced Trefftz type approach when the trial functions are so chosen as to satisfy the natural boundary conditions of the problem under consideration.

Tab 2.1: Interrelationship between Trefftz-concept, reduced Trefftz-concept, displacement, stress and hybrid concepts

$$\int_{t_1}^{t_2} \Big\{ \int_{s_\sigma} (\sigma^{ij} n_i - \overline{T}^i)\,\delta u_i \,ds + \int_{s_u} (u_i - \overline{u}_i)\,\delta\sigma^{ij} n_j \,ds \Big\}\, dt = 0$$

$$\int_{t_1}^{t_2} \Big\{ \int_{s_\sigma} (\sigma^{ij} n_i - \overline{T}^i)\,\delta u_i \,ds \Big\}\, dt = 0 \qquad\qquad \int_{t_1}^{t_2} \Big\{ \int_{s_u} (u_i - \overline{u}_i)\,\delta\sigma^{ij} n_j \,ds \Big\}\, dt = 0$$

$$\int_{t_1}^{t_2} \Big\{ \int_V \rho\, \ddot{u}^i\, \delta u_i \,dV + \int_V \sigma^{ij}\, \delta\varepsilon_{ij} \,dV - \int_V \overline{f}^i \delta u_i \,dV - \int_{s_\sigma} \overline{T}^i \delta u_i \,ds \Big\} dt = 0$$

$$\int_{t_1}^{t_2} \Big\{ \int_V \rho\, \delta\ddot{u}^i\, u_i \,dV + \int_V \varepsilon_{ij}\, \delta\sigma^{ij} \,dV - \int_{s_u} \overline{u}_i\, \delta\sigma^{ij} n_j \,ds \Big\} dt = 0$$

$$\int_{t_1}^{t_2} \Big\{ \int_V \rho\, \delta\ddot{u}^i\, u_i \,dV + \int_V \varepsilon_{ij}\, \delta\sigma^{ij} \,dV - \int_{s_u} \overline{u}_i\, \delta\sigma^{ij} n_j \,ds - \delta \int_{s_\sigma} u_i (\sigma^{ij} n_j - \overline{T}^i)\,ds \Big\} dt = 0$$

Displacement concept

$$\int_{t_1}^{t_2} \Big\{ \int_V \rho\, \ddot{u}^i\, \delta u_i \,dV + \int_V \sigma^{ij}\, \delta\varepsilon_{ij} \,dV - \int_V \overline{f}^i \delta u_i \,dV - \int_{s_\sigma} \overline{T}^i \delta u_i \,ds + \delta \int_{s_u} \sigma^{ij} n_j (u_i - \overline{u}_i)\,ds \Big\} dt = 0$$

Stress concept

# 3 Finite element approximation basis

## 3.1 Displacement technique

The reduced Trefftz type approach exploits the basic concept of the Trefftz method in the use of assumed internal displacement fields that solve the Lagrangian equation. What distinguishes this approach is the technique used to approximate the essential boundary conditions or the natural boundary conditions. In the displacement version of the reduced Trefftz type approach the global approximation basis operats on the natural boundary conditions enforced in an integral form. The essential boundary conditions may be used in the approximation at the finite element level in form of a modified interpolation technique to construct 'interpolation' functions over an element that approximate the internal displacement field. The modified interpolation technique, which corresponds to the requirement of the reduced Trefftz type approach was formulated in /56-58/ and first published in /59, 60/. For the convenience of the reader, this interpolation technique will be first recalled.

In the following, Latin indices in brackets range over the nodal points where indices without brackets identify the Cartesian coordinates. For example, $^{i}$ ranges over $x^i\ (i = 1,2,3)$, where $^{(i)}$ denotes the number of the nodal points.
Consider the variational problem governed by the differential equation (3.1) and boundary conditions (3.2) and (3.3)

$$\Delta^{ij} u_i = \bar{q}^j \tag{3.1}$$

$$u_i = \bar{u}_i \quad \text{on } s_u \subseteq s \tag{3.2}$$

$$\sigma^{ij} n_j = \bar{T}^i \quad \text{in } s_\sigma \quad s / s_u \tag{3.3}$$

In (3.1) $\Delta^{ij}$ is a matrix of differential operators, $u_i$ is the displacement function, and $\bar{q}^j$ is the load function. In (3.2), (3.3), $s_u$ is the surface of the domain on which the displacements are prescribed and $s_\sigma$ is the surface of the domain on which the forces are prescribed.

After dividing the domain into finite elements, in addition to the requirements on continuity, linearly independence and completeness of the assumed displacement functions, the essential boundary conditions of the element require further that

$$[u_k]_{x^k = x^k_{(e)}} = u_{k\,(e)} \tag{3.4}$$

Where $x^k_{(e)}$ are the coordinates of the element nodes and $u_{k\,(e)}$ are their displacements.
In order to describe the internal displacement field $u_i$, the usual parametric form (3.5) can be used, where the terms of matrix $M_i^{\ n}$ are trial approximating functions of cordinates $x^i\ (i = 1,2,3)$, and $c_n$ represents the undetermined parameters,

$$u_i = M_i^{\ n} c_n \tag{3.5}$$

The number of parameters $c_n$ is now chosen to be larger than the usual (nodal) number, in order to identify the load functions. Substituting equation. (3.5) into the differential equation (3.1) yields the relationship between the free parameters and the load functions

$$\Delta^{ij} M_i^{\ n} c_n = \overline{q}^{\,j} \tag{3.6}$$

Equation (3.6) shows that a subset of the free parameters $c_n$ can be expressed in terms of the element loading, $\overline{q}^{\,j}$ , as stated by (3.7), where $N_{(p)}$ are the loading shape functions and $\overline{q}^{\,j\,(p)}$ are the corresponding nodal values.

$$\Delta^{ij} M_i^{\ n} c_n = \overline{q}^{\,j} = N_{(p)}\ \overline{q}^{\,j\,(p)} \tag{3.7}$$

A suitable solution of (3.6) or (3.7) enables the separation of the trial approximating functions presents in equation (3.5) into a homogeneous part, with the same dimension as the number of degrees of freedom of the element, and a particular part dependent on the element loading

$$u_i = M_i^{\ j\,(m)}\ c_{j\,(m)} + \overline{M}_{i\,(p)}\ \overline{q}^{\,(p)} \tag{3.8}$$

The homogeneous set of the trial functions fulfils the homogeneous part of Lagrangian equation and the particular set fulfils the particular part (Trefftz-type).

As stated at the beginning of this section, the natural boundary conditions are enforced on average and their global integral form constitutes the variational expression (reduced Trefftz-type approach of displacement type).

Since the essential boundary conditions and the inter-element conditions have been ignored in constructing the global variational expression of the displacement version of the reduced Trefftz type approach, the undetermined parameters must be adjusted to match these conditions. In order to achieve this, the essential boundary conditions may be used at the finite element level to relate the undetermined parameters to the nodal degrees of freedom of the element in different ways. Three procedures are possible:
The first procedure is a "modified geometrical interpolation technique", in which a direct relationship between the displacement shape functions and the element loading at the finite element level is established.

This relationship can be established simply by directly linking the free parameters $c_{j\,(m)}$ to the nodal degrees of freedom using the property (3.4), to yield,

$$u_{k(e)} = A_{k(e)}^{\,j(m)}\ c_{j(m)} + \overline{A}_{k(e)\,(p)}\ \overline{q}^{\,(p)} \tag{3.9}$$

$$c_{j\,(m)} = B_{j\,(m)}^{\,k(e)}(u_{k(e)} - \overline{A}_{k\,(e)\,(p)}\ \overline{q}^{\,(p)}) \tag{3.10}$$

In equations (3.9), (3.10), $A_{k(e)}^{j(m)}$ and $\overline{A}_{k(e)(p)}$ are matrices derived from $M_k^{j(m)}$ and $\overline{M}_{k(p)}$ respectively, by substituting the coordinates of the element nodes and $B_{j(m)}^{k(e)}$ is the inverse matrix of $A_{k(e)}^{j(m)}$,

Substituting into equation (3.8) the free parameters defined by (3.10), the following relationship between the internal displacements $u_i$ and the nodal displacements $u_{k(e)}$ as well as the loads $\overline{q}^j$ is obtained

$$u_i = M_i^{j(m)} B_{j(m)}^{k(e)} (u_{k(e)} - \overline{A}_{k(e)(p)} \overline{q}^{(p)}) + \overline{M}_{i(p)} \overline{q}^{(p)} \tag{3.11}$$

$$u_i = N_i^{k(e)} u_{k(e)} + \overline{N}_{i(p)} \overline{q}^{(p)} \tag{3.12}$$

In equation (3.12), the following definitions apply:

$$N_i^{k(e)} = M_i^{j(m)} B_{j(m)}^{k(e)} \tag{3.13}$$

$$\overline{N}_{i(p)} = -M_i^{j(m)} B_{j(m)}^{k(e)} \overline{A}_{k(e)(p)} + \overline{M}_{i(p)} = -N_i^{k(e)} \overline{A}_{k(e)(p)} + \overline{M}_{i(p)} \tag{3.14}$$

In the equations above, $N_i^{k(e)}$ are the homogeneous shape functions and represents, weighted by the nodal displacements the homogeneous part of $u_i$, $\overline{N}_{i(p)}$ are the non-homogeneous shape functions and represents, weighted by the element nodal loading the non-homogeneous part of $u_i$. The relationship between both shape functions is expressed by equation (3.14).

The displacement approximation basis (3.12) , constructed as suggested above, can be used in the application of the displacement version of the finite element method. As the displacement trial functions defined by (3.8) satisfy the non-homogeneous differential equation that governs the problem under analysis, they satisfy also the non-homogeneous equilibrium equations. Therefore, they can be used to derive the force functions needed to implement hybrid stress version of the finite element method /61,62/ and they can, also, be used directly in the application of the Trefftz type approach /63/.

The second and third procedure for linking the free parameters to the nodal degrees of freedom of the element are hybrid and boundary technique forms, respectively. They are discussed in the following two subsections.

### 3.2 Hybrid technique

The modified geometrical interpolation technique described above may not comply with the boundary conditions and the continuity conditions. In this case, a hybrid technique or a boundary technique at the finite element level can be used to enforce the essential boundary conditions.

The hybrid technique consists simply in using the Gauss divergence theorem for the stress variation at the finite element level, after imposing the essential boundary conditions and the equations of equilibrium in order to eliminate the undetermined parameters. As we use displacement field functions that satisfy the governing differential equation, and consequently, the equilibrium equation, the Gauss divergence theorem for the stress variation reduces to (3.15), where the integral over surface s is uncoupled in the sum of two integrals on the static and kinematic boundaries $s_\sigma$ and $s_u$ ($s = s_u + s_\sigma$), respectively.

$$\int_V u_{i,j}\,\delta\sigma^{ij} dV = \int_s (\delta\sigma^{ij} n_j) u_i ds = \int_{s_\sigma} (\delta\sigma^{ij} n_j) u_i ds + \int_{s_u} (\delta\sigma^{ij} n_j) u_i ds \tag{3.15}$$

The last equation can be formulated after introducing the essential boundary conditions ($u_i = \bar{u}_i$ on $s_u$) as follows:

$$\int_V u_{i,j}\,\delta\sigma^{ij} dV = \int_s (\delta\sigma^{ij} n_j) u_i ds = \int_{s_\sigma} (\delta\sigma^{ij} n_j) u_i ds + \int_{s_u} (\delta\sigma^{ij} n_j) \bar{u}_i ds \tag{3.16}$$

Thus, the first procedure for linking the free parameters to the nodal degrees of freedoms at the element level is a hybrid technique form results in from using the following equation

$$\int_V u_{i,j}\,\delta\sigma^{ij} dV = \int_{s_\sigma} (\delta\sigma^{ij} n_j) u_i ds + \int_{s_u} (\delta\sigma^{ij} n_j) \bar{u}_i ds \tag{3.17}$$

### 3.3 Boundary technique

The second optional form for linking the free parameters to the nodal degrees of freedom is encoded by an alternative but equivalent boundary technique and results from the following identity:

$$\int_s (\delta\sigma^{ij} n_j)\ u_i ds = \int_{s_\sigma} (\delta\sigma^{ij} n_j)\ u_i ds + \int_{s_u} (\delta\sigma^{ij} n_j)\ \bar{u}_i ds \tag{3.18}$$

As equation (3.17) and (3.18) are valid both for the whole domain and for every sub-domain, they can be used also at finite element level to enforce the inter-element continuity using the so-called "frame function" concept in the selection of an independent inter-element displacement field for $\bar{u}_i$ in the customary way.

### 3.4 Notes on the various techniques

The application of the trial approximating functions $M_i^{j(m)}$, $\overline{M}_{i(p)}$ given in (3.8) in association with the variational statement of the extended natural boundary conditions (2.10) and the equivalence (3.18) enables the encodement of the method using boundary integral expressions only. The undetermined parameters may then be eliminated from equation (3.18) and the degrees of freedom of the system can be calculated from the variational statement of the extended natural boundary conditions (2.10) after collocation at the nodal points.

Furthermore, the application of the trial approximating functions (3.8) in association with the variational statement of the extended natural boundary conditions (2.10) and the reduced equilibrium equation under using the Gauss divergence theorem at the finite element level (3.17) enables the encodement of the method using integral expressions in a hybride sence. The undetermined parameters may then be eliminated from equation (3.17) and the degrees of freedom of the system can be calculated from the variational statement of the extended natural boundary conditions (2.10) after collocation at the nodal points

Another possibility for encoding the method consists in applying the trial aproximating functions in association with the variational statement of the modified principle of complementary virtual work (2.15). This leads to the usual well-known conventional hybrid technique.

The simplest way of implementing the method in the framework of the finite element method consists in applying the shape functions (3.12), which involve the homogeneous part (3.13) and the non-homogeneous part (3.14), in association with the variational expression (2.13). This enables the implementation the method following a computational path similar to that one of the conventional finite element displacement method.

It is noted that the application of the modified interpolation technique to the solution of beam bending problems produces shape functions that satisfy the Lagrangian equation and the inter-element continuity. The same results are recovered by applying the chosen displacement functions with any of the variational statement described above, provided that the 'actual' essential boundary conditions are strictly observed, see for example /63/.

# 4 Finite elements of a reduced Trefftz type

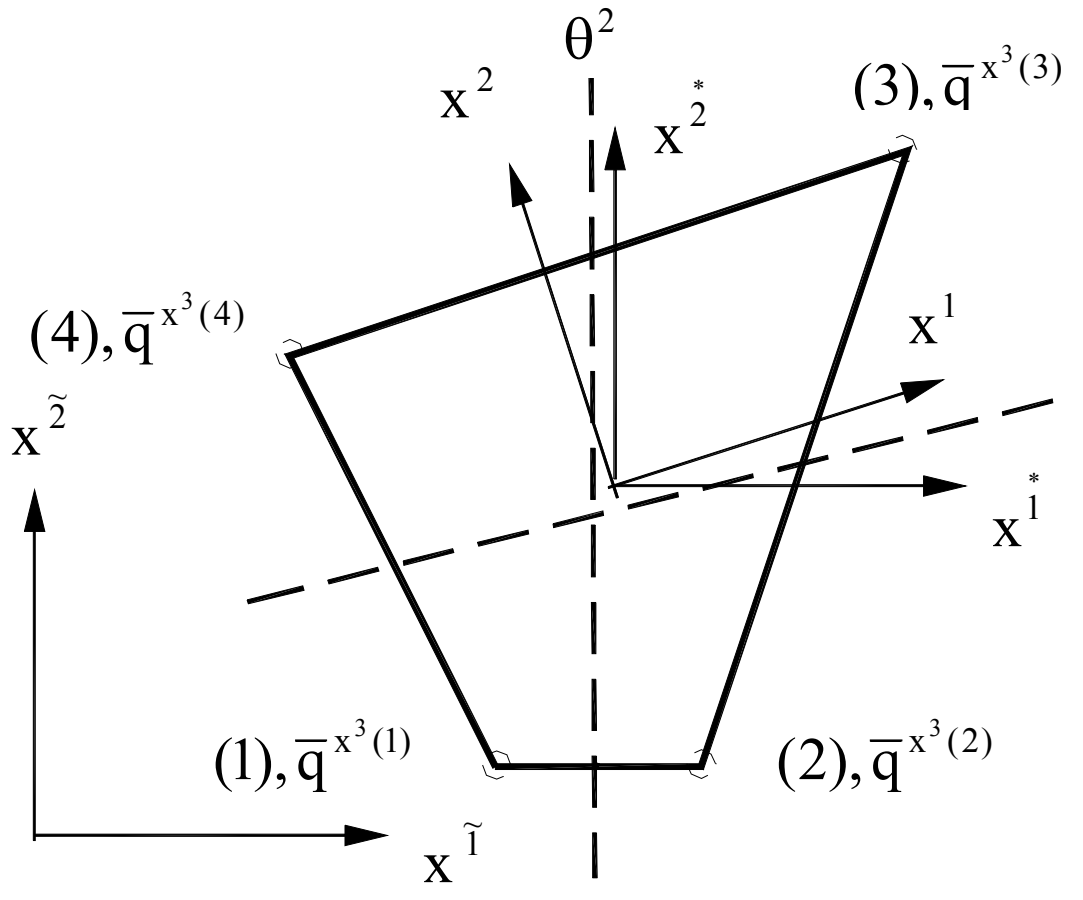

Fig. 4.1: Quadrilateral finite element for plate bending
Coordinate systems and element loading

A general procedure for formatting finite elements of general geometric form should be presented by means of the quadrilateral Kirchhoff plate bending element shown in Fig. 4.1. The four-node elements have three degrees of freedom per nodal point, namely the transverse displacement and the two rotations. The application allows the element to be loaded by a distributed load $\overline{q}^{x^3}(x^1, x^2)$ with different nodal values $\overline{q}^{x^3(1)}$, $\overline{q}^{x^3(2)}$, $\overline{q}^{x^3(3)}$, $\overline{q}^{x^3(4)}$ at the element nodes $^{(1),\,(2),\,(3),\,(4)}$, respectively (see Fig. 4.1). The homogeneous approximation basis is taken from the 12-DOF rectangular plate bending element. The approximation basis is extended to involve a particular part, which satisfy the differential equation of the plate problem. Thus, the application can be classified under the displacement version of the reduced Trefftz-type approach.

Various difficulties may be encountered in developing such element forms in association with a reduced Trefftz type approach. The first difficulty, encountered, consists in satisfying the requirement of the approach in constructing the approximation basis, so that the Lagrangian equation is a priori fullfiled. In addition, difficulties encounterd in formatting finite elements of general geometric form in association with displacement and hybrid versions of the finite element method in what concerns convergence, invariance and nodal point numbering insensitivity are to be overcome.

Further diffuculties appear in applying the Trefftz-type approach for dynamic problems. In this case, approximation functions, which satisfy the Lagrangian equation in the dynamic case, are required. Such requirement can not be easily achieved in general for Lagrangian equations formulated in a natural coordinate system.

The use of Cartesian coordinate system enhances basically the application of the reduced Trefftz type approach in the framework of the finite element method. For further simplicity, the trial functions for the displacement applied in a static application can be used in case of a dynamic application together with variational concept involving the inertia forces. In such case, it is necessary to evaluate domain integral which means that the approach is only used in a “static Trefftz sense”.
In order to circumvent the difficulties associated with the application of the reduced Trefftz type approach and the formatting of finite elements of general geometric form for other version of finite elements (i. e. displacement and hybrid versions), the following steps of the general application procedure are adopted:

1. Beside global Cartesian coordinate system $(x^{\tilde{1}}, x^{\tilde{2}})$, parallel local coordinate system located at the element centroid $(x^{\overset{*}{1}}, x^{\overset{*}{2}})$ and the natural coordinate system $(\theta^1, \theta^2)$, a suitable Cartesian coordinate system $(x^1, x^2)$ located at the element centroid is defined in the following way:

Firstly, the coordinate system $(x^{\overset{*}{1}}, x^{\overset{*}{2}})$ is defined by translating the origin of the global coordinate system into the element centroid. The coordinates of an arbitrary point in the element related to system $(x^{\overset{*}{1}}, x^{\overset{*}{2}})$ can be rewritten as:

$$x^{\overset{*}{i}} = x^{\tilde{i}} - x^{\tilde{i}}_{(c)} \tag{4.1}$$

$x^{\tilde{i}}$ are the global Cartesian coordinates of the arbitrary point and $x^{\tilde{i}}_{(c)}$ are the global coordinates of the element centroid defined as

$$x^{\tilde{i}}_{(c)} = \int_A x^{\tilde{i}}\, dA \Big/ \int_A dA \tag{4.2}$$

Now the following differential geometry properties of the element are defined:
* Coordinates of an arbitrary point of the element

$$\begin{bmatrix} x^{\overset{*}{1}} & x^{\overset{*}{2}} \end{bmatrix} = \frac{1}{4} \begin{bmatrix} (1-\theta^1)(1-\theta^2) & (1+\theta^1)(1-\theta^2) & (1+\theta^1)(1+\theta^2) & (1-\theta^1)(1+\theta^2) \end{bmatrix} * \begin{bmatrix} x^{\overset{*}{1}}_{(1)} & x^{\overset{*}{2}}_{(1)} \\ x^{\overset{*}{1}}_{(2)} & x^{\overset{*}{2}}_{(2)} \\ x^{\overset{*}{1}}_{(3)} & x^{\overset{*}{2}}_{(3)} \\ x^{\overset{*}{1}}_{(4)} & x^{\overset{*}{2}}_{(4)} \end{bmatrix} \tag{4.3}$$

* Covariant base vectors

$$g_\alpha = g^{\overset{*}{i}}_\alpha e_{\overset{*}{i}} \; ; \; g^{\overset{*}{i}}_\alpha = \begin{bmatrix} g_1^{x^{\overset{*}{1}}} & g_1^{x^{\overset{*}{2}}} \\ g_2^{x^{\overset{*}{1}}} & g_2^{x^{\overset{*}{2}}} \end{bmatrix} = \frac{1}{4} \begin{bmatrix} -(1-\theta^2) & (1-\theta^2) & (1+\theta^2) & -(1+\theta^2) \\ -(1-\theta^1) & -(1+\theta^1) & (1+\theta^1) & (1-\theta^1) \end{bmatrix} \begin{bmatrix} x^{\overset{*}{1}}_{(1)} & x^{\overset{*}{2}}_{(1)} \\ x^{\overset{*}{1}}_{(2)} & x^{\overset{*}{2}}_{(2)} \\ x^{\overset{*}{1}}_{(3)} & x^{\overset{*}{2}}_{(3)} \\ x^{\overset{*}{1}}_{(4)} & x^{\overset{*}{2}}_{(4)} \end{bmatrix} \tag{4.4}$$

* Covariant metric coefficients

$$g_{\alpha\beta} = g_{\alpha}^{\overset{*}{i}} \; g_{\beta}^{\overset{*}{j}} \; e_{\underset{i}{*}} \; e_{\underset{j}{*}} \tag{4.5}$$

* Contravariant metric coefficients

$$g^{\alpha\beta} \; g_{\beta\gamma} = \delta_{\gamma}^{\alpha} \tag{4.6}$$

* Contravariant base vectors

$$g^{\beta} = g^{\alpha\beta} \; g_{\alpha} = g_{\underset{i}{*}}^{\alpha} \; e^{\overset{*}{i}} = \begin{bmatrix} g^{1}_{\underset{x^1}{*}} & g^{1}_{\underset{x^2}{*}} \\ g^{2}_{\underset{x^1}{*}} & g^{2}_{\underset{x^2}{*}} \end{bmatrix} \begin{bmatrix} e^{\overset{*}{x^1}} \\ e^{\overset{*}{x^2}} \end{bmatrix} \tag{4.7}$$

The covariant base vectors and the contravariant base vectors are orthogonal and its directions construct naturaly a suitable Cartesian coordinate basis for defining the local coordinate system

$$\begin{aligned} & g_{\alpha}\, g^{\beta} = \delta_{\beta}^{\alpha} \;; \\ & g_1\, g^1 = 1 \;;\; \underline{\underline{g_1\, g^2 = 0}}; \\ & \underline{\underline{g_2\, g^1 = 0}} \;;\; g_2\, g^2 = 1; \end{aligned} \tag{4.8}$$

Thus, the local coordinate system $(x^1, x^2)$ or $(x_1, x_2)$ with the unit vectors $(e_{x^1}, e_{x^2})$ or $(e^{x^1}, e^{x^2})$ is, for example, defined from the directions of the covariant base vectors and the perpendicular contravariant base vector of the geometric center of the element where $(\theta^1 = 0,\ \theta^2 = 0)$ as follows,

$$e_{g_1} = \frac{1}{\left|g_{\underset{1}{*}}\right|}\left(g_1^{\overset{*}{x^1}} \; e_{\underset{x^1}{*}} + g_1^{\overset{*}{x^2}} \; e_{\underset{x^2}{*}}\right) \;;\; e_{g_1} \,//\, g_{\underset{1}{*}}$$

$$e^{g^2} = \frac{1}{\left|g^{\overset{*}{2}}\right|}\left(g^{2}_{\underset{x^1}{*}} \; e^{\overset{*}{x^1}} + g^{2}_{\underset{x^2}{*}} \; e^{\overset{*}{x^2}}\right) \;;\; e^{g^2} \,//\, g^{\overset{*}{2}}$$

(4.9a)

$$e_{g^1} = \frac{1}{\left|g^{\overset{*}{1}}\right|}\left(g^{1}_{\underset{x^1}{*}} \; e^{\overset{*}{x^1}} + g^{1}_{\underset{x^2}{*}} \; e^{\overset{*}{x^2}}\right) \;;\; e_{g^1} \,//\, g^{\overset{*}{1}}$$

$$e_{g_2} = \frac{1}{\left|g_{\underset{2}{*}}\right|}\left(g_2^{\overset{*}{x^1}} \; e_{\underset{x^1}{*}} + g_2^{\overset{*}{x^2}} \; e_{\underset{x^2}{*}}\right) \;;\; e_{g_2} \,//\, g_{\underset{2}{*}}$$

(4.9a)

$$
s_1 = e_{g_1} + e_{g^1} \quad ; \quad e_{x^1} = \frac{s_1}{|s_1|} \quad ;
$$
$$
s_2 = e^{g^2} + e_{g_2} \quad ; \quad e_{x^2} = \frac{s_2}{|s_2|} \tag{4.10}
$$

The coordinates in the local system and the unit vectors can be calculated and the transformation relation between both systems can be uniquely defined

$$
x_i = a_i^{\overset{*}{j}} x_{\underset{j}{*}} \quad ; \quad x^i = a^i_{\underset{j}{*}} x^{\overset{*}{j}} ; \quad e_i = a_i^{\overset{*}{j}} e_{\underset{j}{*}} \quad ; \quad e^i = a^i_{\underset{j}{*}} e^{\overset{*}{j}} \tag{4.11}
$$

2. The displacement approximation basis is constructed in the defined local coordinate system so that the Lagrangian equation is satisfied. In such a way, the difficulties in constructing trial functions, which satisfy the Lagrangian equation for applying the Trefftz method can be overcome.

3. The advantages of using natural coordinate systems are exploited. All quantities needed in the application are expressed in the natural coordinate system by using the isoparametric transformation.

4. The integration over the interval [-1, +1] is used to derive the element matrices and load vectors.

### 4.1 Plate bending elements

#### 4.1.1 Modified non-conforming plate bending element (ZDEQ)

This element is denoted by the abbreviation ZDEQ for further reference, see also /44/. The internal displacement field is approximated in parametric form. The following symmetric sixteen-term polynomial is selected in order to accommodate the load functions,

$$
\begin{aligned}
&u^0_{x^3}(x^1, x^2) = M^n c_n \\
M^n = \big[\, 1 \quad x^1 \quad x^2 \quad (x^1)^2 \quad x^1x^2 \quad (x^2)^2 \quad (x^1)^3 \quad (x^1)^2x^2 \quad x^1(x^2)^2 \quad (x^2)^3 \\
&(x^1)^3x^2 \quad x^1(x^2)^3 \quad (x^1)^2(x^2)^2 \quad (x^1)^4x^2 \quad x^1(x^2)^4 \quad (x^1)^3(x^2)^3 \,\big]
\end{aligned} \tag{4.12}
$$

Applying the differential equation (4.13) of the plate bending problem on the internal displacement field (4.12) provides the condition (4.14) for satisfying the differential equation in each point of the element.

$$
u^0_{x^3,ijkl}\delta^{jk}\delta^{il} = \bar{q}^{x^3} / D \tag{4.13}
$$

$$\begin{bmatrix} 8 & 24\,x^1 & 24\,x^2 & 72\,x^1 x^2 \end{bmatrix} \begin{bmatrix} c_{13} \\ c_{14} \\ c_{15} \\ c_{16} \end{bmatrix} = \overline{q}^{x^3}(x^1, x^2)/D \tag{4.14}$$

Approximating the load function by the four-term polynomial, as stated in equation (4.15), a subset of the undetermined parameters (the four parameters added to the displacement function of the first 12-DOF plate-bending element) can be determined depending on the load parameter by comparing its coefficients with the condition (4.14). This yields the possible solution (4.16).

$$\overline{q}^{x^3}(x^1, x^2) = \overline{M}_1^{(q)}\; c_{(q)}$$

$$\overline{q}^{x^3}(x^1, x^2) = \begin{bmatrix} 1 & x^1 & x^2 & x^1 x^2 \end{bmatrix} \begin{bmatrix} c_{\overline{q}^1} \\ c_{\overline{q}^2} \\ c_{\overline{q}^3} \\ c_{\overline{q}^4} \end{bmatrix} \tag{4.15}$$

$$\begin{bmatrix} c_{13} \\ c_{14} \\ c_{15} \\ c_{16} \end{bmatrix} = \frac{1}{D} \begin{bmatrix} c_{\overline{q}^1}/8 \\ c_{\overline{q}^2}/24 \\ c_{\overline{q}^3}/24 \\ c_{\overline{q}^4}/72 \end{bmatrix} \tag{4.16}$$

Substituting the coordinates of the element nodes into equation (4.15) and inverting the result (4.17), the load parameters can be eliminated depending on their nodal values

$$\overline{q}^{(p)} = \overline{A}_1^{(p)(q)}\, c_{(q)}$$

$$\begin{bmatrix} \overline{q}^{x^3(1)} \\ \overline{q}^{x^3(2)} \\ \overline{q}^{x^3(3)} \\ \overline{q}^{x^3(4)} \end{bmatrix} = \begin{bmatrix} 1 & x^1_{(1)} & x^2_{(1)} & x^1_{(1)} x^2_{(1)} \\ 1 & x^1_{(2)} & x^2_{(2)} & x^1_{(2)} x^2_{(2)} \\ 1 & x^1_{(3)} & x^2_{(3)} & x^1_{(3)} x^2_{(3)} \\ 1 & x^1_{(4)} & x^2_{(4)} & x^1_{(4)} x^2_{(4)} \end{bmatrix} \begin{bmatrix} c_{\overline{q}^1} \\ c_{\overline{q}^2} \\ c_{\overline{q}^3} \\ c_{\overline{q}^4} \end{bmatrix} \tag{4.17}$$

$$c_{(q)} = \overline{A}_{1(p)(q)}\, \overline{q}^{(p)}$$

$$\begin{bmatrix} c_{\overline{q}^1} \\ c_{\overline{q}^2} \\ c_{\overline{q}^3} \\ c_{\overline{q}^4} \end{bmatrix} = \begin{bmatrix} \overline{A}_{11} & \overline{A}_{12} & \overline{A}_{13} & \overline{A}_{41} \\ \overline{A}_{21} & \overline{A}_{22} & \overline{A}_{23} & \overline{A}_{42} \\ \overline{A}_{31} & \overline{A}_{32} & \overline{A}_{33} & \overline{A}_{43} \\ \overline{A}_{41} & \overline{A}_{42} & \overline{A}_{34} & \overline{A}_{44} \end{bmatrix} \begin{bmatrix} \overline{q}^{x^3(1)} \\ \overline{q}^{x^3(2)} \\ \overline{q}^{x^3(3)} \\ \overline{q}^{x^3(4)} \end{bmatrix} \tag{4.18}$$

In (4.18) $\overline{A}_{1(p)(q)}$ is the inverse matrix of $\overline{A}_{1}^{(p)(q)}$ .

Substituting the eliminated load parameters from (4.18) into (4.16) and after that substituting the result in (4.12), the internal displacement field may be separated in a homogeneous part with the same dimension as the degrees of freedom of the element and a particular part dependent on the element loading:

$$u^{0}_{x^3}(x^1,x^2) = M^{m(m)}\ c_{m(m)} + \overline{M}_{(p)}\ \overline{q}^{(p)} \tag{4.19a}$$

$$M^{m(m)} = \left[1 \quad x^1 \quad x^2 \quad (x^1)^2 \quad x^1x^2 \quad (x^2)^2 \quad (x^1)^3 \quad (x^1)^2x^2 \quad x^1(x^2)^2 \quad (x^2)^3 \ (x^1)^3x^2 \ x^1(x^2)^3\right] \tag{4.19b}$$

$$\overline{M}_{(p)} = \frac{1}{D}\left[\frac{(x^1)^2(x^2)^2}{8} \quad \frac{(x^1)^4x^2}{24} \quad \frac{x^1(x^2)^4}{24} \quad \frac{(x^1)^3(x^2)^3}{72}\right]\begin{bmatrix} \overline{A}_{11} & \overline{A}_{12} & \overline{A}_{13} & \overline{A}_{41} \\ \overline{A}_{21} & \overline{A}_{22} & \overline{A}_{23} & \overline{A}_{42} \\ \overline{A}_{31} & \overline{A}_{32} & \overline{A}_{33} & \overline{A}_{43} \\ \overline{A}_{41} & \overline{A}_{42} & \overline{A}_{34} & \overline{A}_{44} \end{bmatrix} \tag{4.19c}$$

As may be seen, the homogeneous part of the internal displacement field of the current quadrilateral non-conforming thin-plate bending element is equivalent to the non-conforming displacement function of the conventional 12-DOF rectangular element. The non-conforming rectangular 12-DOF element was originally developed by Melosh /64, 65/ and others and thoroughly studied by numerous authors /52, 66/. The generalization of refined version of the 12-DOF element was carried out in /55/.

In the following equations, the Latin indices range over the three degrees of freedom each nodal point and the indices between round brackets range over the nodal point of the element.

Now the 12-nodal degrees of freedom $u_{k(e)}$ (i. e. the nodal displacement and slopes at the four element nodes) will be used to form the element shape functions following the modified geometrical interpolation technique explained in section 3.1.

By adopting the steps (3.9) to (3.14) of the interpolation procedure, the internal displacement field can be formulated related to the nodal degrees of freedom of the element and to the element loading. This can be rewritten using different index notations as follows

$$u^{0}_{x^3} = N^{m(m)}\ u_{m(m)} + \overline{N}_{(p)}\ \overline{q}^{(p)} = N^{n(n)}\ u_{n(n)} + \overline{N}_{(q)}\ \overline{q}^{(q)} \tag{4.20}$$

The global variational basis for applying the current approach is represented by the following expression (Trefftz-method only in a "static sense").

$$\int_{t_1}^{t_2}\{\int_V \rho\ddot{u}^i\,\delta u_i dV + \int_V \sigma^{ij}\ \delta\varepsilon_{ij} dV - \int_V \overline{f}^i\delta u_i\, dV - \int_{s_\sigma}\overline{T}^i\delta u_i\, ds\}dt = 0 \tag{4.21}$$

Applying the internaldisplacement field (4.20) in the potential form of (4.21) leads after performing the variation to the following standard FEM-relation (in the absence of damping effects)

$$k^{m(m)n(n)}\ u_{n(n)} + m^{m(m)n(n)}\ \ddot{u}_{n(n)} = \bar{f}2^{m(m)} - \bar{f}1^{m(m)} \tag{4.22}$$

In equation (4.22), the following definitions apply:

Stiffness matrix

$$k^{m(m)n(n)} = \int_A N^{m(m)}_{,ij}\ E^{ijkl}\ N^{n(n)}_{,kl}\ dA \tag{4.23}$$

Mass matrix

$$m^{m(m)n(n)} = \int_A N^{m(m)}_{i}\ \rho^{ij}\ N^{n(n)}_{j}\ dA \tag{4.24}$$

Equivalent nodal forces to distributed load resulting in from the application of the conventional displacement finite element model

$$\bar{f}2^{m(m)} = \int_A N^{m(m)}\ \overline{N}_{(p)}\ \bar{q}^{(p)}\ dA \tag{4.25}$$

Equivalent nodal forces associated with the added particular term

$$\bar{f}1^{m(m)} = \int_A N^{m(m)}_{,ij}\ E^{ijkl}\ \overline{N}_{(p),kl}\bar{q}^{(p)}\ dA \tag{4.26}$$

The constant term (4.27) appears only in the energy expression:

$$c = \int_A \bar{q}^{(p)}\overline{N}_{(p),ij}\ E^{ijkl}\ \overline{N}_{(q),kl}\bar{q}^{(p)}\ dA \tag{4.27}$$

$E^{ijkl}$ : Matrix of force curvature dependency

In the equations above, $N^{m(m)}_{i}$ are the shape functions, in which the displacements and rotations are included and $\rho^{ij}$ is the corresponding mass density matrix. $E^{ijkl}$ is the matrix of the force-curvature dependency.

$$E^{ijkl} = \frac{E\,t^3}{12(1-\upsilon^2)} \begin{bmatrix} 1 & 0 & 0 & \upsilon \\ 0 & (1-\upsilon)/2 & (1-\upsilon)/2 & 0 \\ 0 & (1-\upsilon)/2 & (1-\upsilon)/2 & 0 \\ \upsilon & 0 & 0 & 1 \end{bmatrix} \tag{4.28}$$

E is the modulus of elasticity, t is the plate thickness and $\upsilon$ the Poissom's ratio. $\rho^{ij}$ is defined by the following matrix where $\rho$ is the material density.

$$\rho^{ij} = \begin{bmatrix} \rho & 0 & 0 \\ 0 & \rho\, t^2/12 & 0 \\ 0 & 0 & \rho\, t^2/12 \end{bmatrix} \tag{4.29}$$

The advantages of using natural coordinate systems can be exploited to perform the integration. All quantities needed in the application may be expressed in the natural coordinate system by using the isoparametric transformation. Then, the integration over the interval [-1, +1] may be used to derive the element matrices and load vectors.

### 4.1.2 Hybrid element (TFEQ)

This element is denoted by the abbreviation TFEQ for further reference, see also /44/. The implementation of this element is currently confined to a static application of the plate bending problem. The global variational approximation is based on the static version of the extended variational form of the natural boundary conditions (4.30)

$$\delta I = \int_s \sigma^{ij} n_j\, \delta u_i\, ds - \int_{s_u} \overline{T}^i\, \delta u_i\, ds = 0 \tag{4.30}$$

The internal displacement field is approximated by the trial function (4.19). The alternative procedure for linking the free parameters to the nodal degrees of freedom of the element consists in applying the following reduced form of the equilibrium conditions at the finite element level, in which the Gauss divergence theorem is imposed.

$$\int_V u_{i,j}\, \delta\sigma^{ij} dV = \int_{s_\sigma} (\delta\sigma^{ij} n_j) u_i ds + \int_{s_u} (\delta\sigma^{ij} n_j) \overline{u}_i ds \tag{4.31}$$

In order to enforce conformity and inter-element continuity we make use of the 'frame function' concept developed by Jirousek /29/ in selecting at the boundary an independent displacement field for $\overline{u}_i$ which ensures inter-element continuity in the customary way.
At the element boundary the conjugate vector of boundary tractions is determined from (4.19) and can be written using different index notations as follows

$$T^{i((e)(b))} = R^{i((e)(b))m(m)}\, c_{m(m)} + \overline{R}^{i((e)(b))}_{(p)}\, \overline{q}^{(p)} \tag{4.32a}$$

$$T^{i((e)(b))} = R^{i((e)(b))n(n)}\, c_{n(n)} + \overline{R}^{i((e)(b))}_{(p)}\, \overline{q}^{(p)} \tag{4.32b}$$

$$T^{i((e)(b))} = R^{i((e)(b))\ell(\ell)}\, c_{\ell(\ell)} + \overline{R}^{i((e)(b))}_{(p)}\, \overline{q}^{(p)} \tag{4.32c}$$

Assuming that the variation of the load terms is zero, the variation of the boundary tractions is then given by (4.33)

$$\delta T^{i((e)(b))} = R^{i((e)(b))n(n)}\, \delta c_{n(n)} \tag{4.33}$$

The index $^{((e)(b))}$ varies over the four element boundaries, along which we assume for $\overline{u}_i$ a displacement field that ensures the inter-element continuity in the customary way

$$\overline{u}_{i((e)(b)} = L^{m(m)}_{i((e)(b))}\, u_{m(m)} \;;\; \overline{u}_{i((e)(b)} = L^{k(k)}_{i((e)(b))}\, u_{k(k)} \tag{4.34}$$

Matrix $L^{m(m)}_{i((e)(b))}$ or $L^{k(k)}_{i((e)(b))}$ contains shape functions on the element boundaries. In some Trefftz type formulations the conformity between the internal displacement field (4.19) and the boundary displacement (4.34) is enforced in a weak residual sense Jirousek /20, 29, 33/, and in others, in a least square sense Jirousek /33/, Piltner /39/.

Here, the undetermined parameters $c_{m(m)}$ (or $c_{\ell(\ell)}$) may now be eliminated using (4.31) using a hybrid technique. By substituting the conjugate vector of boundary tractions (4.32) and the prescribed displacement field $\overline{u}_i$ into (4.31) the relation (4.35) can be evaluated and the relationship (4.36) between the undetermined parameters and the nodal degrees of freedom can be established.

$$\delta c_{m(m)} H^{m(m)n(n)} c_{n(n)} + \delta c_{m(m)}\, \overline{H}^{m(m)}{}_{(p)}\, \overline{q}^{(p)} = \delta c_{m(m)}\, T^{m(m)n(n)}\, u_{n(n)} \tag{4.35}$$

This leads straightforwardly to the following relation, where $H_{\ell(\ell)n(n)}$ is the inverse matrix of (4.37)

$$c_{\ell(\ell)} = H_{\ell(\ell)m(m)} (T^{m(m)n(n)} u_{n(n)} - \overline{H}^{m(m)}{}_{(p)}\, \overline{q}^{(p)}) \tag{4.36}$$

$$H^{m(m)n(n)} = \int_A M^{m(m)}_{,ij}\, E^{ijkl}\, M^{n(n)}_{,kl}\, dA \tag{4.37}$$

The remaining terms in Equation (4.36) are defined as follows

$$T^{m(m)n(n)} = \int_s R^{i((e)(b))\, n(n)}\, L^{m(m)}_{i((e)(b))}\, ds \tag{4.38}$$

$$\overline{H}_{(p)}^{\ m(m)} = \int_A \overline{M}_{(p),ij}\, E^{ijkl}\, M^{m(m)}_{,kl}\ dA \tag{4.39}$$

Definitions (4.32) and (4.34) for the boundary tractions and the boundary displacements, respectively, can be used to derive the following relation by evaluating the variational expression (4.30)

$$\delta u_{k(k)}\{\int_s \left( R^{i((e)(b))\ell(\ell)}\, c_{\ell(\ell)} + \overline{R}^{i((e)(b))}_{(p)}\ \overline{q}^{(p)} \right) L^{k(k)}_{i((e)(b))}\ ds - \int_{s_\sigma} L^{k(k)}_{i((e)(b))}\ \overline{T}^{i((e)(b)}ds\} = 0 \tag{4.40}$$

The last equation can be recast under observing (4.38) and introducing the definitions (4.42) and (4.43) in the form (4.41)

$$\delta u_{k(k)}\,(T^{k(k)\ell(\ell)}\ c_{\ell(\ell)} + \ \overline{T}^{k(k)}_{(p)}\ \ \overline{q}^{(p)} \ - \ \overline{r}^{0k(k)}) = 0 \tag{4.41}$$

$$\overline{r}^{0\ k(k)} = \int_{s_\sigma} \overline{T}^{i\ ((e)(b))}\ L^{k(k)}_{i((e)(b))} ds \tag{4.42}$$

$$\overline{T}^{k(k)}_{(p)} = \int_{s_\sigma} \overline{R}^{i((e)(b))}_{(p)}\ L^{k(k)}_{i((e)(b))} ds \tag{4.43}$$

Substituting the free parameters $c_{\ell(\ell)}$ defined by equation (4.36) into equation (4.41) leads to the following equation

$$\begin{aligned} &\delta u_{k(k)}\,(T^{k(k)\ell(\ell)} H_{\ell(\ell)m(m)} T^{m(m)n(n)})\ u_{n(n)} + \ \delta u_{k(k)}\,(T^{k(k)\ell(\ell)} H_{\ell(\ell)m(m)} \overline{H}^{m(m)}_{(p)}\ \overline{q}^{(p)}) \\ &+ \delta u_{k(k)}\, \overline{T}^{k(k)}_{(p)}\ \ \overline{q}^{(p)} - \delta u_{k(k)}\, \overline{r}^{0k(k)}) = 0 \end{aligned} \tag{4.44}$$

This leads by recasting the previous equation to the following “force-displacement” finite element relationship

$$k^{k(k)n(n)} u_{n(n)} - \overline{r}^{\ 0k(k)} = \overline{r}^{\ k(k)} \tag{4.45}$$

In this equation the symmetric finite element stiffness matrix and the equivalent nodal force vectors are defined by the equations below

$$k^{k(k)n(n)} = T^{k(k)\ell(\ell)} H_{\ell(\ell)m(m)} T^{m(m)n(n)} \tag{4.46}$$

$$\bar{r}^{k(k)} = -T^{k(k)\ell(\ell)} H_{\ell(\ell)m(m)} \bar{H}^{m(m)}_{(p)} \bar{q}^{(p)} + \bar{T}^{k(k)}_{(p)} \bar{q}^{(p)} \tag{4.47}$$

### 4.1.3 Boundary element (JFEQ)

This element is denoted by the abbreviation JFEQ for further reference, see also /44/. The element has the same computational basis as the previous element. It differs only in relation to the elimination of the undetermined parameters $c_{m(m)}$ (or $c_{\ell(\ell)}$), using (4.48) in form of hybrid boundary technique equivalent to the hybrid technique described in the previous section .

$$\int_{s} (\delta\sigma^{ij} n_j)\, u_i ds = \int_{s_\sigma} (\delta\sigma^{ij} n_j)\, u_i ds + \int_{s_u} (\delta\sigma^{ij} n_j)\, \bar{u}_i ds \tag{4.48}$$

In order to apply this technique, first we derive from $u^0_{x^3}$ as given in (4.19), the corresponding displacement functions $u_{i((e)(b)}$ on the element boundary

$$u_{i((e)(b)} = M^{m(m)}_{i((e)(b))} c_{m(m)} + \bar{M}_{i((e)(b))(p)} \bar{q}^{(p)} \tag{4.49}$$

Using these functions and the boundary tractions (4.32) we can evaluate the boundary integrals present in (4.48). This leads to the boundary $H^{m(m)n(n)}$ and $\bar{H}^{m(m)}_{(p)}$ matrices, which are quasi equivalent to the hybrid $H^{m(m)n(n)}$ and $\bar{H}^{m(m)}_{(p)}$ matrices defined by equations (4. 37) and (4.39).

$$H^{m(m)n(n)} = \int_{s} M^{m(m)}_{i((e)(b))} R^{i((e)(b))n(n)} ds \tag{4.50}$$

$$\bar{H}^{m(m)}_{(p)} = \int_{s} \bar{M}_{i((e)(b))(p)} R^{i((e)(b))m(m)} ds \tag{4.51}$$

The matrix $H^{m(m)n(n)}$ is quasi symmetric. It is possible to evaluate the integral (4.52) instead of the integral (4.50) to ensure the symmetry (see for example /67/)

$$H^{m(m)n(n)} = \frac{1}{2} \left( \int_{s} M^{m(m)}_{i((e)(b))} R^{i((e)(b))n(n)} ds + \int_{s} R^{i((e)(b))n(n)} M^{m(m)}_{i((e)(b))} ds \right) \tag{4.52}$$

Elimination of the undetermined parameters $c_{m(m)}$ (or $c_{\ell(\ell)}$) using equation (4.48) recovers relation (4.36), which means that the application may be implemented in an analogous way.

## 4.2 Plain stress element

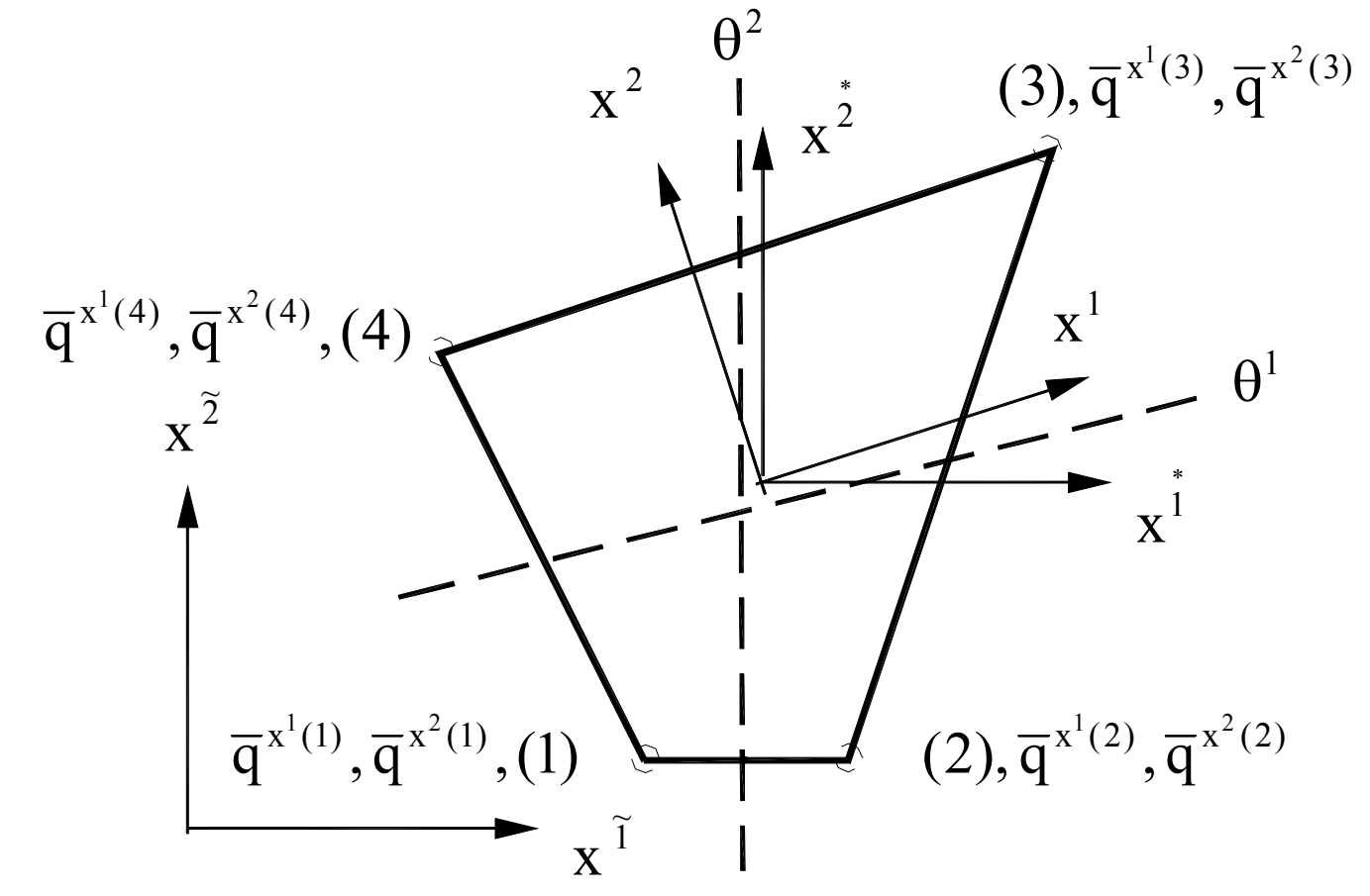

Fig. 4.2:Quadrilateral finite element for plain stress
Coordinate systems and element loading

The four-node quadrilateral plain stress element shown in Fig.4.2 has three degrees of freedom per nodal point, these are the displacement components in the directions of the defined local axes $x^1, x^2$ and the rotation about the third axis $x^3$ normal to the element plane, that means $\{ u^0_{x^1} \;,\; u^0_{x^2} \;,\; \varphi_{x^3} \}$. The local axes $x^1, x^2$ are defined using the directions of the element basis vectors in an analogous procedure used in section 4. The approximation basis is constructed using a stress function $F(x^1, x^2)$ approximated in parametric form

$$F(x^1, x^2) = M^n \, c_n \tag{5.53a}$$

$$M^n = \begin{bmatrix} 1 & x^1 & x^2 & (x^1)^2 & x^1x^2 & (x^2)^2 & (x^1)^3 & (x^1)^2x^2 & x^1(x^2)^2 & (x^2)^3 \\ & (x^1)^4 & (x^1)^3x^2 & (x^1)^2(x^2)^2 & x^1(x^2)^3 & (x^2)^4 \end{bmatrix} \tag{5.53b}$$

$$c_n = \{ c_1 \; c_2 \; .......... \; c_{14} \; c_{15} \} \tag{4.53c}$$

In order to satisfy the differential equations (4.54) the stress function (4.53) may be substituted into the differential equations to yield the relationship (4.55) between a subset of undetermined parameters

$$\frac{\partial^4 F(x^1,x^2)}{(\partial x^1)^4} + 2\frac{\partial^4 F(x^1,x^2)}{(\partial x^1)^2(\partial x^2)^2} + \frac{\partial^4 F(x^1,x^2)}{(\partial x^2)^4} = 0 \tag{4.54}$$

$$24\, c_{11} + \; 8\, c_{13} + 24\, c_{15} = 0 \tag{4.55}$$

The possible solution (4.56) of equation (4.55) enables the rearranging of equation (4.53) in the form (4.57)

$$c_{13} = 3\ c_{11} - 3\ c_{15} \tag{4.56}$$

$$\begin{aligned} F(x^1,x^2) = {} & c_1 + c_2 x^1 + c_3 x^2 + c_4 (x^1)^2 + c_5 x^1 x^2 + c_6 (x^2)^2 + c_7 (x^1)^3 + c_8 (x^1)^2 x^2 + c_9 x^1 (x^2)^2 + c_{10} (x^2)^3 \\ & + c_{11}((x^1)^4 - 3\,(x^1)^2 (x^2)^2) + c_{12}(x^1)^3 x^2 + c_{14} x^1 (x^2)^3 + c_{15}((x^2)^4 - 3\,(x^1)^2 (x^2)^2) \end{aligned} \tag{4.57}$$

The conjugate force vector to the previous stress function can be derived using the following relation

$$n_{x^1x^1} = \frac{\partial^2 F(x^1,x^2)}{(\partial x^2)^2}$$

$$n_{x^1x^2} = \frac{\partial^2 F(x^1,x^2)}{\partial x^1 \partial x^2} - x^2 \bar{q}^{x^1} - x^1 \bar{q}^{x^2} \ ; \ n_{x^2x^1} = n_{x^1x^2} \tag{4.58}$$

$$n_{x^2x^2} = \frac{\partial^2 F(x^1,x^2)}{(\partial x^2)^2}$$

This yields the following approximation basis for the force functions

$$\begin{bmatrix} n_{x^1x^1} \\ n_{x^2x^1} \\ n_{x^1x^2} \\ n_{x^2x^2} \end{bmatrix} = \begin{bmatrix} 0 & 0 & 2 & 0 & 0 & 2x^1 & 6x^2 & -6(x^1)^2 & 0 & 6x^1x^2 & (12(x^2)^2 - 6(x^1)^2) & \\ 0 & -1 & 0 & 0 & -2x^1 & -2x^2 & 0 & 0 & 12x^1x^2 & -3(x^1)^2 & -3(x^1)^2 & 12x^1x^2 \\ 0 & -1 & 0 & 0 & -2x^1 & -2x^2 & 0 & 0 & 12x^1x^2 & -3(x^1)^2 & -3(x^1)^2 & 12x^1x^2 \\ 2 & 0 & 0 & 6x^1 & 2x^2 & 0 & 0 & (12(x^1)^2 - 6(x^2)^2) & 6x^1x^2 & 0 & -6(x^2)^2 & \end{bmatrix} \begin{bmatrix} c_4 \\ c_5 \\ . \\ . \\ c_{12} \\ c_{14} \\ c_{15} \end{bmatrix} + \begin{bmatrix} 0 & 0 \\ -x^2 & -x^1 \\ -x^2 & -x^1 \\ 0 & 0 \end{bmatrix} * \begin{bmatrix} \bar{q}^{x^1} \\ \bar{q}^{x^2} \end{bmatrix} \tag{4.59}$$

The strains follow from the forces (4.59) using the strain-force dependency

$$\varepsilon_{ij} = E_{ijk\ell}\, n^{k\ell}$$

$$\begin{bmatrix} \varepsilon_{x^1x^1} \\ \varepsilon_{x^2x^1} \\ \varepsilon_{x^1x^2} \\ \varepsilon_{x^2x^2} \end{bmatrix} = \frac{1}{E\,t} \begin{bmatrix} 1 & 0 & 0 & -\upsilon \\ 0 & (1+\upsilon)/2 & (1+\upsilon)/2 & 0 \\ 0 & (1+\upsilon)/2 & (1+\upsilon)/2 & 0 \\ -\upsilon & 0 & 0 & 0 \end{bmatrix} \begin{bmatrix} n_{x^1x^1} \\ n_{x^2x^1} \\ n_{x^1x^2} \\ n_{x^2x^2} \end{bmatrix} \tag{4.60}$$

E is the elastic modulas, $\upsilon$ is the Poisson's ratio.
For the displacement $\bar{u}_i$ along the four element boundaries we assume a displacement field that ensure the inter-element continuity in the customary way

$$\bar{u}_i = L^{m(m)}_{i((e)(b))}\, u_{m(m)} \tag{4.61}$$

The matrix $L^{m(m)}_{i((e)(b))}$ containes the shape functions of the element boundaries.
Now the implementation of the element can be carried out in an analogy to TFEQ-element described in section 4.2.

The element matrices can be derived by performing the integration over the biunit interval [-1,+1]. The use of 2×2 Gaussian integration formula leads to singular $H^{m(m)n(n)}$ matrix. The exact integration of $H^{m(m)n(n)}$ matrix shows its regularity. The use of at least 3×3 Gaussian integration formula is necessary.

#### 4.3 Space frame elements of Trefftz type

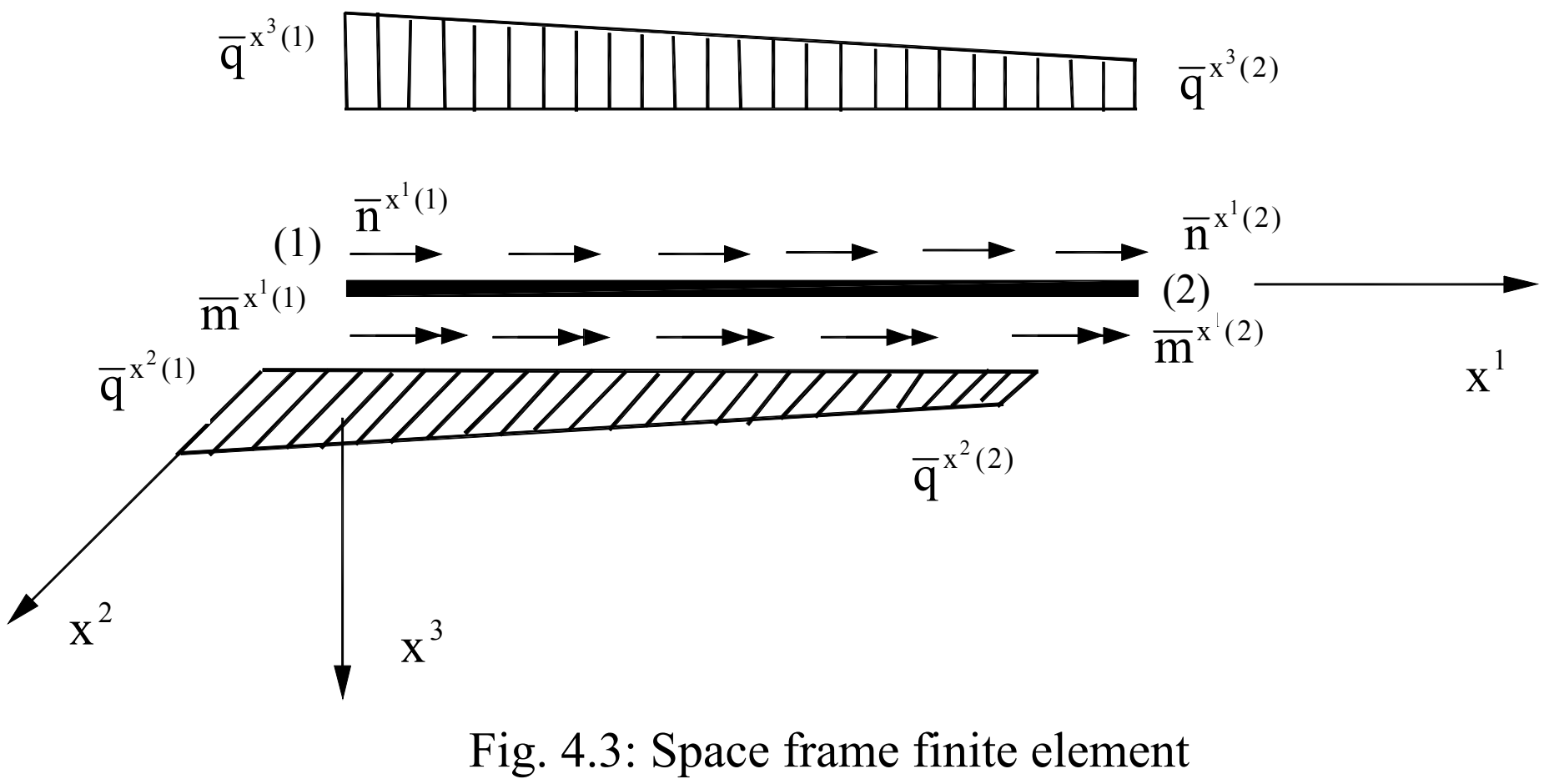

Fig. 4.3: Space frame finite element
Coordinate systems and element loading

The space frame element with the nodal points [(1), (2)] has six degrees of freedom per node represented by the vector $\{ u^0_{x^1},\ u^0_{x^2},\ u^0_{x^3},\ \varphi_{x^1},\ \varphi_{x^2},\ \varphi_{x^3} \}$. Figure 4.3 shows the element geometry, the element loading and the coordinate system used. The differential equations governing this problem are

$$EA\frac{\partial^2 u^0_{x^1}}{(\partial x^1)^2} = -\bar{n}^{x^1}(x^1) \tag{4.62a}$$

$$E\,I_{x^3}\,\frac{\partial^4 u^0_{x^2}}{(\partial x^1)^4} = \overline{q}^{x^2}(x^1) \tag{4.62b}$$

$$E\,I_{x^2}\,\frac{\partial^4 u^0_{x^3}}{(\partial x^1)^4} = \overline{q}^{x^3}(x^1) \tag{4.62c}$$

$$G\,I_D\,\frac{\partial^2 \varphi^0_{x^1}}{(\partial x^1)^2} = -\,\overline{m}^{x^1}(x^1) \tag{4.62d}$$

The following approximation function for the displacements is constructed such that the differential equations are identically satisfied

$$\begin{bmatrix} u^0_{x^1} \\ u^0_{x^2} \\ u^0_{x^3} \\ \varphi_{x^1} \end{bmatrix} = \begin{bmatrix} 1 & x^1 & 0 & 0 & 0 & 0 & 0 & 0 & 0 & 0 & 0 & 0 \\ 0 & 0 & 1 & x^1 & (x^1)^2 & (x^1)^3 & 0 & 0 & 0 & 0 & 0 & 0 \\ 0 & 0 & 0 & 0 & 0 & 0 & 1 & x^1 & (x^1)^2 & (x^1)^3 & 0 & 0 \\ 0 & 0 & 0 & 0 & 0 & 0 & 0 & 0 & 0 & 0 & 1 & x^1 \end{bmatrix} \begin{bmatrix} c_1 \\ . \\ . \\ c_{12} \end{bmatrix}$$

$$+ \begin{bmatrix} \overline{M}_{11} & \overline{M}_{12} & 0 & 0 & 0 & 0 & 0 & 0 \\ 0 & 0 & \overline{M}_{23} & \overline{M}_{24} & 0 & 0 & 0 & 0 \\ 0 & 0 & 0 & 0 & \overline{M}_{35} & \overline{M}_{36} & 0 & 0 \\ 0 & 0 & 0 & 0 & 0 & 0 & \overline{M}_{47} & \overline{M}_{48} \end{bmatrix} \begin{bmatrix} \overline{n}^{x^1(1)} \\ \overline{n}^{x^1(2)} \\ \overline{q}^{x^2(1)} \\ \overline{q}^{x^2(2)} \\ \overline{q}^{x^3(1)} \\ \overline{q}^{x^3(1)} \\ \overline{m}^{x^1(1)} \\ \overline{m}^{x^1(2)} \end{bmatrix} \tag{4.63}$$

Where

$$\overline{M}_{11} = \frac{1}{E\,A}\left[-\frac{(x^1)^2}{2} + \frac{(x^1)^3}{6\ell}\right] \;;\; \overline{M}_{12} = \frac{1}{E\,A}\left[-\frac{(x^1)^3}{6\ell}\right] \tag{4.64a}$$

$$\overline{M}_{23} = \frac{1}{E\,I_{x^3}}\left[\frac{(x^1)^4}{24} - \frac{(x^1)^5}{120\ell}\right] \;;\; \overline{M}_{24} = \frac{1}{E\,I_{x^3}}\left[\frac{(x^1)^5}{120\ell}\right] \tag{4.64b}$$

$$\overline{M}_{35} = \frac{1}{E I_{x^2}}\left[\frac{(x^1)^4}{24} - \frac{(x^1)^5}{120\ell}\right] \;\; ; \;\; \overline{M}_{36} = \frac{1}{E I_{x^2}}\left[\frac{(x^1)^5}{120\ell}\right] \qquad (4.64c)$$

$$\overline{M}_{47} = \frac{1}{G I_D}\left[-\frac{(x^1)^2}{2} + \frac{(x^1)^3}{6\ell}\right] \;\; ; \;\; \overline{M}_{48} = \frac{1}{G I_D}\left[-\frac{(x^1)^3}{6\ell}\right] \qquad (4.64d)$$

G is the shear modulus ($G = E/2(1+\upsilon)$) , $I_D$ is the St. Venant torosional moment of inertia, $I_{x^2}$ and $I_{x^3}$ are the moments of intertia about the corresponding axis.

Applying the displacement functions in association with the displacement version of the reduced Trefftz type approach can be also carried out in many forms.

### 4.3.1 Modified displacement element

The displacement form can be applied following a modified geometrical interpolation procedure similar to that adopted in section 4.1. The procedure used for linking the free parameters to the nodal degrees of freedom produces shape functions with a homogeneous part satisfying the homogeneous differential equations and a particular part satisfying the non-homogeneous differential equations. These functions satisfy, also, the inter-element continuity. Applying the shape functions in the corresponding variational form and evaluating the element integrals produces "exact" element matrices and vectors, see /63/ for details. The added particular term to the displacement functions does not effect the element stiffness matrix or the element mass matrix. It effects only the equivalent nodal forces corresponding to the distributed loading which now involve compared with the conventional displacement technique an extra term associated with the added particular shape functions. This extra term is expressed through the integral (4.26).

Substituting the interpolated homogeneous shape functions and the particular shape functions in the integrand (4.26) and performing the integration produce a zero force vector associated with the particular term. This means that the interpolated shape functions are in this case orthogonal to the homogeneous shape functions. Hereupon, the result for the nodal displacement remain unchanged but the results for the displacement and the forces inside the finite element will be changed and corrected corresponding to the non-homogeneous part. The results for the forces both inside the element and the nodal points are effected by the added particular term. The results obtained using the explained displacement element shows exact accuracy both for displacements and forces.

### 4.3.2 Hybrid element

The conventional hybrid form applying the generated displacement functions is briefly discussed in /62/. The integrated element matrices and vectors are also "exact". The produced numerical results are identical to that produced by the explained displacement element.

Now the trefftz hybrid and boundary formulation of the space frame problem should be next explained.

### 4.3.3 Hybrid Trefftz element

The force vector inside the frame element conjugated to the displacement functions defined by equation (4.63) and (4.64) can be derived using the force-displacement dependency (4.65)

$$\begin{bmatrix} N^{x^1} \\ M^{x^2} \\ M^{x^3} \\ M^{x^1} \end{bmatrix} = \begin{bmatrix} EA & 0 & 0 & 0 \\ 0 & EI_{x^2} & 0 & 0 \\ 0 & 0 & EI_{x^3} & 0 \\ 0 & 0 & 0 & GI_D \end{bmatrix} \begin{bmatrix} \dfrac{\partial u^0_{x^1}}{\partial x^1} \\ -\dfrac{\partial^2 u^0_{x^3}}{(\partial x^1)^2} \\ \dfrac{\partial^2 u^0_{x^2}}{(\partial x^1)^2} \\ \dfrac{\partial \varphi_{x^1}}{\partial x^1} \end{bmatrix} \tag{4.65}$$

This yields the following force functions inside the element

$$\begin{bmatrix} N^{x^1} \\ M^{x^2} \\ M^{x^3} \\ M^{x^1} \end{bmatrix} = \begin{bmatrix} 0 & EA & 0 & 0 & 0 & 0 & 0 & 0 & 0 & 0 & 0 & 0 \\ 0 & 0 & 0 & 0 & 0 & 0 & 0 & 0 & -2EI_{x^2} & -6EI_{x^2}x^1 & 0 & 0 \\ 0 & 0 & 0 & 0 & 2EI_{x^3} & 6EI_{x^3}x^1 & 0 & 0 & 0 & 0 & 0 & 0 \\ 0 & 0 & 0 & 0 & 0 & 0 & 0 & 0 & 0 & 0 & 0 & GI_D \end{bmatrix} \begin{bmatrix} c_1 \\ . \\ . \\ c_{12} \end{bmatrix}$$

$$+ \begin{bmatrix} -x^1 + \dfrac{(x^1)^2}{2\ell} & -\dfrac{(x^1)^2}{2\ell} & 0 & 0 & 0 & 0 & 0 & 0 \\ 0 & 0 & 0 & 0 & -\dfrac{(x^1)^2}{2} + \dfrac{(x^1)^3}{6\ell} & -\dfrac{(x^1)^3}{6\ell} & 0 & 0 \\ 0 & 0 & \dfrac{(x^1)^2}{2} - \dfrac{(x^1)^3}{6\ell} & \dfrac{(x^1)^3}{6\ell} & 0 & 0 & 0 & 0 \\ 0 & 0 & 0 & 0 & 0 & 0 & -x^1 + \dfrac{(x^1)^2}{2\ell} & -\dfrac{(x^1)^2}{2\ell} \end{bmatrix} \begin{bmatrix} \overline{n}^{x^1 (1)} \\ \overline{n}^{x^1 (2)} \\ \overline{q}^{x^2 (1)} \\ \overline{q}^{x^2 (2)} \\ \overline{q}^{x^3 (1)} \\ \overline{q}^{x^3 (1)} \\ \overline{m}^{x^1 (1)} \\ \overline{m}^{x^1 (2)} \end{bmatrix} \tag{4.66}$$

The boundary tractions at the finite element boundary (in this case the nodal points) follow from the previous force functions with the specific coordinates of the nodal points, i. e. $x^1 = 0$ at the left nodal point and $x^1 = \ell$ at the right nodal point.

Substituting these coordinates in equation (4.66) and constructing the boundary tractions involving in addition to the axial forces and moments the shear forces leads to

$$\begin{bmatrix} -N^{x^1(1)} \\ -Q^{x^2(1)} \\ -Q^{x^3(1)} \\ -M^{x^1(1)} \\ -M^{x^2(1)} \\ -M^{x^3(1)} \\ N^{x^1(1)} \\ Q^{x^2(1)} \\ Q^{x^3(1)} \\ M^{x^1(1)} \\ M^{x^2(1)} \\ M^{x^3(1)} \end{bmatrix} = \begin{bmatrix} 0 & -EA & 0 & 0 & 0 & 0 & 0 & 0 & 0 & 0 & 0 & 0 \\ 0 & 0 & 0 & 0 & 0 & 6EI_{x^2} & 0 & 0 & 0 & 0 & 0 & 0 \\ 0 & 0 & 0 & 0 & 0 & 0 & 0 & 0 & 0 & 6EI_{x^3} & 0 & 0 \\ 0 & 0 & 0 & 0 & 0 & 0 & 0 & 0 & 0 & 0 & 0 & -GI_D \\ 0 & 0 & 0 & 0 & 0 & 0 & 0 & 0 & 2EI_{x^3} & 6EI_{x^3}\ell & 0 & 0 \\ 0 & 0 & 0 & 0 & -2EI_{x^2} & -6EI_{x^2}\ell & 0 & 0 & 0 & 0 & 0 & 0 \\ 0 & EA & 0 & 0 & 0 & 0 & 0 & 0 & 0 & 0 & 0 & 0 \\ 0 & 0 & 0 & 0 & 0 & -6EI_{x^2} & 0 & 0 & 0 & 0 & 0 & 0 \\ 0 & 0 & 0 & 0 & 0 & 0 & 0 & 0 & 0 & -6EI_{x^3} & 0 & 0 \\ 0 & 0 & 0 & 0 & 0 & 0 & 0 & 0 & 0 & 0 & 0 & GI_D \\ 0 & 0 & 0 & 0 & 0 & 0 & 0 & 0 & -2EI_{x^3} & -6EI_{x^3}\ell & 0 & 0 \\ 0 & 0 & 0 & 0 & 2EI_{x^2} & 6EI_{x^2}\ell & 0 & 0 & 0 & 0 & 0 & 0 \end{bmatrix} \begin{bmatrix} c_1 \\ . \\ . \\ c_{12} \end{bmatrix} \tag{4.67}$$

$$+ \begin{bmatrix} 0 & 0 & 0 & 0 & 0 & 0 & 0 & 0 \\ 0 & 0 & 0 & 0 & 0 & 0 & 0 & 0 \\ 0 & 0 & 0 & 0 & 0 & 0 & 0 & 0 \\ 0 & 0 & 0 & 0 & 0 & 0 & 0 & 0 \\ 0 & 0 & 0 & 0 & 0 & 0 & 0 & 0 \\ 0 & 0 & 0 & 0 & 0 & 0 & 0 & 0 \\ -\frac{\ell}{2} & -\frac{\ell}{2} & 0 & 0 & 0 & 0 & 0 & 0 \\ 0 & 0 & -\frac{\ell}{2} & -\frac{\ell}{2} & 0 & 0 & 0 & 0 \\ 0 & 0 & 0 & 0 & -\frac{\ell}{2} & -\frac{\ell}{2} & 0 & 0 \\ 0 & 0 & 0 & 0 & 0 & 0 & -\frac{\ell}{2} & -\frac{\ell}{2} \\ 0 & 0 & 0 & 0 & -\frac{(\ell)^2}{3} & -\frac{(\ell)^2}{6} & 0 & 0 \\ 0 & 0 & \frac{(\ell)^2}{3} & \frac{(\ell)^2}{6} & 0 & 0 & 0 & 0 \end{bmatrix} \begin{bmatrix} \overline{n}^{x^1\,(1)} \\ \overline{n}^{x^1\,(2)} \\ \overline{q}^{x^2\,(1)} \\ \overline{q}^{x^2\,(2)} \\ \overline{q}^{x^3\,(1)} \\ \overline{q}^{x^3\,(1)} \\ \overline{m}^{x^1\,(1)} \\ \overline{m}^{x^1\,(2)} \end{bmatrix}$$

The vector of boundary displacements conjugated to the boundary tractions is defined by the vector

$$\{ u^0_{x^1(1)} \; u^0_{x^2(1)} \; u^0_{x^3(1)} \; \varphi_{x^1(1)} \; \varphi_{x^2(1)} \; \varphi_{x^3(1)} \; u^0_{x^1(2)} \; u^0_{x^2(2)} \; u^0_{x^3(2)} \; \varphi_{x^1(2)} \; \varphi_{x^2(2)} \; \varphi_{x^3(2)} \}. \tag{4.68}$$

This vector is identical with the nodal degrees of freedom of the space frame element. Therefore, we define the prescribed boundary displacement $\bar{u}_i$ through the identity matrix.

Now all the quantities needed to eliminate the undetermined parameters and to evaluate the element matrices defined in a similar way as given in the equation (4.34) to (4.47) are prepared. The matrices $H^{m(m)n(n)}$, $\overline{H}^{m(m)}_{(p)}$, $T^{k(k)\ell(\ell)}$, $\overline{T}^{k(k)}_{(p)}$ and the vectors $\bar{r}^{k(k)}$ can be easily evaluted by integration over the beam length and the stiffness matrix $k^{k(k)m(m)}$ and the right side $\bar{r}^{k(k)}$ can be calculated.

### 4.3.4 Boundary Trefftz element

In order to implement the element as a boundary element as described in section 4.3 we derive the boundary displacements, needed to evaluate the left side of integral (4.48), which is used in order to eliminate the undetermined parameters, from the internal displacement field (4.63) by substituting the specific coordinates of the nodal points, i. e. $x^1 = 0$ at the left nodal point and $x^1 = \ell$ at the right nodal point, and arranging the boundary displacements as in (4.68), which corresponds to the boundary tractions (4.66). This yields the boundary displacement field given in (4.69).

Now we can evaluate the boundary integral (4.48) using the difinitions (4.69) and (4.66), which leads directly to the matrices $H^{m(m)n(n)}$, $\overline{H}^{m(m)}_{(p)}$. Note that the integral over the element boundary reduces to a simple multiplication of the defined forces by the displacement. In the same way we evaluate the matrices $T^{k(k)\ell(\ell)}$ and $\overline{T}^{k(k)}_{(p)}$.

Both hybrid and boundary technique described above lead to the same matrices $H^{m(m)n(n)}$, $\overline{H}^{m(m)}_{(p)}$, $T^{k(k)\ell(\ell)}$, $\overline{T}^{k(k)}_{(p)}$ and the vectors $\bar{r}^{k(k)}$, which means that the element stiffness matrix and the load vectors are the same in both cases. Finally, the stiffness matrix and load vectors are “exact” in both cases.

$$
\begin{bmatrix} u^0_{x^1(1)} \\ u^0_{x^2(1)} \\ u^0_{x^3(1)} \\ \varphi_{x^1(1)} \\ \varphi_{x^2(1)} \\ \varphi_{x^3(1)} \\ u^0_{x^1(2)} \\ u^0_{x^2(2)} \\ u^0_{x^3(2)} \\ \varphi_{x^1(2)} \\ \varphi_{x^2(2)} \\ \varphi_{x^3(2)} \end{bmatrix}
=
\begin{bmatrix}
1 & 0 & 0 & 0 & 0 & 0 & 0 & 0 & 0 & 0 & 0 & 0 \\
0 & 0 & 1 & 0 & 0 & 0 & 0 & 0 & 0 & 0 & 0 & 0 \\
0 & 0 & 0 & 0 & 0 & 0 & 1 & 0 & 0 & 0 & 0 & 0 \\
0 & 0 & 0 & 0 & 0 & 0 & 0 & 0 & 0 & 0 & 1 & 0 \\
0 & 0 & 0 & 0 & 0 & 0 & 0 & -1 & 0 & 0 & 0 & 0 \\
0 & 0 & 0 & 1 & 0 & 0 & 0 & 0 & 0 & 0 & 0 & 0 \\
1 & \ell & 0 & 0 & 0 & 0 & 0 & 0 & 0 & 0 & 0 & 0 \\
0 & 0 & 1 & \ell & (\ell)^2 & (\ell)^3 & 0 & 0 & 0 & 0 & 0 & 0 \\
0 & 0 & 0 & 0 & 0 & 0 & 1 & \ell & (\ell)^2 & (\ell)^3 & 0 & 0 \\
0 & 0 & 0 & 0 & 0 & 0 & 0 & 0 & 0 & 0 & 1 & \ell \\
0 & 0 & 0 & 0 & 0 & 0 & 0 & -1 & -2\ell & -3(\ell)^2 & 0 & 0 \\
0 & 0 & 0 & 1 & 2\ell & 3(\ell)^2 & 0 & 0 & 0 & 0 & 0 & 0
\end{bmatrix}
\begin{bmatrix} c_1 \\ . \\ . \\ c_{12} \end{bmatrix}
$$

$$
+ \begin{bmatrix}
0 & 0 & 0 & 0 & 0 & 0 & 0 & 0 \\
0 & 0 & 0 & 0 & 0 & 0 & 0 & 0 \\
0 & 0 & 0 & 0 & 0 & 0 & 0 & 0 \\
0 & 0 & 0 & 0 & 0 & 0 & 0 & 0 \\
0 & 0 & 0 & 0 & 0 & 0 & 0 & 0 \\
0 & 0 & 0 & 0 & 0 & 0 & 0 & 0 \\
-\frac{\ell^2}{3EA} & -\frac{\ell^2}{6EA} & 0 & 0 & 0 & 0 & 0 & 0 \\
0 & 0 & \frac{\ell^4}{30EI_{x^3}} & \frac{\ell^4}{120EI_{x^3}} & 0 & 0 & 0 & 0 \\
0 & 0 & 0 & 0 & \frac{\ell^4}{30EI_{x^2}} & \frac{\ell^4}{120EI_{x^2}} & 0 & 0 \\
0 & 0 & 0 & 0 & 0 & 0 & -\frac{\ell^2}{3GI_D} & -\frac{\ell^2}{6GI_D} \\
0 & 0 & 0 & 0 & -\frac{\ell^3}{8EI_{x^2}} & -\frac{\ell^3}{24EI_{x^2}} & 0 & 0 \\
0 & 0 & \frac{\ell^3}{8EI_{x^3}} & \frac{\ell^3}{24EI_{x^3}} & 0 & 0 & 0 & 0
\end{bmatrix}
\begin{bmatrix} \overline{n}^{x^1(1)} \\ \overline{n}^{x^1(2)} \\ \overline{q}^{x^2(1)} \\ \overline{q}^{x^2(2)} \\ \overline{q}^{x^3(1)} \\ \overline{q}^{x^3(1)} \\ \overline{m}^{x^1(1)} \\ \overline{m}^{x^1(2)} \end{bmatrix}
\tag{4.69}
$$

# 5 Element investigations

## 5.1 Preliminary test of rectangular elements

### 5.1.1 Square plate with various boundary conditions and loads

The first set of numerical examples concerns the analysis of a square plate with various boundary conditions and loads. The testing problem defined on the square plate in Fig. 5.1 are: a) a simply supported square plate subjected to a hydrostatic pressure that varies along the $x^1$-axis; b). a plate with two simply supported edges and two clamped edges subjected to sinusoidal load and c) a simply supported square plate subjected to pyramid-type loading.
The following normalized values of the transverse displacements, bending moments, torsion moments and shear forces are used below

$$w(x^1,x^2) = u^0_{x^3}(x^1,x^2)\,D/(\overline{q}^0/a^4) \tag{5.1}$$

$$m^{ij}(x^1,x^2) = M^{x^ix^j}(x^1,x^2)\;/(\overline{q}^0/a^2) \tag{5.2}$$

$$q^i\,(x^1,x^2) = Q^{x^i}\,(x^1,x^2)\;/(\overline{q}^0/a) \tag{5.3}$$

The results obtained for the plate tests defined in Fig. 5.1 with different elements and meshes are summarised in Tables 5.1 to 5.5. For a comparison the analytical solutions obtained by Reddy /68/ are also given . The results show very good agreement between the finite element calculations and the analytical solutions for all test problems.

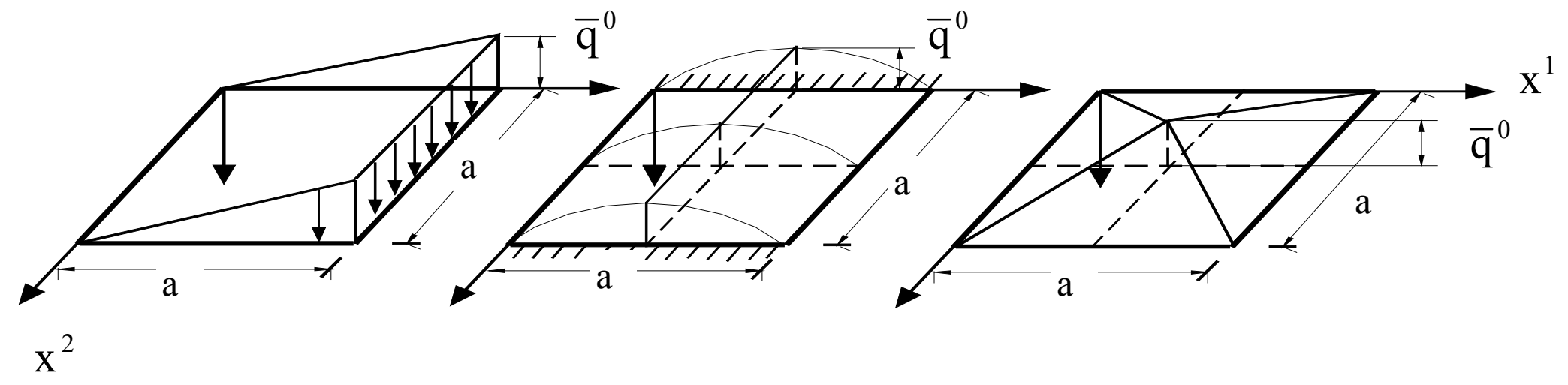

a) simply supported under hydrostatic pressure

b) two edges simply supported and two edges clamped under sinusoidal loading

c) simply supported under pyramid-type loading

Fig. 5.1: Square plate under various types of loading with different boundary conditions

The elements presented here are assessed also using the results obtained with the hybrid Trefftz elements HTQI1 and HTQI2 due to Jirousek /29/. These quadrilateral elements have mid-side nodes and corner nodes, for a total of 24 degrees of freedom and 21 trial functions. The frame functions are polynomials of degree five for the transverse displacement and of the second degree for the slopes. This relative assessment is based on the number of degrees of freedom.

Table 5.4 shows the results obtained for the displacement and the bending moment at the centre of the square plate, simply supported and subject to a uniform load $\overline{q}^0$ . It shows also the drill-moment at the plate corner and the shear force at the mid-side boundary point, see Fig. 5.2 a.

The percentage errors in the same displacement and force components measured now in the clamped square plate, subjected also to a uniform load $\overline{q}^0$ , shown in Fig. 5.2 b, are presented in table 5.5. The error is computed for the solutions found in /29/. Bearing in mind the simplicity of the elements presented here, their numerical performance seems to be quite satisfactory.

Table 5.1: Results at the centre of a simply supported square plate subjected to a hydrostatic pressure

| Analytical | $u^0_{x^3}(a/2,a/2)/(\overline{q}^0 a^4/D) \approx 0.00203$ | | | $M^{x^1x^1}(a/2,a/2)/(\overline{q}^0 a^2) \approx 0.0235$ | | |
|---|---|---|---|---|---|---|
| Mesh | ZDE | TFE | JFE | ZDE | TFE | JFE |
| 2×2 | 0.002023 | 0.001953 | 0.001935 | 0.030834 | 0.023698 | 0.023700 |
| 4×4 | 0.002038 | 0.002026 | 0.002026 | 0.025063 | 0.023923 | 0.023966 |
| 6×6 | 0.002036 | 0.002031 | 0.002031 | 0.024409 | 0.023961 | 0.023963 |
| 8×8 | 0.002034 | 0.002031 | 0.002031 | 0.024187 | 0.023946 | 0.023941 |
| 10×10 | 0.002033 | 0.002031 | 0.002032 | 0.024121 | 0.023951 | 0.023973 |
| 12×12 | 0.002032 | 0.002031 | 0.002032 | 0.024023 | 0.023925 | 0.023941 |
| 14×14 | 0.002032 | 0.002031 | 0.002031 | 0.024060 | 0.023934 | 0.023945 |

Table 5.2: Results at the centre of an $a \times a$ square plate, with two edges simply supported $(x^1 = 0; x^1 = a)$ and the other two edges clamped $(x^2 = 0; x^2 = a)$ subjected to sinusoidal load, $\overline{q} = \overline{q}^0 \sin(\pi x^1/a)$.

| Analytical | $u^0_{x^3}(a/2,a/2)/(\overline{q}^0 a^4/D) \approx 0.00154$ | | | $M^{x^1x^1}(a/2,a/2)/(\overline{q}^0 a^2) \approx 0.0268$ | | |
|---|---|---|---|---|---|---|
| Mesh | ZDE | TFE | JFE | ZDE | TFE | JFE |
| 2×2 | 0.001490 | 0.000868 | 0.000936 | 0.040118 | 0.025190 | 0.022615 |
| 4×4 | 0.001550 | 0.001353 | 0.001359 | 0.029862 | 0.024854 | 0.024652 |
| 6×6 | 0.001549 | 0.001460 | 0.001462 | 0.028341 | 0.026130 | 0.026121 |
| 8×8 | 0.001546 | 0.001496 | 0.001497 | 0.027889 | 0.026609 | 0.026601 |
| 10×10 | 0.001544 | 0.001513 | 0.001513 | 0.027626 | 0.026857 | 0.026820 |
| 12×12 | 0.001543 | 0.001522 | 0.001522 | 0.027516 | 0.026954 | 0.026938 |
| 14×14 | 0.001543 | 0.001527 | 0.001528 | 0.027396 | 0.027011 | 0.027057 |

Table 5.3 : Results at the centre of a simply supported square plate subjected to a pyramid-type loading

| Analytical | $u^0_{x^3}(a/2,a/2)/(\bar{q}^0 a^4/D) \approx 0.002083$ | | | $M^{x^1x^1}(a/2,a/2)/(\bar{q}^0 a^2) \approx 0.0271$ | | |
|---|---|---|---|---|---|---|
| Mesh | ZDE | TFE | JFE | ZDE | TFE | JFE |
| 4×4 | 0.001979 | 0.001776 | 0.001776 | 0.0266 | 0.0228 | 0.0228 |
| 6×6 | 0.002069 | 0.001975 | 0.001975 | 0.0273 | 0.0254 | 0.0254 |
| 8×8 | 0.002057 | 0.002003 | 0.002004 | 0.0271 | 0.0259 | 0.0260 |
| 10×10 | 0.002062 | 0.002027 | 0.002027 | 0.0271 | 0.0263 | 0.0263 |

Table 5.4: Simply supported square plate subjected to uniformly distributed load

| Elements and nodes | $w(a/2,a/2)$ | $m^{11}(a/2,a/2)$ | $-m^{11}(0,0)$ | $q^1(0,a/2)$ |
|---|---|---|---|---|
| ZDE (45) | 0.004071 | 0.048829 | 0.033685 | 0.30845 |
| TFE (45) | 0.004061 | 0.047933 | 0.036033 | 0.29890 |
| JFE (45) | 0.004061 | 0.047937 | 0.036064 | 0.29931 |
| HTQI1 (40) | 0.00406 | 0.0479 | 0.0348 | 0.341 |
| Analytical | 0.00406 | 0.0479 | 0.0325 | 0.338 |

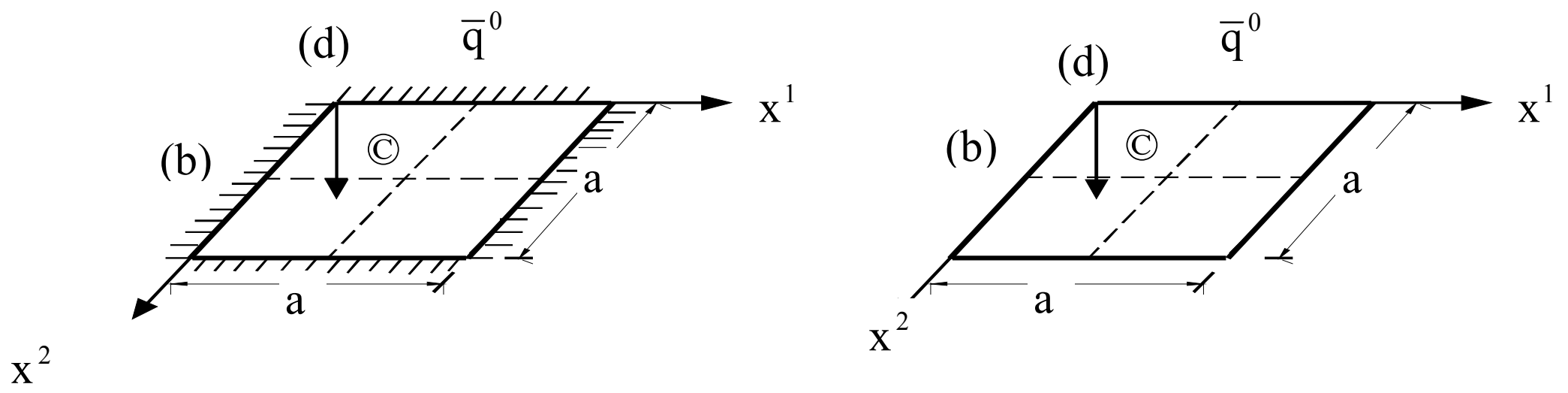

b) clamped plate under uniformly distributed load

a) simply supported plate under uniformly distributed load

Fig.5.2: Square plate under various types of boundary conditions

Table 5.5: Clamped square plate subjected to uniformly distributed load (percent errors)

| Elements and nodes | $w(a/2,a/2)$ | $m^{11}(a/2,a/2)$ | $-m^{11}(0,0)$ | $q^1(0,a/2)$ |
|---|---|---|---|---|
| ZDE | -1.9 | +3.1 | -1.1 | +9.9 |
| TFE | +0.2 | +0.08 | -0.5 | +8.3 |
| JFE ( | +0.2 | +0.13 | -0.7 | +9.2 |
| CIQ | -4.4 | +4.9 | -8.2 | - |
| HTQI1 | 0.0 | +0.3 | -0.6 | +1.7 |
| HTQI2 | 0.0 | +0.3 | -0.7 | -1.3 |
| Analytical | 0.001265 | 0.02291 | 0.0513 | 0.441 |

### 5.1.2 Simply supported rectangular plate

The last preliminary test is on a 4×6 rectangular plate with thickness h = 0.2, unit modulus of elasticity and Poisson ratio $\nu = 0.3$. The plate is simply supported and solved for a unit uniform load. The results for the centre transverse displacement and bending moments obtained with a 4×4 element mesh (symmetry is not used) and eight boundary elements are presented in table 5.6 . Shown in the same table are the solutions obtained with Trefftz elements /41/, and the analytical solutions presented in /69/. The quality of the results obtained with the current elements compare favourably with those obtained with the relatively more complex approach proposed by Piltner /41/.

Table 5.6: Simply supported rectangular plate subjected to a unit uniform load

| Solution | ZDE | TFE | JFE | Piltner | Timoshenko |
|---|---|---|---|---|---|
| $u^0_{x^3}$ | 2672.88 | 2696.10 | 2694.66 | 2698.70 | 2697.60 |
| $M^{x^1x^1}$ | 1.3567 | 1.2950 | 1.2916 | 1.2984 | 1.2992 |
| $M^{x^2x^2}$ | 0.8061 | 0.7737 | 0.7770 | 0.7974 | 0.7984 |

### 5.1.3 Cantilevered beam problem

For the clamped-free beam shown in Fig. 5.3 of dimensions $0.9 \times 0.2 \times 0.036$ the analytical deflection at the free end, neglecting the shear deformation effect, is given by

$$u^0_{x^3} = p\,L^3 / 3\,E\,I \tag{5.4}$$

In (5.4), p is the concentrated load, E is the elasticity modulus, I is the second moment of inertia and L is the length of the beam. The beam is modelled by six elements with a distortion ratio one to five. The result obtained for the deflection using the undistorted mesh is exact compared with the analytical value. In Fig 5.3 the solution obtained for the deflection is plotted in 3-D representation (continuous line). Table 5.7 shows a comparison of the result of the present solution with that obtained by QUAD4 /70/, MITC4 /71/ and HSQK1 /54/.

Table 5.7: Percentage error in the deflection at the free end of the beam $w_{anal.}$ =15.625

| a / b | MITC4 | QUAD4 | HSQK1 | ZDEQ |
|---|---|---|---|---|
| 1 | 0.96 | 0.0 | 0.12 | 0.0 |
| 5 | 1.58 | -0.96 | 0.13 | 0.18 |

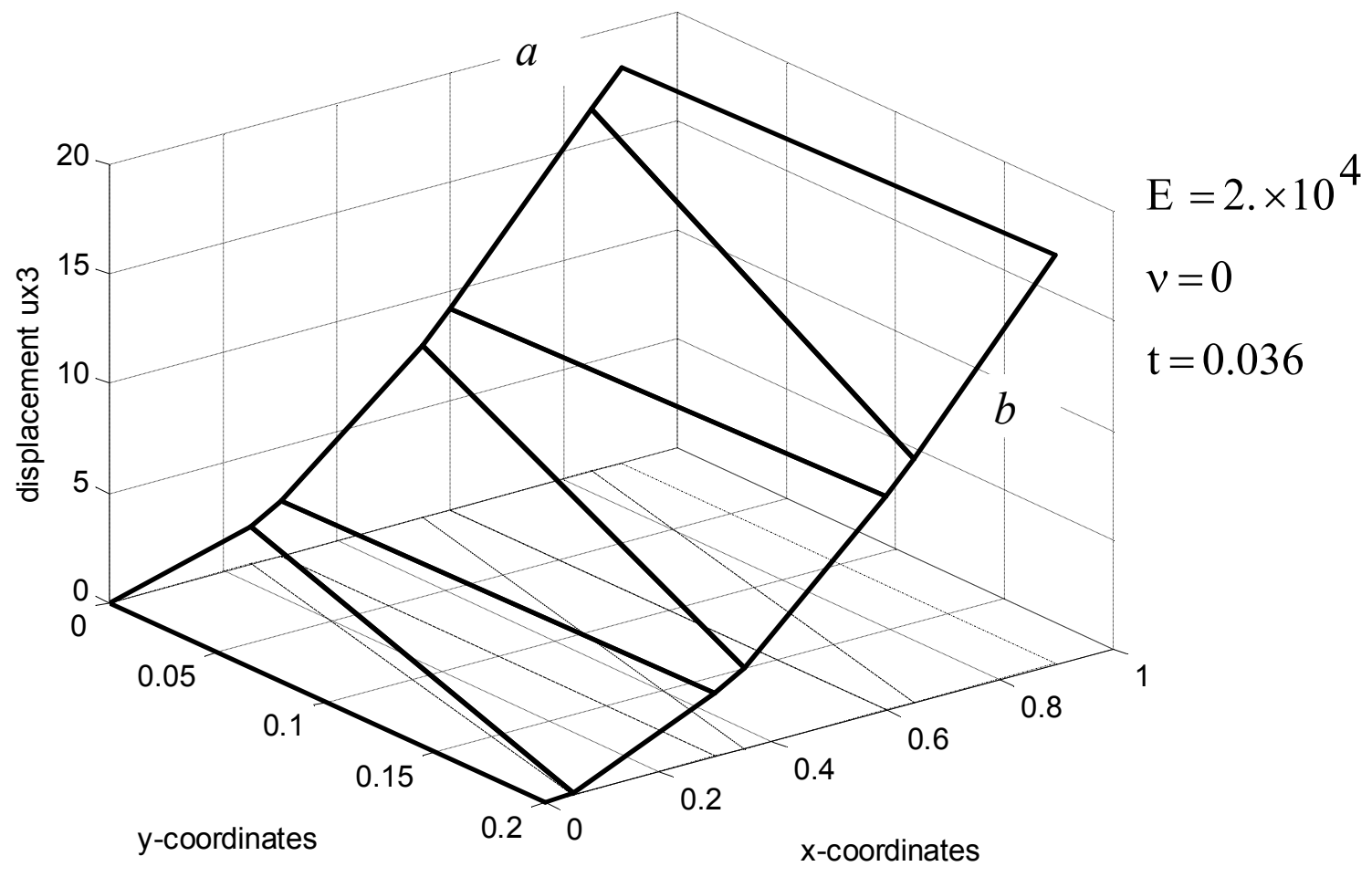

Fig. 5.3: 3-D representation of beam deflection

## 5.2 Invariance properties

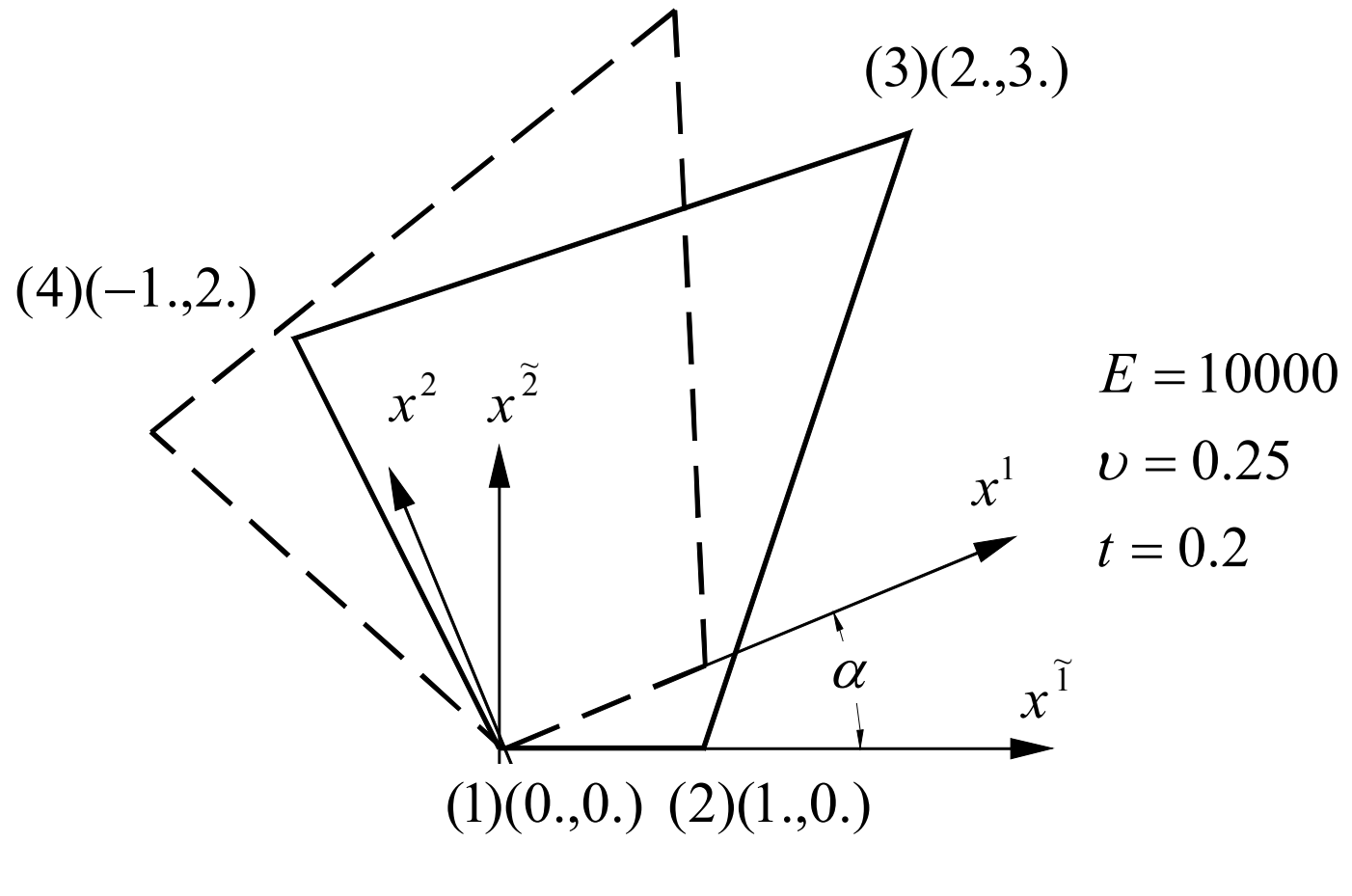

Fig. 5.4: Single quadrilateral element, coordinates, geometry properties

In the first case the stiffness matrix is numerically tested by an eigenvalue analysis of the single quadrilateral element structure presented in Fig. 5.4. The element coordinates and the geometry properties are shown in the same figure. The eigenvalue analysis of the stiffness matrix of the quadrilateral element shows that the only zero energy modes are the three rigid-body modes, which means that these elements are numerically stable.

The invariance study of the elements is performed using the single element structure also shown in Fig. 5.4. The global co-ordinate system $x^{\tilde{1}}, x^{\tilde{2}}$ is located in node (1). The element is subjected to a unit force in the positive direction of the $x^{\tilde{3}}$ co-ordinate axis at node (3). The boundary conditions are chosen such that the structure is clamped at node (1). The element is rotated about node (1) for different rotation angles $\alpha$. The structure is then computed for each rotation angle. The computed values $u^0_{x^{\bar{3}}}, \varphi_{x^{\bar{1}}}, \varphi_{x^{\bar{2}}}$ are transformed to a local system $x^1, x^2$ located in node (1). These axes are inclined at an angle $\alpha$ relative to the global axes. The local $x^1$ axis corresponds with the direction of the element boundary (1), (2). The global values $u^0_{x^{\bar{3}}}, \varphi_{x^{\bar{1}}}, \varphi_{x^{\bar{2}}}$ are transformed to the local values $u^0_{x^3}, \varphi_{x^1}, \varphi_{x^2}$ using the following rotation matrix:

$$\begin{bmatrix} u^0_{x^3} \\ \varphi_{x^1} \\ \varphi_{x^2} \end{bmatrix} = \begin{bmatrix} 1 & 0 & 1 \\ 0 & \cos\alpha & \sin\alpha \\ 0 & -\sin\alpha & \cos\alpha \end{bmatrix} \begin{bmatrix} u^0_{x^{\bar{3}}} \\ \varphi_{x^{\bar{1}}} \\ \varphi_{x^{\bar{2}}} \end{bmatrix} \qquad (5.5)$$

The computed values at nodal point (3) listed in table 5.8 show the desired invariance property of the element (ZDEQ). By performing an eigenvalue analysis on the rotated structure for all the rotation angles listed in Tab 5.8 it is found that the eigenvalues remain unchanged by varying the rotation angle (see Table 5.9). The computed values for the bending moments and the shear forces in the defined local coordinate systems remain also unchanged (see Table 5.10)..

A standard patch of elements is also studied and the element stiffness matrices are assembled to construct the global stiffness matrix. The patch is subjected to prescribed boundary displacements and rotations corresponding to rigid-body motions. The remaining displacements are calculated by solving the algebraic system of equations for the patch. It is found that the element is able to represent rigid-body motions as well as constant strains.

Table 5.8: Deflection and rotation at nodal point (3) under unit force applied in (3), computed using the element ZDEQ (exactly integrated stiffness matrix)

| $\alpha$ | $u^0_{x^3}$ | $\varphi_{x^{\bar{1}}}$ | $\varphi_{x^{\bar{2}}}$ | $\varphi_{x^1}$ | $\varphi_{x^2}$ |
|---|---|---|---|---|---|
| $0^\circ$ | 1.845160 | 0.701552 | -0.442977 | 0.701552 | -0.442977 |
| $22.5^\circ$ | 1.845160 | 0.817669 | -0.140785 | 0.701552 | -0.442977 |
| $45^\circ$ | 1.845160 | 0.809304 | 0.182840 | 0.701552 | -0.442977 |
| $67.5^\circ$ | 1.845160 | 0.677730 | 0.478629 | 0.701552 | -0.442977 |
| $90^\circ$ | 1.845160 | 0.442977 | 0.701552 | 0.701552 | -0.442977 |

Table 5.9: Frequencies of the rotated structure for all the rotation angles listed in Tab 5.8 Obtained using the ZDEQ-element (exactly integrated stiffness and mass matrices)

| $\omega_1$ | $\omega_2$ | $\omega_3$ | $\omega_4$ | $\omega_5$ | $\omega_6$ | $\omega_7$ | $\omega_8$ | $\omega_9$ |
|---|---|---|---|---|---|---|---|---|
| 0.666902 | 2.016901 | 4.637127 | 8.129742 | 14.14990 | 17.17961 | 20.89449 | 25.92336 | 38.80187 |

Table 5.10: Results of the ZDEQ-Element using 3x3 Gauss-integration formula

```
INFO:Eigenvalues D and Eigenvectors V
the 1. normalized angular frequency w is=            0.664051
the 2. normalized angular frequency w is=            2.001157
the 3. normalized angular frequency w is=            4.441355
the 4. normalized angular frequency w is=            7.758092
the 5. normalized angular frequency w is=           14.726824
the 6. normalized angular frequency w is=           18.905644
the 7. normalized angular frequency w is=           27.730588
the 8. normalized angular frequency w is=           33.413430
the 9. normalized angular frequency w is=           51.644230
statical calculation of the plate
nodal displacements
node    ux        phix          phiy
1   0.00000000   0.00000000   0.00000000
2  -0.00773080   0.30584045   0.01663497
3   1.84516050   0.70155173  -0.44297734
4   0.36000582   0.37358474  -0.26097220
Forces pro element in global Cartesian system:
El. node mxx          mxy          myx          myy          Qx          Qy
1   1   -0.77189831  -1.48654605  -1.48654605  -3.45489125   1.75814241   1.85381377
1   2   -0.25894691  -1.79341139  -1.79341139  -1.59082312   2.04508885   0.70602803
1   3   -0.90491222  -0.65850258  -0.65850258  -1.47982036  -1.11132195  -1.30259702
1   4    0.23660692  -0.53721770  -0.53721770   0.30489971  -0.82437551   2.42770664

Forces pro element in the defined local system:
1 4
1 1  -1.17249242  -1.76752219  -1.76752219  -3.05429715   1.97147465   1.62510860
1 2  -0.71379631  -1.90137833  -1.90137833  -1.13597372   2.11603095   0.45086252
1 3  -1.07320522  -0.70855919  -0.70855919  -1.31152736  -1.26215098  -1.15705245
1 4   0.10733202  -0.51289600  -0.51289600   0.43417462  -0.52157380   2.51024208
```

## 5.3 Sensitivity to mesh distortion

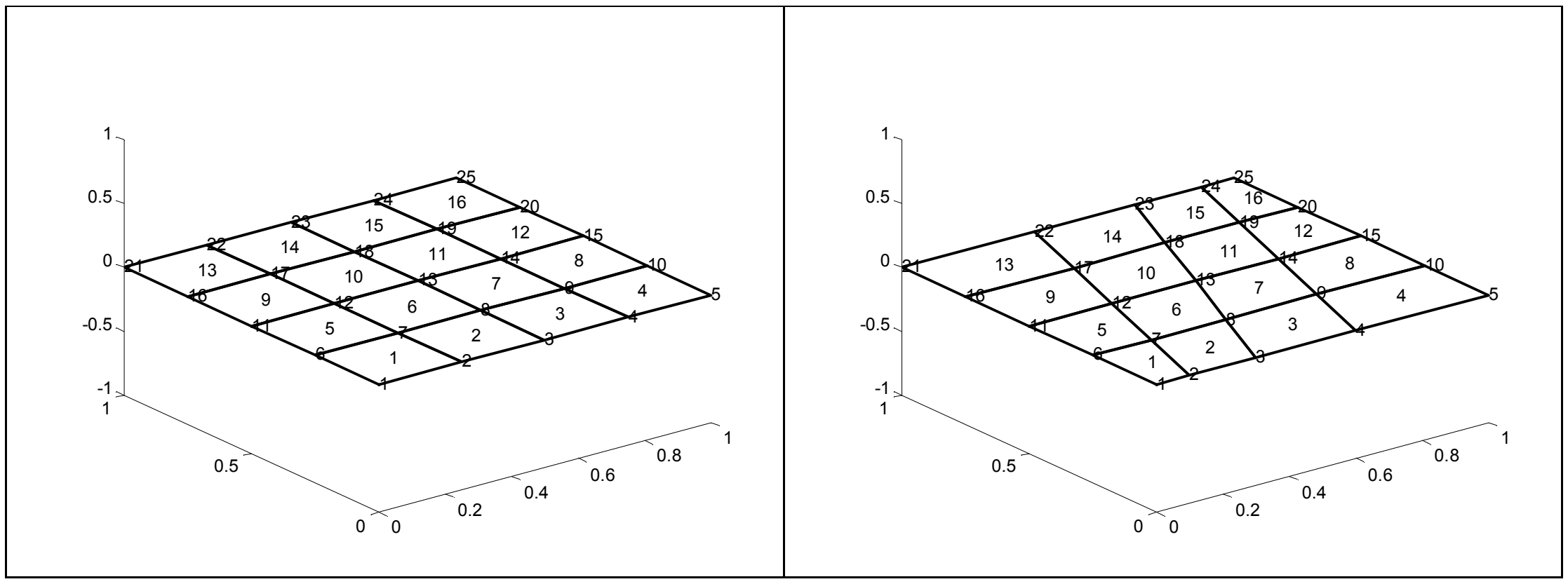

a) Regular mesh    b) Distorted mesh

Fig. 5.5: Simply supported square plate meshed by 4×4 elements for studying the influence of mesh distortion.

The influence of mesh distortion on the results is studied with the aid of a simply supported square plate under uniformly distributed load. The plate is first meshed by 4×4 regular elements, and secondly by the highly distorted 4×4 quadrilateral elements shown in Fig. 5.5. Tab. 5.11 shows the result of the central displacement and the percentage error for the irregular mesh compared to the distorted mesh. It also represents the same percentage error for various known finite elements such as HTQ3 /29/, HSQK /54/, DKQ /72/ and QUAD4 /70/.

Tab 5.11: Percentage difference in central deflection for distorted mesh over the simply supported square plate (Fig. 5.5) compared to the regular mesh.

| ZDEQ | TFEQ | JFEQ | QUAD4 Ref. /70/ | DKQ Ref. /72/ | HTQ3 Ref. /29/ | HSQK1 Ref. /54/ |
|---|---|---|---|---|---|---|
| 0.61 | -0.40 | -0.40 | 5.94 | 0.45 | -0.04 | 0.17 |

## 5.4 Preliminary test of the plain stress element

### 5.4.1 Cantilevered plain stress problem

For the clamped-free beam meshed in Fig. 5.6 of dimensions 12×48×1, the analytical deflection at the free end, considering the shear deformation effect, is given by

$$u^{0}_{x^1} = p\,L^3 / 3\,E\,I + pL / GA_D \tag{5.6}$$

In (5.6), p is the concentrated load, E is the elasticity modulus, I is the second moment of inertia, G the shear modulus $A_D$ the effective shear area and L is the length of the beam. The beam is modelled by 4×16 rectangular plain stress elements. The result obtained for the deflection using this element agree with the analytical beam solution. Table 5.12 shows the displacements at the nodal points of the free end of the beam computed firstly using stiffness matrix of the element integrated numerically with 3×3 Gaussian integration formula for the $H^{m(m)n(n)}$ and secondly stiffness matrix integrated exactly using MATLAB language code /95 /. Table 5.13 shows a comparison of the result of the present solution with that obtained by FALT-FEM elements based on hybrid formulation / 75-90/.

Table 5.12: Free-end nodal point displacements

| Nodal point | (81) | (82) | (83) | (84) | (85) |
|---|---|---|---|---|---|
| Gauss integration | 0.35437484 | 0.35293386 | 0.35269853 | 0.35293386 | 0.35437484 |
| Exact integration | 0.35513325 | 0.35410994 | 0.35380746 | 0.35410994 | 0.35513325 |

Table 5.13: Percentage error in the deflection at the free end of the beam $u_{x^1}^0 = 0.35583$

| Mesh 4×16 | $u_{x^1}^0 = 0.35583$ |
|---|---|
| Plain stress element | 0.99432 |
| FALT-FEM-HSM | 0.9974 |
| FALT-FEM-HSO | 0.9951 |

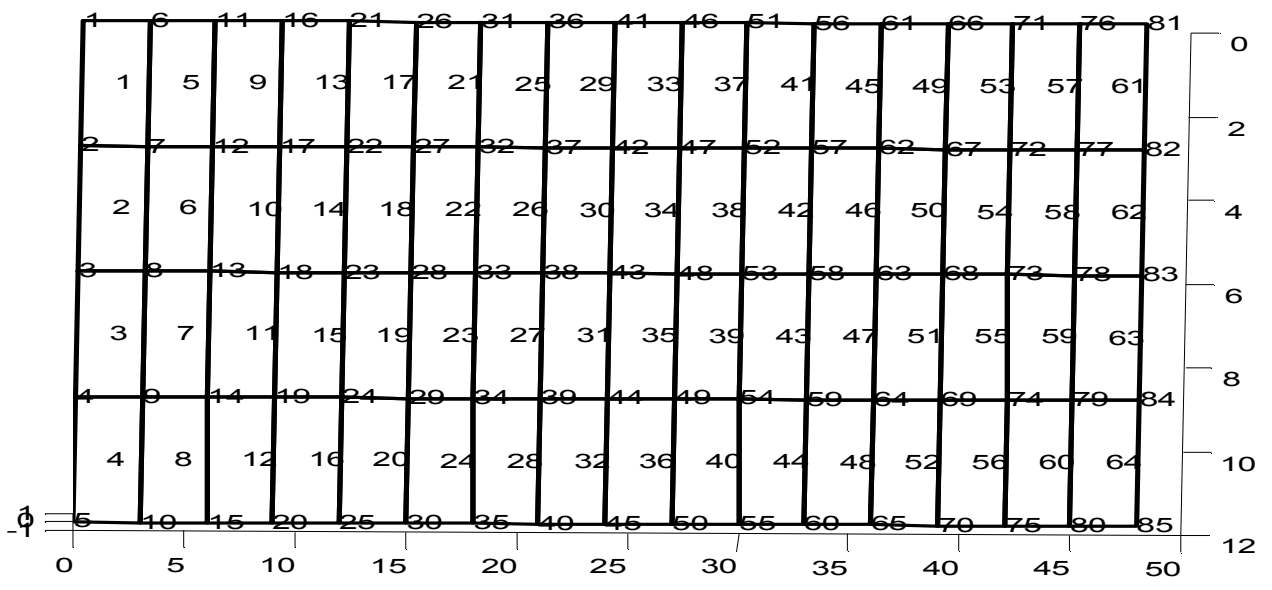

$p = 40$

$E = 30000$

$\nu = 0.25$

$t = 1$

Fig. 5.6: Finite element mesh of the cantilevered beam

## 5.4.2 Cook's membrane problem: plain stress structure (see 4.2)

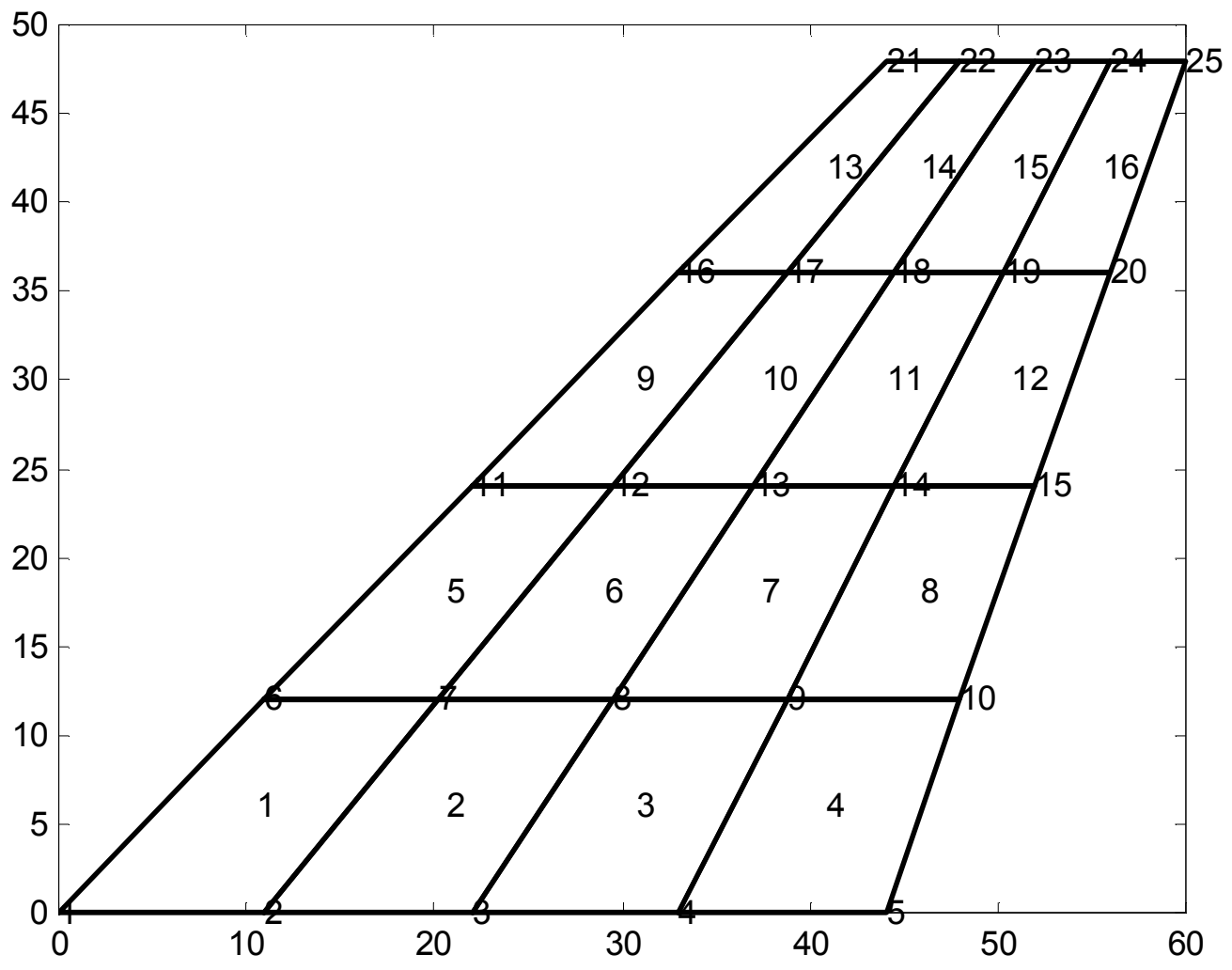

$E = 1$

$\nu = 1/3$

$P = 1$

$t = 1$

Fig. 5.7: Plain stress structure

The Cook's membrane problem: plain stress structure meshed in Fig. 5.7 with 4×4 quadrilateral plain stress elements is clamped along its lower boundary. The upper boundary is subjected to unity force distributed in five concentrated forces act in the nodal points parallel to the boundary . The free-end nodal point displacements $u^{0}_{x^1}$ are listed in Table 5.14. The stiffness matrix of the current calculation is numerically obtained using 3×3 Gaussian integration formula for the $H^{m(m)n(n)}$. These results can be compared favourably with the results obtained by FALT-FEM solution and the results obtained by various finite element solutions listed in /73/ such as Pian-Sumihara element, EAS4, Eas7, etc ...

Table 5.14: displacement at the free end of the plain stress problem $u^{0}_{x^1}(23) = 23.91$

| Nodal point | (21) | (22) | (23) | (24) | (25) |
|---|---|---|---|---|---|
| Gauss integration | 20.6598520 | 18.5898156 | 20.6135840 | 19.4872702 | 24.2071733 |

### 5.5 Preliminary test of frame elements

The framed structure shown in Fig. 5.8 a is analysed using plain versions of the space frame elements introduced in section 4.3. All the element types introduced produce the same results which agree with the exact solution. The framed structure analysed consists of four finite elements with the same material properties $EA = 2\times10^{6}\,kN$ , $EI = 2\times10^{4}\,kN.m^{2}$. The structure is clamped in all the nodal points $^{(2),\,(3),\,(4),\,(5)}$. The elements with the nodal points $^{(2),\,(1)}$ and $^{(1),\,(4)}$ are subjected to linearly distributed loading with the intensity $20\ kN/m$ in the nodal points $^{(2),\,(4)}$ and $40\ kN/m$ in the nodal point $^{(1)}$. The element with the nodal points (3), (1) is subjected to linearly distributed axial force with the intensity $10\ kN/m$ in the nodal points $^{(3)}$ and $20\ kN/m$ in the nodal point $^{(1)}$. Table 5.15 lists the frequencies of the framed structure, the displacements at the nodal point $^{(1)}$, the axial forces, shear forces and bending moments at the nodal points of the structure. Fig.5.8 b, c, d shows the force digrams generated automaticly in the loaded elements.

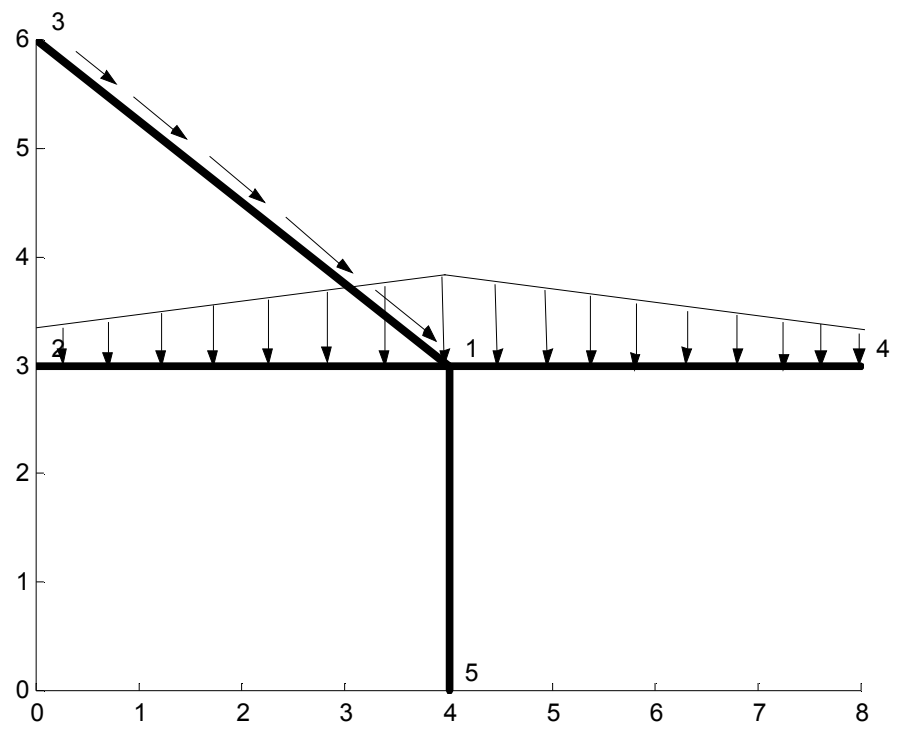

Fig. 5.8 a: Framed structure

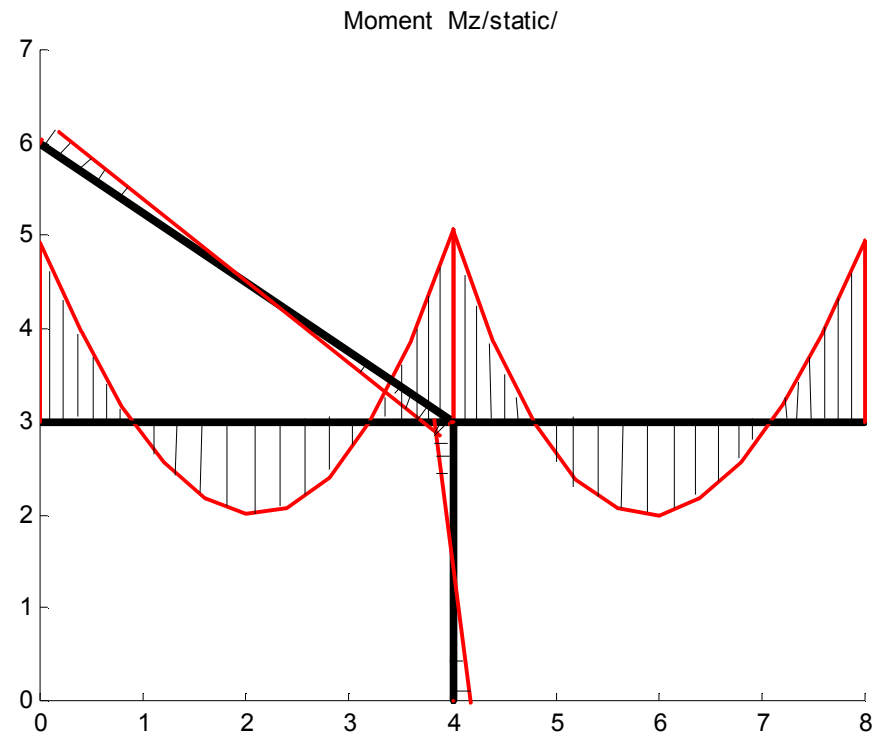

Fig. 5.8 b: Bending moments

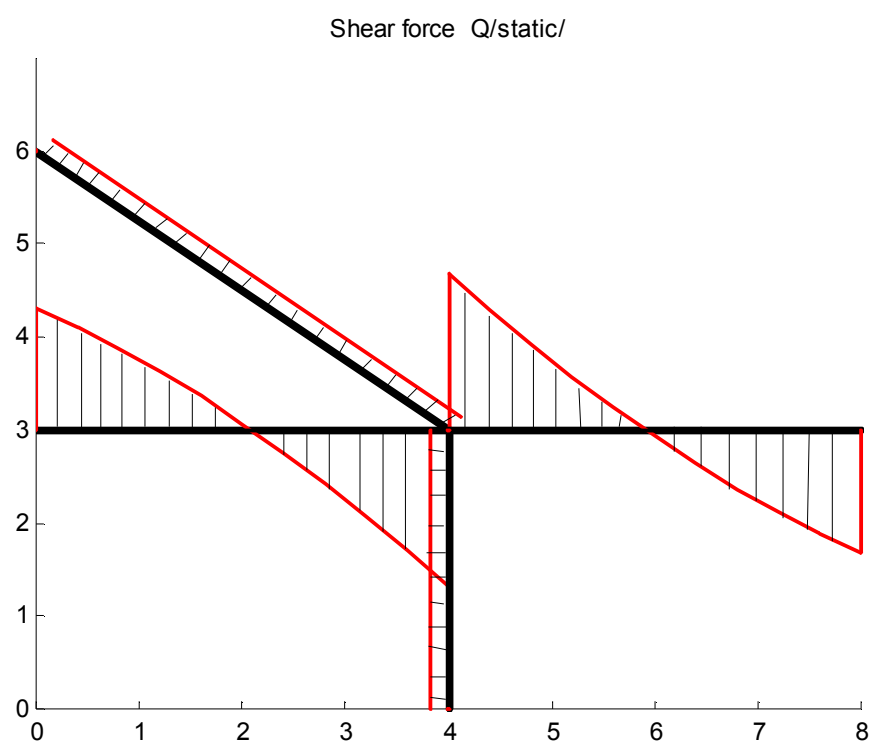

Fig. 5.8 c: Shear forces

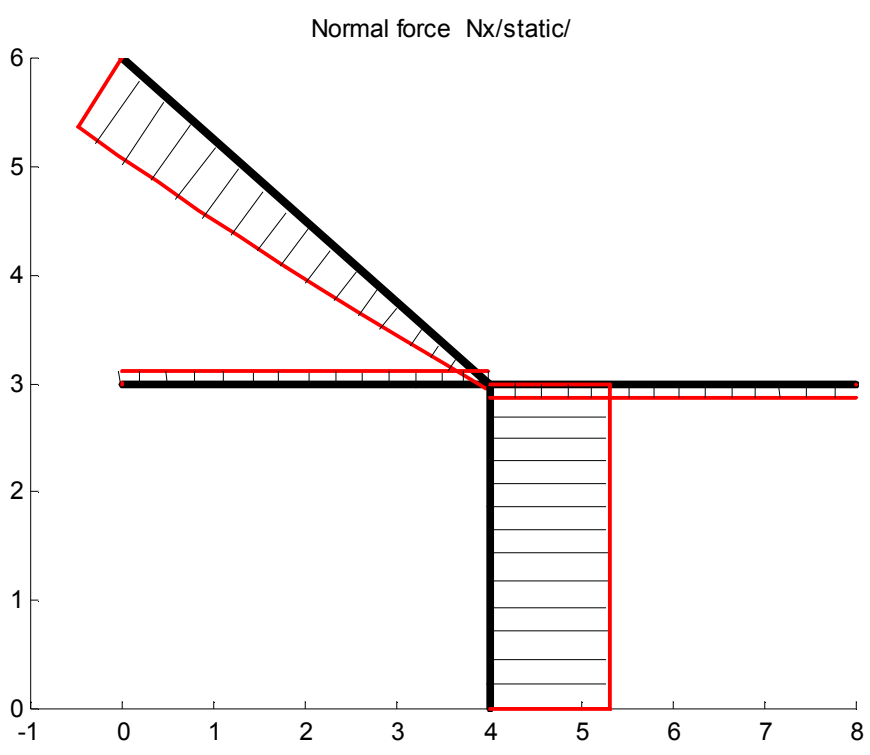

Fig. 5.8 d: Axial forces

Fig. 5.8: Framed structure, geometry, loads and force diagrams

Table 5.15: frequencies, displacements and forces of framed structure.

Eigenvalues D and Eigenvectors V
the 1. angular frequency is= 353.733
the 2. angular frequency is= 772.146
the 3. angular frequency is= 1007.02
statical calculation of the framed structure
nodal displacements

| node | ux | wy | phiz |
|---|---|---|---|
| 1 | -0.00000338 | -0.00019732 | -0.00000874 |
| 2 | 0.00000000 | 0.00000000 | 0.00000000 |
| 3 | 0.00000000 | 0.00000000 | 0.00000000 |
| 4 | 0.00000000 | 0.00000000 | 0.00000000 |
| 5 | 0.00000000 | 0.00000000 | 0.00000000 |

Forces pro element:

| El. | (i) | (k) | Nx(i) | Qy(i) | Mz(i) | Nx(k) | Qy(k) | Mz(k) |
|---|---|---|---|---|---|---|---|---|
| 1 | 2 | 1 | -1.690158 | 52.674397 | -38.725818 | -1.690158 | -67.325603 | -41.361564 |
| 2 | 3 | 1 | 79.607362 | 0.265028 | -0.697522 | 4.607362 | 0.265028 | 0.627616 |
| 3 | 1 | 4 | 1.690158 | 67.194530 | -41.012037 | 1.690158 | -52.805470 | -38.900582 |
| 4 | 1 | 5 | -131.543694 | -0.146556 | 0.278089 | -131.543694 | -0.146556 | -0.161580 |

# 6 Numerical performance and test

## 6.1 Static analysis

### 6.1.1 Morley's skew plate

An a×a simply supported rhombic plate with a skew angle of $30^\circ$ (Moreley's skew plate) is analysed using different finite element meshes. The plate is subjected to a uniformly distributed load $\overline{q}^0 = 1.0$. The geometry properties are assumed to be constant and are chosen such that the plate rigidity is kept equal to unity. As reported by Jirousek (1997), this example is very sensitive to the specified boundary conditions in the final systems of equations of the finite element calculation, especially when a crude mesh is used. The boundary conditions are specified, here, by omitting the transverse displacement along the plate boundary. Table 6.1 shows the convergence of the displacement and the principal moments at the centre of the plate compared with the analytical solution for various finite element meshes. The principal moments are calculated as a nodal average. Comparing the result of the ZDEQ element for the rate of convergence and accuracy with that of its counterparts: HTQ3 /29/, HSQK1 /54/ and the results obtained in /50/, the presented element is found to be satisfactory. Figure 6.1 shows a 3-D plot of the finite element solution of the plate deflection over the plate area obtained by a developed MATLAB Version / 95/ of the program, which was originally written in "C" language.

Table 6.1: Results at the plate centre for a uniformly loaded Moreley's $30^\circ$ rhombic plate

| Mesh | $u^0_{x^3} / \overline{q}^0 a^4 / D$ | $m^{x^1x^1} / \overline{q}^0 a^2 / D$ | $m^{x^2x^2} / \overline{q}^0 a^2 / D$ |
|---|---|---|---|
| 2×2 | 0.00062105 | 0.0224 | 0.0326 |
| 4×4 | 0.00046224 | 0.0158 | 0.0230 |
| 6×6 | 0.00044143 | 0.0139 | 0.0207 |
| 8×8 | 0.00043238 | 0.0127 | 0.0198 |
| 10×10 | 0.00042894 | 0.0124 | 0.0195 |
| 12×12 | 0.00042658 | 0.0122 | 0.0193 |
| 14×14 | 0.00042509 | 0.0121 | 0.0192 |
| Analytical | 0.000408 | 0.0109 | 0.0191 |

Table 6.2: Results at the plate centre for a uniformly loaded M Moreley's $30^\circ$ rhombic plate (TFEQ element)

| Mesh | $u^0_{x^3} / \overline{q}^0 a^4 / D$ | $m^{x^1x^1} / \overline{q}^0 a^2 / D$ | $m^{x^2x^2} / \overline{q}^0 a^2 / D$ |
|---|---|---|---|
| 2×2 | 0.00056517 | 0.0139 | 0.0295 |
| 4×4 | 0.00044737 | 0.0122 | 0.0215 |
| 6×6 | 0.00043235 | 0.0136 | 0.0200 |
| 8×8 | 0.00042483 | 0.0121 | 0.0193 |
| 10×10 | 0.00042172 | 0.0124 | 0.0192 |
| 12×12 | 0.00041976 | 0.0120 | 0.0190 |
| 14×14 | 0.00041853 | 0.0120 | 0.0190 |
| Analytical | 0.000408 | 0.0109 | 0.0191 |

Table 6.3: Results at the plate centre for a uniformly loaded Moreley's $30^\circ$ rhombic plate (JFEQ element)

| Mesh | $u^0_{x^3}/\bar{q}^0a^4/D$ | $m^{x^1x^1}/\bar{q}^0a^2/D$ | $m^{x^2x^2}/\bar{q}^0a^2/D$ |
|---|---|---|---|
| 2×2 | 0.00060969 | 0.0122 | 0.0287 |
| 4×4 | 0.00044720 | 0.0107 | 0.0211 |
| 6×6 | 0.00043309 | 0.0130 | 0.0197 |
| 8×8 | 0.00042527 | 0.0117 | 0.0192 |
| 10×10 | 0.00042223 | 0.0122 | 0.0191 |
| 12×12 | 0.00042022 | 0.0118 | 0.0190 |
| 14×14 | 0.00041898 | 0.0119 | 0.0189 |
| Analytical | 0.000408 | 0.0109 | 0.0191 |

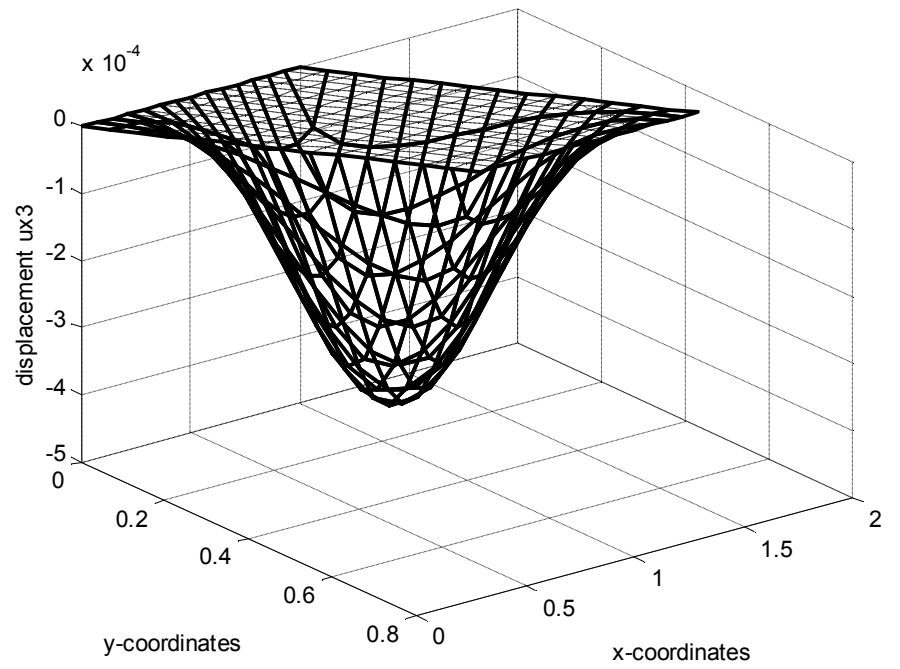

Fig. 6.1: simply supported uniformly loaded rhombic plate meshed by 14×14 elements,
3-D representation of the plate deflection over the plate area

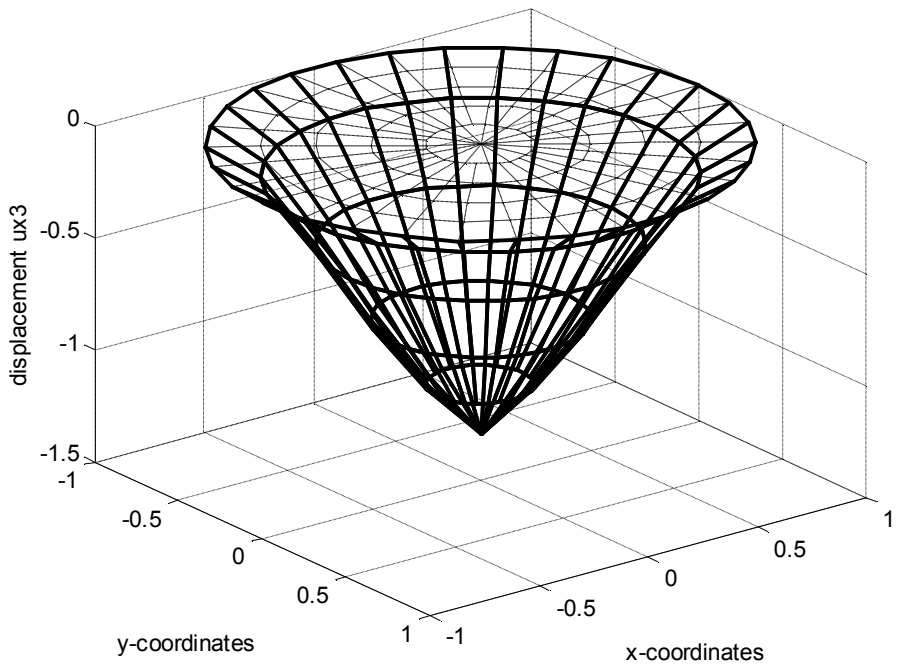

Fig. 6.2: Clamped circular plate uniformly loaded plate meshed by 5×30 elements,
3-D representation of the plate deflection over the plate area

### 6.1.2 Clamped circular plate under uniformly distributed load

A circular plate clamped along its boundary is subjected to a uniformly distributed load. The analytical solution for the Kirchhoff theory is available /74/. The analytical solution is presented beside the current finite element solution for different meshes in table 6.4 The finite element mesh is automatically generated. The radial refinement in the last ring contains only a half number of elements of the other rings . Fig. 6.2 also shows a 3-D plot of the finite element solution of the plate deflection over the plate area.

Table 6.4: Results at the plate centre for a uniformly loaded circular plate

| Mesh | deflection | Deflection(last ring) |
|---|---|---|
| 3×18 | 1.35958580 | 0.97543936 |
| 5×30 | 1.29131897 | 1.07420426 |
| 7×42 | 1.19204559 | 1.06759427 |
| 9×54 | 1.13178675 | 1.05356621 |
| 11×66 | 1.09570914 | 1.04246758 |
| Analytical | 1.00000000 | - |

### 6.1.3 Cantilevered beam with varying cross section

For a clamped trapezoidal free beam with varying cross section of dimensions (1.× (0.05 and 0.1)×0.01) the analytical deflection at the free end of the beam subjected to unity uniformly distributed, neglecting the shear deformation effect, is given by w = 0.00505116. The beam is modelled by different finite element meshes, Fig. 6.3. The results obtained for the deflection at the free end and the principal moments at the fixed end are listed in Tab 6.5a beside finite element solutions obtained using a mixed hybrid formulation (FALT-FEM-Quadrilateral elements), The mechanic basis of FALT-FEM is published in various works /75-90/. The current solutions for both deflection and fixed end bending moment are close to the analytical solutions obtained by integrating the differential equation of an equivalent beam problem. The FALT-FEM solutions are also close in what concerns the bending moments and the displacement. Bearing in mind that the ZDEQ element is a displacement element, its numerical results seem to quite satisfactory.

Tab 6.5 a: Convergence study of cantilevered trapezoidal plate subjected to uniform load

| Beam solution | $u^0_{x^3}$ = 0.00505116 | | $M^{x^2x^2}$ = 0.3333333 | |
|---|---|---|---|---|
| | ZDEQ | FALT-FEM | ZDEQ | FALT-FEM |
| 2×2 | 0.00485700 | 0.004814 | 0.328387 | 0.338435 |
| 4×4 | 0.00493098 | 0.004904 | 0.335333 | 0.342568 |
| 6×6 | 0.00495992 | 0.004936 | 0.340277 | 0.345639 |
| 8×8 | 0.00497313 | 0.004944 | 0.345514 | 0.346564 |
| 10×10 | 0.00498010 | 0.004950 | 0.350826 | 0.350828 |
| 12×12 | 0.00498410 | 0.004955 | 0.355924 | 0.353000 |
| 14×14 | 0.00498657 | 0.004958 | 0.360638 | 0.354771 |
| 16×16 | 00.00498815 | 0.004959 | 0.364891 | 0.356188 |
| 18×18 | 0.00498922 | 0.004962 | 0.368673 | 0.357285 |
| 20×20 | 0.00498992 | 0.004961 | 0.372001 | 0.358252 |

Tab 6.5 b: Convergence study of cantilevered trapezoidal plate subjected to varying distributed load

| Beam solution | $u^0_{x^3}$ = 0.00714286 | | $M^{x^2x^2}$ = 0.0500000 | |
|---|---|---|---|---|
| | ZDEQ | TFEQ | ZDEQ | TFEQ |
| 2×2 | 0.00627667 | 0.00615544 | 0.445288 | 0.467696 |
| 4×4 | 0.00636445 | 0.00629112 | 0.459010 | 0.492046 |
| 6×6 | 0.00640257 | 0.00632255 | 0.466635 | 0.490489 |
| 8×8 | 0.00642031 | 0.00633736 | 0.474084 | 0.485755 |
| 10×10 | 0.00642971 | 0.00634830 | 0.481464 | 0.481879 |
| 12×12 | 0.00643514 | 0.00635747 | 0.488488 | 0.479206 |
| 14×14 | 0.00643850 | 0.00636539 | 0.494957 | 0.477465 |
| 16×16 | 0.00644065 | 0.00637228 | 0.500781 | 0.476375 |
| 18×18 | 0.00644210 | 0.00637830 | 0.505952 | 0.475719 |
| 20×20 | 0.00644306 | 0.00638353 | 0.510498 | 0.475377 |

Table 6.5 b shows a convergence study of the cantilevered trapezoidal plate when the plate is subjected to varying distributed load with the intensity 2.0 kN/m$^2$ at the fixed end and 1.0 kN/m² at the free end. The numerical convergence can be easily observed for both displacements and moments. The obtained solution for the deflection and the moment of the plate meshed by 8×8 elements is presented for the last loading case as 3-D plot in figure 6.3 a and 6.3 b, respectively.

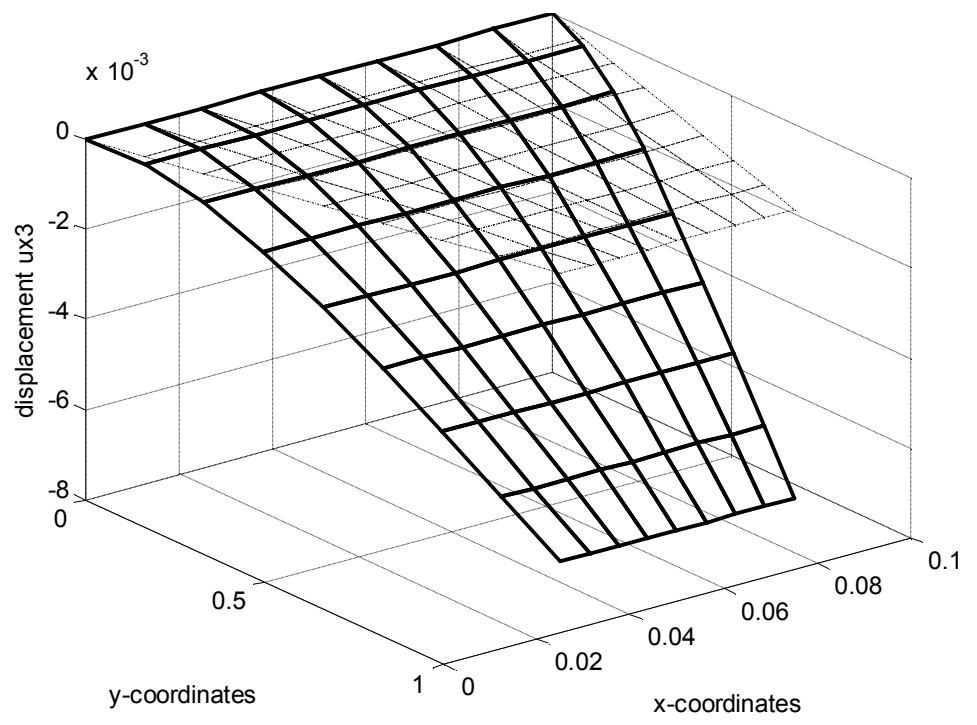

6.3 a: Displacement distribution over the plate area

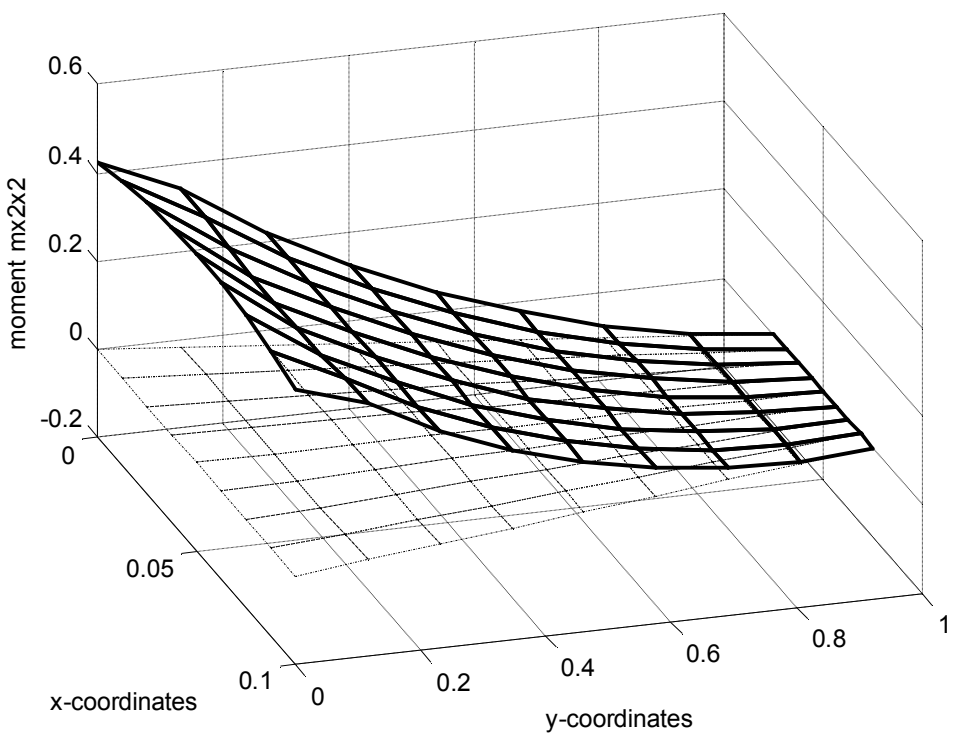

Fig 6.3 b: Moment distribution over the plate area

### 6.2 Eigenvalue analysis

#### 6.2.1 Clamped rhombic plate

A rhombic plate with dimension a×a is fully clamped along its all four boundaries. The plate is subjected to eigenvalue-analysis using the current element for two skew angles ($\alpha = 15^\circ, 45^\circ$ ). The first six normalized natural frequencies ($\omega\ (a/\pi)^2 / \sqrt{\rho h / D}$) are calculated and listed in Tab. 6.6 a and 6.6 b. For a comparison, the result obtained by Rayleih-Ritz method in /91/ and by the isoparametric strip distributed transfer functions method (ISDTFM) in /92/ for a number of strips equal to eight are also included. The Rayleigh-Ritz solution and the ISDTFM-predictions are in good agreement with the current solution. The convergence properties of the finite element solution can be observed easily even for the high frequencies.

#### 6.2.2 Circular plate

A circular plate clamped along its boundary is subjected to eigenvalue analysis using various finite element meshes. The solution produced by the current element is compared for the square root values of the nine normalised angular frequencies ($(\lambda)^2 = \omega / R^2 \sqrt{\rho h / D}$ ) with the analytical solution given in /93/. The results are listed in Tab 6.7. For a comparison purpose, finite element solutions based on a mixed hybrid formulation are listed, also, in same table ( FALT-FEM-elements /83/). The repeated angular frequencies are given only once. Figure 6.4 shows the mode shape corresponds to some eigenvalues of the plate.

Tab 6.6 a: Normalized natural frequencies of a clamped rhombic plate with skew angle, $\alpha = 15^\circ$

| The first six normalized natural frequencies $(\omega\,(a/\pi)^2/\sqrt{\rho h/D})$ | | | | | | |
|---|---|---|---|---|---|---|
| | $\lambda_1$ | $\lambda_2$ | $\lambda_3$ | $\lambda_4$ | $\lambda_5$ | $\lambda_6$ |
| 4×4 | 3.638155 | 6.959558 | 7.802662 | 9.871103 | 13.39054 | 13.99544 |
| 6×6 | 3.746417 | 7.126638 | 7.999162 | 10.33567 | 13.56985 | 14.09434 |
| 8×8 | 3.795381 | 7.222491 | 8.131333 | 10.61506 | 13.71028 | 14.24467 |
| 10×10 | 3.820370 | 7.275413 | 8.207348 | 10.77131 | 13.81488 | 14.36945 |
| 12×12 | 3.834628 | 7.305553 | 8.253148 | 10.86436 | 13.88428 | 14.45521 |
| 14×14 | 3.843476 | 7.326585 | 8.282432 | 10.92348 | 13.93085 | 14.51378 |
| Ritz /91/ | 3.870 | 7.388 | 8.377 | 11.12 | 14.09 | 14.72 |
| ISDTFM | 3.8696 | 7.3888 | 8.3749 | 11.1095 | 14.0998 | 14.7256 |

Tab 6.6b: Normalized natural frequencies of a clamped rhombic plate with skew angle $\alpha = 45^\circ$

| The first six normalized natural frequencies $(\omega\,(a/\pi)^2/\sqrt{\rho h/D})$ | | | | | | |
|---|---|---|---|---|---|---|
| | $\lambda_1$ | $\lambda_2$ | $\lambda_3$ | $\lambda_4$ | $\lambda_5$ | $\lambda_6$ |
| 4×4 | 5.905763 | 9.756622 | 12.57157 | 13.39730 | 18.77089 | 21.09772 |
| 6×6 | 6.281911 | 10.17367 | 13.76718 | 14.45491 | 18.02721 | 21.45991 |
| 8×8 | 6.434682 | 10.40988 | 14.26752 | 15.01516 | 18.72718 | 21.98451 |
| 10×10 | 6.509364 | 10.53501 | 14.52233 | 15.31546 | 19.12232 | 22.35738 |
| 12×12 | 6.551183 | 10.60779 | 14.66791 | 15.49099 | 19.35373 | 22.59592 |
| 14×14 | 6.576902 | 10.65355 | 14.75866 | 15.60154 | 19.49955 | 22.75304 |
| Ritz /91/ | 6.680 | 10.8 | 15.1 | 16.1 | 20.2 | 23.5 |
| ISDTFM | 6.6603 | 10.8125 | 15.1024 | 15.9832 | 20.1130 | 23.3721 |

Table 6.7: The values ( $\lambda = \sqrt{\omega / R^2 \sqrt{\rho h / D}}$ ) of the nine normalised angular frequencies of a clamped circular plate

| | $\lambda^1$ | $\lambda^2, \lambda^3$ | $\lambda^4, \lambda^5$ | $\lambda^6$ | $\lambda^7, \lambda^8$ | $\lambda^9, \lambda^{10}$ | $\lambda^{11}, \lambda^{12}$ | $\lambda^{13}$ | $\lambda^{14}, \lambda^{15}$ |
|---|---|---|---|---|---|---|---|---|---|
| Exact | 3.196 | 4.611 | 5.906 | 6.306 | 7.144 | 7.799 | 8.347 | 9.197 | 9.256 |
| 24×5 | 3.109 | 4.5651 | 5.818 | 6.015 | 7.000 | 7.684 | 8.118 | 8.978 | 9.025 |
| 36×7 | 3.144 | 4.589 | 5.859 | 6.110 | 7.074 | 7.737 | 8.234 | 9.068 | 9.095 |
| 24×5 /83/ | 3.217 | 4.642 | 5.948 | 6.371 | 7.208 | 7.889 | 8.428 | 9.305 | 9.585 |
| 36×7 /83/ | 3.205 | 4.624 | 5.924 | 6.332 | 7.171 | 7.831 | 8.380 | 9.239 | 9.501 |

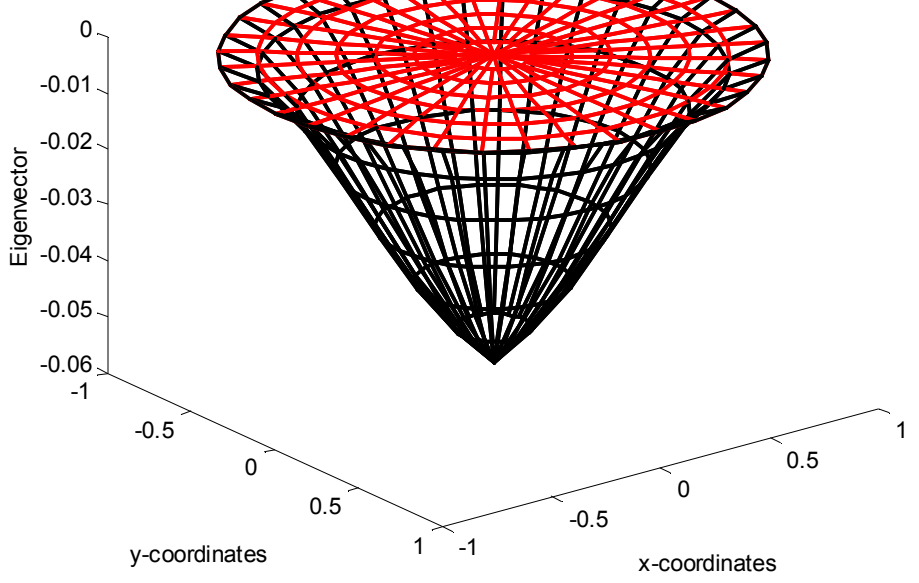

The mode shape to the 1.eigenvalue

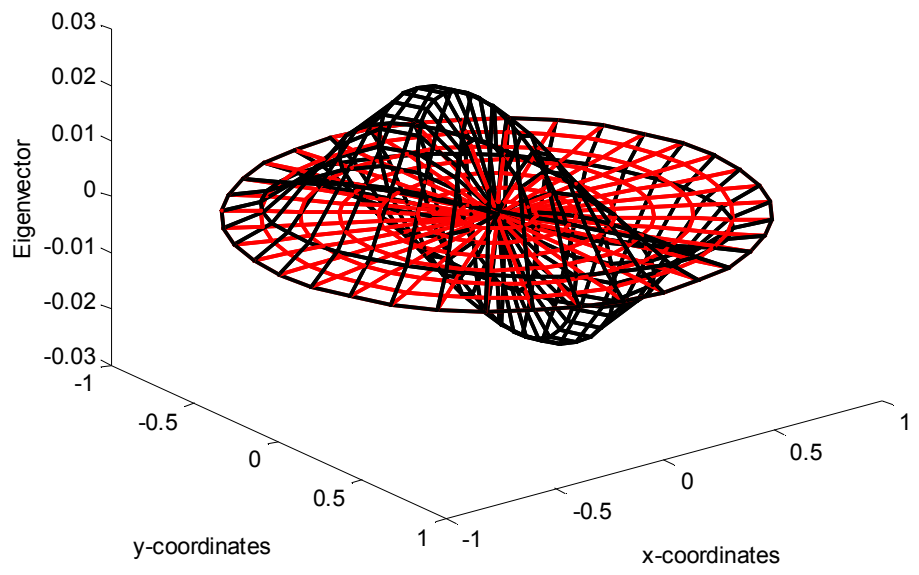

The mode shape to the 2.eigenvalue.

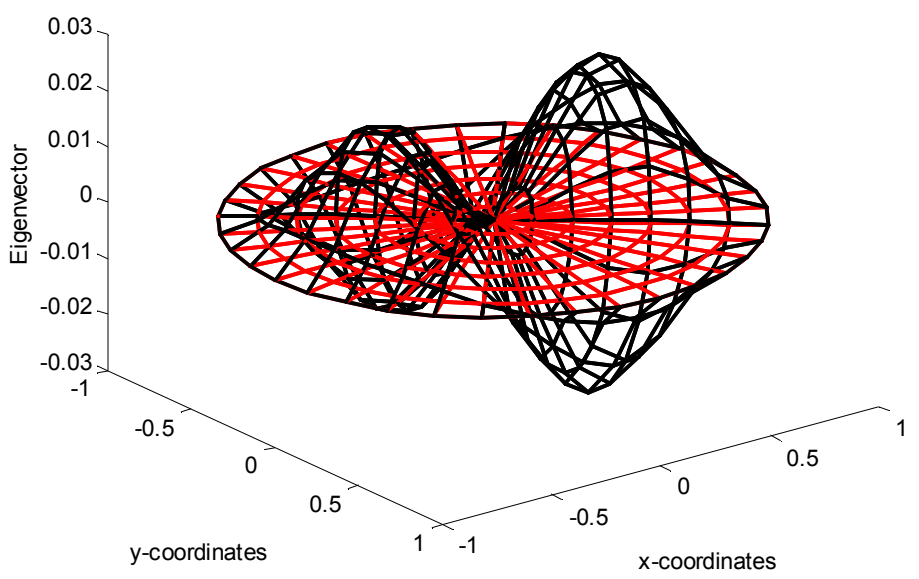

The mode shape to the 4.eigenvalue

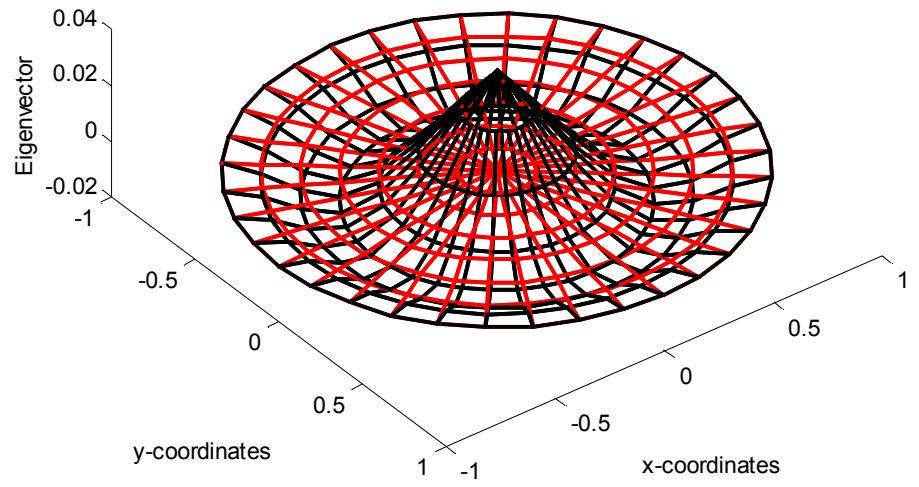

The mode shape to the 6.eigenvalue

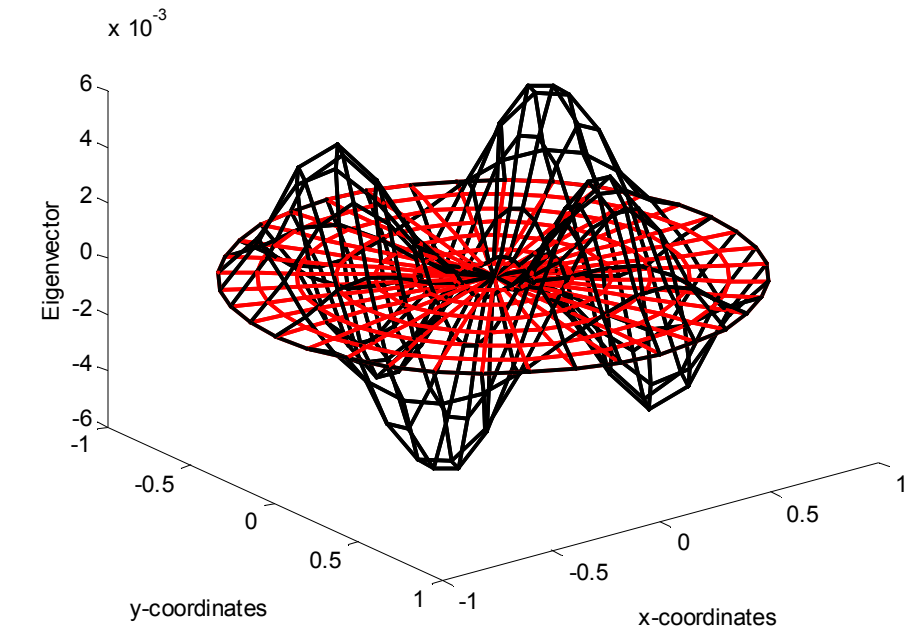

The mode shape to the 7.eigenvalue

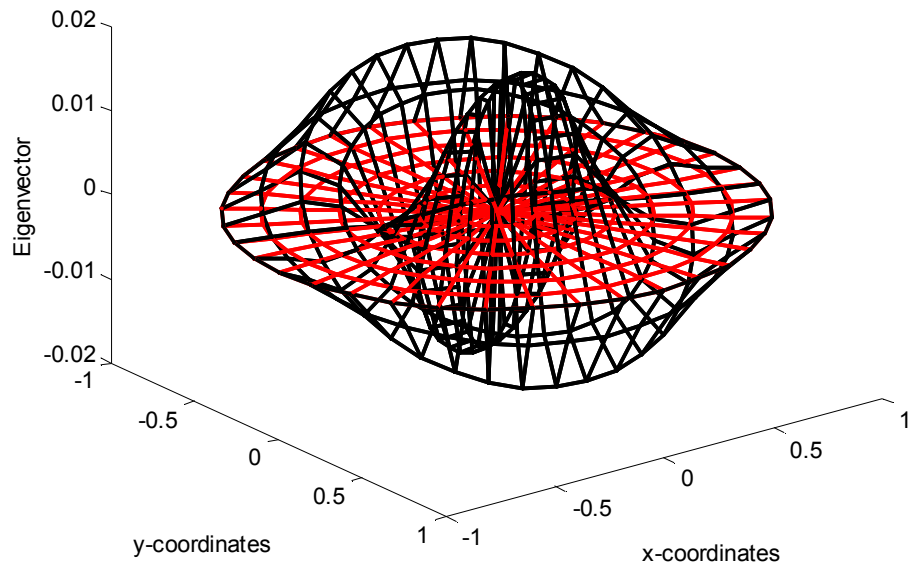

The mode shape to the 9.eigenvalue

Fig. 6.4 : Mode shape corresponding to some chosen eigenvalues of a clamped circular plate

### 6.2.3 Cantilevered trapezoidal plates

Two cantilevered trapezoidal Plates with thickness h=0.2, unit plate-rigidity, Poisson ratio $\upsilon = 0.3$ and mass density $\rho = 5$ as shown in Fig. 6.5 are subjected to eigenvalue analysis using program version written in MATLAB-language code. The results for the normalized angular frequency $\omega / \sqrt{D / \rho h a^4}$ for different element meshes are listed in Tab 6.8 a and Tab 6.8 b. As may be seen, a very fast convergence is obtained. The first three digits do not change practically after the second mesh refinement. The solutions for the same plate models are reported in /93/. They are close to the computed values and, also, obtained numerically using the Ritz-Method. Figure 6.6 shows the first sixth mode shapes of the trapezoidal plate of figure 6.5 b meshed by 14×14 elements.

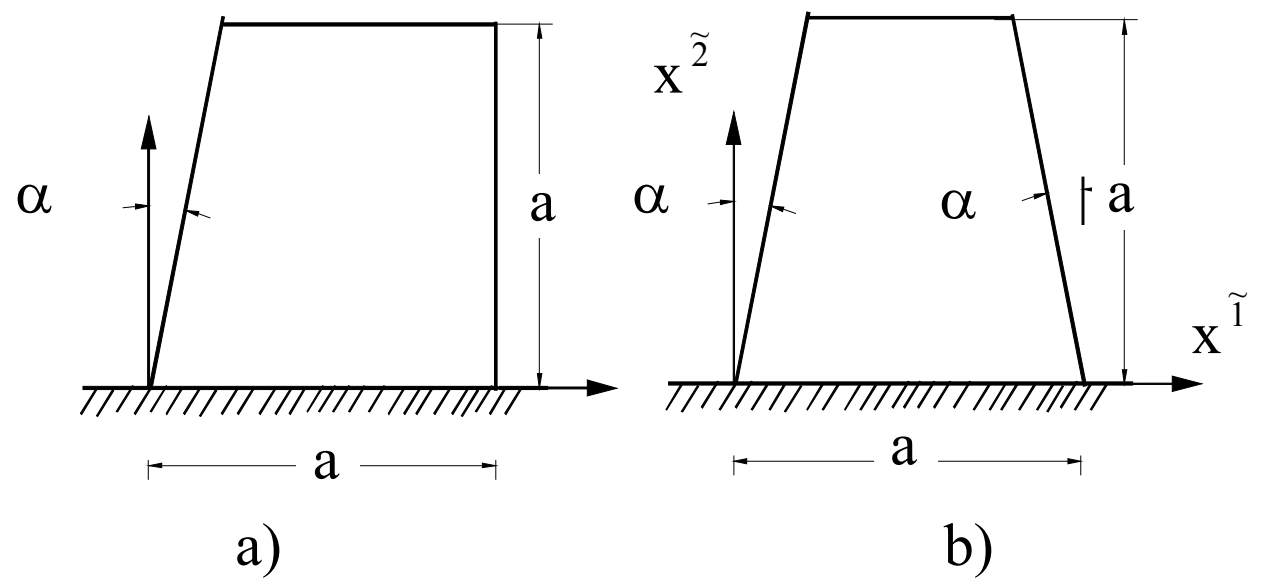

Fig. 6.5: Cantilevered trapezoidal plate

Table 6.8 a : Normalized angular frequency of cantilevered trapezoidal plate of Fig. 6.5 a

| Normalized angular frequency ($\omega / \sqrt{D / \rho h a^4}$) | | | | |
|---|---|---|---|---|
| mesh | $\alpha = 9^\circ$ | $\alpha = 18^\circ$ | $\alpha = 27^\circ$ | $\alpha = 36^\circ$ |
| 2×2 | 3.596789 | 3.773692 | 4.021291 | 4.471740 |
| 4×4 | 3.623460 | 3.806608 | 4.064293 | 4.539906 |
| 6×6 | 3.627078 | 3.811708 | 4.072257 | 4.553765 |
| 8×8 | 3.628170 | 3.813492 | 4.075264 | 4.558931 |
| 10×10 | 3.628638 | 3.814368 | 4.076783 | 4.561456 |
| 12×12 | 3.628882 | 3.814881 | 4.077683 | 4.562896 |
| 14×14 | 3.629028 | 3.815218 | 4.078268 | 4.563806 |
| (Ritz) /93/ | 3.706 | 3.910 | 4.243 | 4.822 |

Table 6.8 b: Normalized angular frequency of cantilevered trapezoidal plate of Fig. 6.5 b

| Normalized angular frequency ($\omega / \sqrt{D / \rho h a^4}$) | | | | |
|---|---|---|---|---|
| Mesh | $\alpha = 6^\circ$ | $\alpha = 12^\circ$ | $\alpha = 18^\circ$ | $\alpha = 24^\circ$ |
| 2×2 | 3.663307 | 3.961824 | 4.441699 | 5.450370 |
| 4×4 | 3.697106 | 4.017142 | 4.534444 | 5.626892 |
| 6×6 | 3.701906 | 4.025737 | 4.550078 | 5.658471 |
| 8×8 | 3.703316 | 4.028496 | 4.555403 | 5.669419 |
| 10×10 | 3.703892 | 4.029744 | 4.557928 | 5.674638 |
| 12×12 | 3.704178 | 4.030432 | 4.559368 | 5.677609 |
| 14×14 | 3.704339 | 4.030863 | 4.560295 | 5.679504 |
| (Ritz) /93/ | 3.718 | 4.153 | 4.750 | 5.995 |

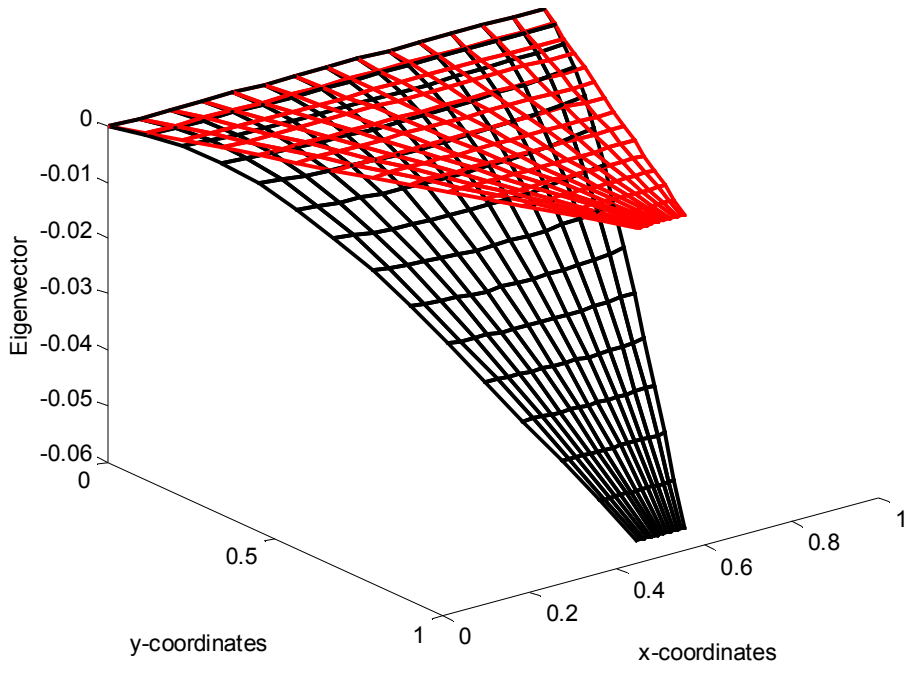

The mode shape to the 1. eigenvalue

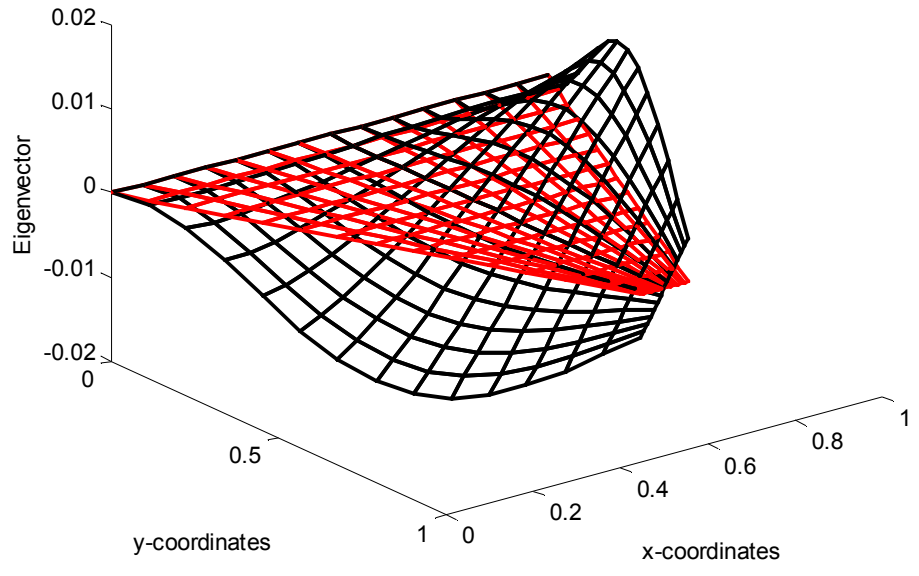

The mode shape to the 2. eigenvalue

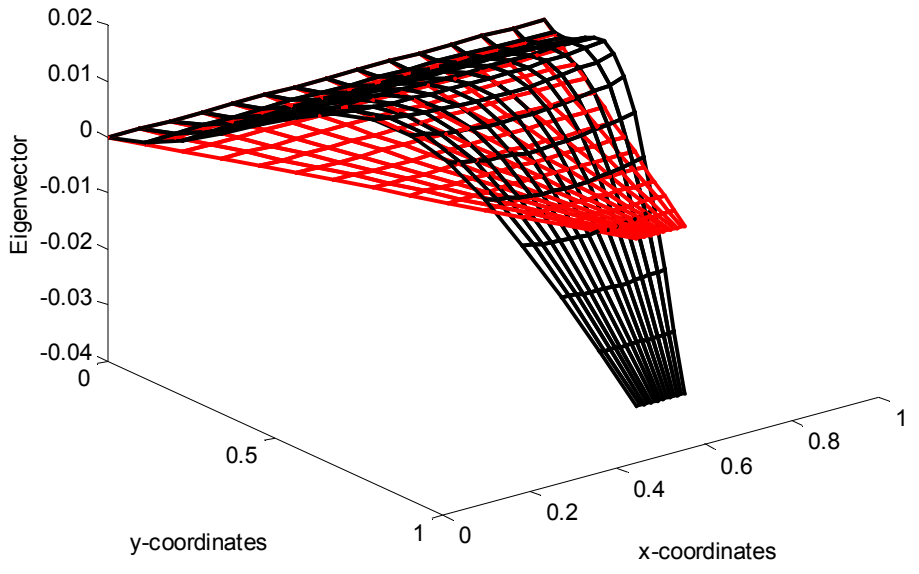

The mode shape to the 3. eigenvalue

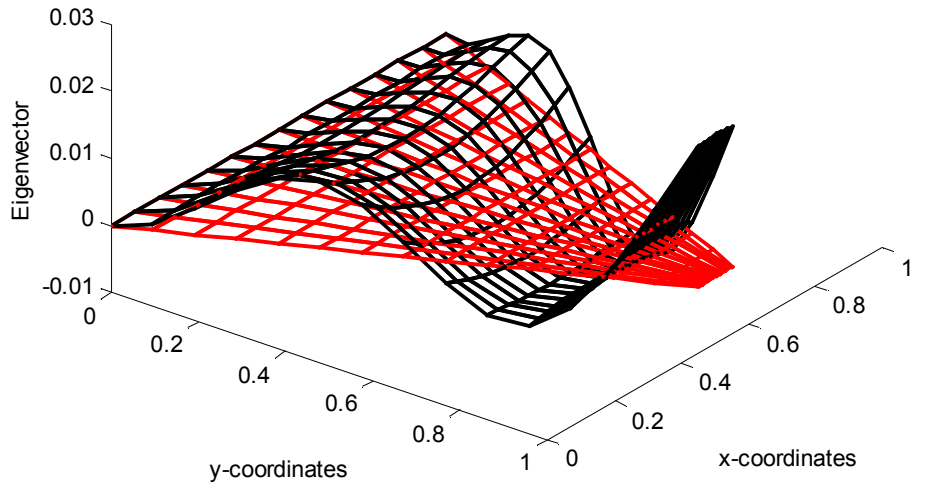

The mode shape to the 4. eigenvalue

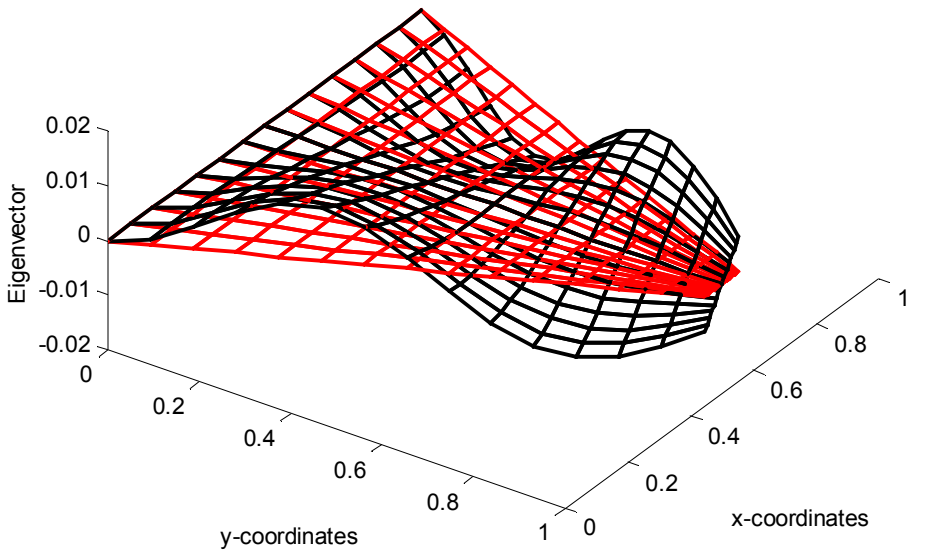

The mode shape to the 5. eigenvalue

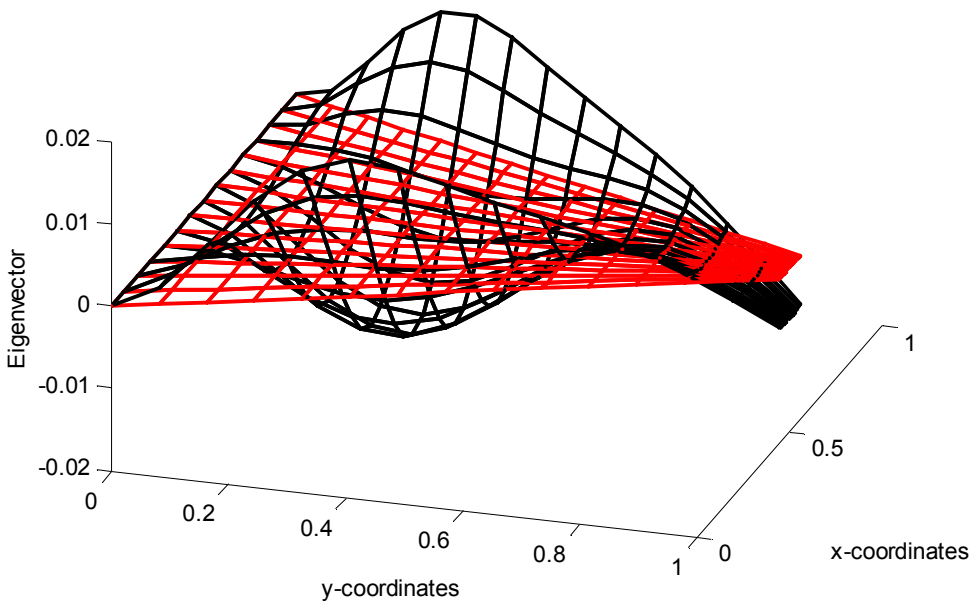

The mode shape to the 6. eigenvalue

Fig. 6.6 : Mode shapes corresponding to the first sixth eigenvalues of a cantilevered trapezoidal plate

## 7 Concluding Remarks

Differnet variational displacement and stress concepts associated with reduced Trefftz type approaches have been presented. The displacement concept operates on the natural boundary conditions enforced in an integral sense. It recovers the conventional displacement and hybrid stress finite element concept. The element approximation basis can be constructed using a modified geometrical interpolation technique or using a weak hybrid technique as well as a boundary technique by applying the Gauss divergence theorem or the equivalency of virtual work between the entire boundary and the static and kinematic boundaries, respectively. The stress concept operates on the essential boundary conditions enforced in an integral sense. It recovers the stress and hybrid displacement finite element concept. The element approximation basis can be constructed in an analogous way to that of the displacement formulation in form of stress components.

A combination of the modified interpolation technique and the use of the Gauss divergence theorem or the equivalency of virtual work is possible.

In order to apply the reduced Trefftz type approach in the framework of the finite element method it is possible to select the approximation functions in the element local Cartesian coordinate system. This facilitates considerably the satisfaction of the differential equation. We can take advantages from the numerical integration by transforming the trial approximating functions and their derivatives to the curvilinear coordinates using the isoparametric transformation and after that performing the element integrals in these coordinates. The origin of the local coordinate system is located at the element centre.

The reduced Trefftz type approach can be applied using a domain technique or hybrid and boundary techniques. Whenever it is possible to satisfy the continuity requirement using the modified geometrical interpolation technique, the essential boundary conditions at finite element level are used to construct 'interpolation' functions over an element that approximate the internal displacement field. This field constitutes the displacement approximation basis that allows the application of the proposed approach using a computational procedure similar to the one used in the conventional displacement finite element method based. Two alternative hybrid and boundary procedures can be applied to enforce conformity and inter-element continuity when the interpolation cannot ensure the continuity requirement. The frame function concept is applied in both procedures.

It is expected, also, that the different computational techniques will yield the same results provided that all the conditions assumed in the variation are strictly observed.

A general procedure for formatting finite elements of general geometric form is presented. The basic idea of this procedure consists in using the natural coordinate system only for defining the geometry of the element and to perform the integration in a mapped biunit square. For defining the approximation function (displacement or stress function) a local coordinate system, defined using the directions of the element covariant and contravariant base vectors, is used. The definition of the coordinate system in such way is essential in order to ensure element invariance, convergence, and insensitivity to nodal point numbering.

## Berichte des Lehrstuhls für Statik der Technischen Universität Dresden

**Summary**

This work presents variational concepts associated with reduced Trefftz type approaches and discusses the interrelationship between various concepts of the displacement, hybrid and Trefftz methods. The basic concept of the displacement version of the reduced Trefftz method operates on the natural boundary conditions enforced in an integral form whereas the stress version of the reduced Trefftz type approach operates on the essential boundary conditions enforced in an integral sense. The application of the method proposed in the framework of the finite element method is briefly outlined. The methods used by the reduced Trefftz type approach for enforcing conformity and interelement continuity between neighboured elements are also discussed. Comparisons with other known methods for the same purpose are performed. General strategy for developing finite elements of general geometric form such as quadrilateral elements with invariance properties is presented. The basic idea of this strategy consists in using the natural coordinate system only for defining the element geometry and performing the element integration in the biunit interval. For defining the approximation functions a local coordinate system defined from the directions of the covariant base vectors and the perpendicular contravariant base vectors computed in the geometric centre of the element is used. This strategy can also be used to implement other versions of finite elements and other forms of finite elements. Different numerical calculations and comparisons in the linear statics and kinetics are performed in order to assess the convergence and the numerical performance of finite elements developed by applying the reduced Trefftz type approach.

**Zusammenfassung**

Die vorliegende Arbeit präsentiert im Zusammenhang mit der reduzierten Trefftz-Methode Variationskonzepte und diskutiert Zusammenhänge zwischen den verschiedenen Variationskonzepten der Verschiebungsmethoden, der hybriden Methoden und der Trefftz-Methoden. Die Verschiebungsversion der reduzierten Trefftz-Methode basiert auf der Integralform der natürlichen Randbedingungen, während die Spannungsversion der reduzierten Trefftz-Methode auf der Integralform der wesentlichen Randbedingungen aufbaut. Der Einsatz der vorgestellten Methode im Rahmen der Finiten Elemente Methode ist ausführlich dargelegt. Die Vorgehensweisen zur Erfüllung der Kontinuität zwischen benachbarten finiten Elementen werden diskutiert. Sie werden mit anderen in der Literatur bekannten Vorgehensweisen verglichen. Eine allgemeine Strategie zur Entwicklung finiter Elemente mit Konvergenz- und Invarianzeigenschaften ist dargestellt. Der Grundgedanke dieser Strategie basiert auf der Verwendung des natürlichen Koordinatensystems allein für die Aufspannung der Geometrie und die Ausführung der Elementintegration. Für die Definition der Approximationsfunktionen wird ein lokales kartesisches Koordinatensystem verwendet, welches durch die Richtungen der kovarianten und kontravarianten Basisvektoren, die im geometrischen Mittelpunkt des Elementes berechnet sind, definiert ist. Dieses Vorgehen ist allgemeingültig und kann zur Entwicklung allgemeiner Elementformen eingesetzt werden. Numerische Ergebnisse aus dem Bereich der linearen Statik und Kinetik demonstrieren die Konvergenz- und Invarianzeigenschaften verschiedener Elementtypen,die auf der Basis der reduzierten Trefftz-Methode entwickelt wurden.